\documentclass{article}
\usepackage[super,sort&compress,comma]{natbib}
\usepackage[version=3]{mhchem}
\usepackage[left=1.5cm, right=1.5cm, top=1.785cm, bottom=2.0cm]{geometry}
\usepackage{balance}
\usepackage{mathptmx}
\usepackage{authblk}
\usepackage{sectsty}
\usepackage{graphicx}
\usepackage{lastpage}
\usepackage[format=plain,justification=justified,singlelinecheck=false,font={stretch=1.125,small,sf},labelfont=bf,labelsep=space]{caption}
\usepackage{float}
\usepackage{fancyhdr}
\usepackage{fnpos}
\usepackage[english]{babel}
\addto{\captionsenglish}{%
  
}
\usepackage{array}
\usepackage{droidsans}
\usepackage{charter}
\usepackage[T1]{fontenc}
\usepackage[usenames,dvipsnames]{xcolor}
\usepackage{setspace}
\usepackage[compact]{titlesec}
\usepackage{hyperref}

\usepackage{epstopdf}

\definecolor{cream}{RGB}{222,217,201}

\title{Model of luminescence and delayed luminescence correlated blinking in single CsPbBr\textsubscript{3} nanocrystals.}

\author[1]{Eduard A. Podshivaylov}
\author[1]{Alexander M. Shekhin}
\author[2,3]{Maria A. Kniazeva}
\author[3]{Alexander O. Tarasevich}
\author[4]{Elizaveta V. Sapozhnikova}
\author[5]{Anatoly P. Pushkarev}
\author[6]{Ivan Yu. Eremchev}
\author[3,6,7]{Andrei V. Naumov}
\author[1,*]{Pavel A. Frantsuzov}

\affil[1]{Voevodsky Institute of Chemical Kinetics and Combustion SB RAS, 630090, Novosibirsk, Russia}
\affil[2]{HSE University, 101000, Moscow, Russia}
\affil[3]{Lebedev Physical Institute of the Russian Academy of Sciences, Troitsk Branch, 108840,
Moscow, Russia}
\affil[4]{School of Physics and Engineering, ITMO University, 197101, St. Petersburg, Russia.}
\affil[5]{Skolkovo Institute of Science and Technology, 30/1 Bolshoy Boulevard, 121205 Moscow, Russia}
\affil[6]{Institute of Spectroscopy of the Russian Academy of Sciences, 108840, Moscow, Russia}
\affil[7]{Moscow Pedagogical State University (MPGU), 119435, Moscow, Russia}

\date{}

\begin{document}
\maketitle

\abstract{Cesium lead halide nanocrystals and quantum dots are prominent materials for different types of applications because of their remarkable photophysical properties. However, they are also known to exhibit the same effects observed for non-perovskite colloidal semiconductor quantum dots such as blinking, photobleaching, delayed luminescence, \textit{etc}. In this study we reveal the correlations between fast and delayed emission components for both the intensity and characteristic decay time for single CsPbBr\textsubscript{3} nanocrystals. In order to explain the phenomena observed, we propose a novel model of single CsPbBr\textsubscript{3} nanocrystals luminescence blinking based on the hypothesis of slow variations in the electron-phonon coupling.} \\

\section{Introduction}
All-inorganic cesium lead halide (CsPbX\textsubscript{3}) perovskite quantum dots (PQDs) and nanocrystals (PNCs) are promising materials for a variety of photonic applications. These materials have a high quantum yield, a tunable emission wavelength with a narrow linewidth throughout the visible range, low trap density, high carrier mobility, and diffusion length. \cite{Swarnkar2015,Raino2022} The broadly applicable CsPbBr\textsubscript{3} perovskite nanocrystals are employed as a material for the production of supercapacitors,\cite{THAKUR2021,Pang2023,Yadav2023} LEDs \cite{Du2017,Yuan2018,LAN2019}, self-assembled superlattices for superradiance sources,\cite{Krieg2021,Rainò2018} photovoltaics,\cite{Zai2018,Ullah2021,Gao2019} microlasers,\cite{Xie2022,Wang2018,Yu2022} \textit{etc}. Colloidal single quantum dots serve as efficient sources of single-photon emission.\cite{Zhu2022,Park2015,Hu2015}

However, single CsPbBr\textsubscript{3} PNCs and PQDs are known to exhibit the well-known phenomenon of luminescence blinking. \cite{Seth2016, Seth2018, Seth2019, Li2018, Palstra2021, Gibson2018, Paul2023} To date, three main mechanisms of single colloidal semiconductor quantum dot and nanocrystal blinking have been reported.

The first mechanism is the Auger mechanism, also called the charging  mechanism or A-type blinking, was first proposed by Efros \textit{et al.}\cite{EfrosPRL1997} The mechanism involves the photoinduced charge exchange of the quantum dot, so that a neutral QD has a high photoluminescence quantum yield, while the luminescence intensity of a charged QD is suppressed, resulting in the appearance of one or two dim states.

The second mechanism is the trapping mechanism (TM), also called the band edge carrier (BC) blinking, which was proposed by Frantsuzov and Marcus.\cite{FrantsuzovPRB2005} The explanation for luminescence blinking in this mechanism is associated with a photoinduced slow change in the rate constant of the nonradiative recombination caused by rearrangements of surface atoms. With this type of blinking, the distribution of luminescence intensity is continuous and proportional to the luminescence lifetime.

The third mechanism is the hot carrier trapping mechanism, also called HC-blinking, which was first proposed by Galland \textit{et al.} \cite{KlimovNature2011} This is a modified trapping mechanism, suggesting that hot carriers can be captured in a metastable trap before they are cooled. This blinking mechanism results in a large change in luminescence intensity, but the luminescence lifetime remains constant.

A generally accepted method for determining which mechanism leads to the blinking in a particular system is the fluorescence lifetime-intensity distribution (FLID). The FLID is calculated by the standard procedure. trajectory bins with a certain level of radiation intensity are identified. They are then divided into several groups.  The luminescence lifetime is determined for each group. In the Auger mechanism, the FLID consists of two spots corresponding to the bright and dim states, sometimes connected by a non-linear curve. In the trapping mechanism, the FLID dependence is linear, while in the hot carrier trapping mechanism the lifetime remains constant with variations in intensity.
 Blinking according to all three mechanisms has been observed in a single CsPbBr\textsubscript{3} PNC. \cite{Ahmed2019}

We have recently proposed a quantitative model based on the electron-phonon coupling variation (EPV model) for the TM blinking of single colloidal cadmium selenide quantum dots.\cite{Podshivaylov2023} The main assumption of the model is that the electron-phonon coupling slowly fluctuates over time, leading to massive changes in the rate of the nonradiative recombination.
The model is free from the shortcomings of the widely accepted  TM based model of multiple recombination centers (MRC)\cite{FrantsuzovPRL2009,VolkanNL2010,FrantsuzovNL2013},
\textit{e.g.} the inability to quantitatively describe the photon distribution function (PDF), as well as the shortcomings of the Frantsuzov-Marcus model \cite{FrantsuzovPRB2005,BusovOS2019}.
 The EPV model allows one to quantitatively fit the PDF, the power spectral density (PSD) of the blinking signal, and the linear FLID dependence. However, the model does not take into account the blinking of the delayed luminescence  that is observed for single colloidal quantum dots\cite{Hinterding2020}.
  A similar model, using slow fluctuations of the electron-phonon interaction, was used to explain spectral diffusion in single cadmium selenide quantum dots\cite{Podshivaylov2019}. The aim of this work is to present a model of CsPbBr\textsubscript{3} PNC luminescence blinking based on the same idea.

\section{Experimental}

\subsection{Synthesis}
CsPbBr\textsubscript{3} PNCs with a mean size of 11-12 nm were obtained according to the “hot injection” protocol reported by Protesescu \textit{et al.}\cite{Protesescu2018} 0.18 mmol of lead bromide (PbBr\textsubscript{2}) and 7 ml of 1-octadecene (ODE) were loaded into a 100 ml flask and stirred under vacuum at 120°C for 40 min. 1.52 mmol of oleylamine (OLAM)  and 1.58 mmol of oleic acid (OA) were then added to the stirred mixture and the flask was filled with N\textsubscript{2} gas. The mixture was heated up to 140 °C to give a clear solution, followed by heating up to 180 °C. When the temperature reached the set up value, 0.05 mmol of preliminary prepared cesium oleate (CsOA) clear solution was swiftly injected. As a result, CsPbBr\textsubscript{3} PNCs were formed within ten seconds, the reaction was quenched by using an ice bath and further purified from byproducts.

To prepare cesium oleate (CsOA) Cs\textsubscript{2}CO\textsubscript{3} (0.814 g) and ODE (40 ml) were loaded into a 100 ml flask and dried under vacuum at 120°C for 1 h. Afterwards, OA (2.5 ml) was injected into the mixture and then heated to 150°C under stirring and N\textsubscript{2} protection, yielding a clear cesium oleate (CsOA) 0.125 M precursor solution. The solution was cooled down to room temperature for storage, and preheated up to 120°C before usage.

To separate CsPbBr\textsubscript{3} PNCs from byproducts, 8 ml of ODE was added to the collected colloidal solution, after which it was centrifuged at 5000 rpm for 5 minutes to give sediment under supernatant solution containing byproducts that was discarded. Thereafter, the sediment was redispersed in toluene (15 ml).
 STEM images of the obtained nanocrystals are provided within the Supplementary Note 1 (ESI\dag)
A toluene solution of CsPbBr\textsubscript{3} PNCs was spin-coated onto a clean non-luminescent cover glass.

\subsection{Experimental setup}
Measurements of individual nanocrystals were carried out using a fluorescence microscope operating in time-correlated single photon counting (TCSPC) mode with pulsed excitation.

A precision three-dimensional piezo-driven stage (NanoScanTechnology) was used to accurately position ($\sim$ 0.6 nm) the sample in our homebuilt luminescent microscope-spectrometer. Focusing of the laser beam and collecting of the fluorescent signal were performed using a high-power immersion microlens Nikon Plan Apo 100X, 1.49 NA. A femtosecond optical parametric oscillator TOPOL (Avesta) was used to excite the studied NCs at a wavelength of 400 nm (second harmonic of the signal wave) at a pulse repetition frequency of 1 MHz. The excitation laser intensity was attenuated using neutral spectral density filters (Thorlabs). The power densities of the excitation radiation were set to low (the average number of electron-hole pairs $\langle N_{e-h} \rangle$ generated per laser pulse is much less than unity) to minimize the undesirable effect of photoinduced degradation leading to changes in the size and photophysical properties of the perovskite PNCs.\cite{Baitova2023} A Semrock 405 nm dichroic beam splitter and a Semrock 409 nm blocking edge long-pass filter were installed in front of the registration system to separate the PNC fluorescence signal from the scattered laser radiation and to increase image contrast.

The arrival times of luminescence single photons and laser synchropulses were registered in time-correlated single photon counting (TCSPC) mode in the Hanbury Brown and Twiss optical scheme, which consisted of a broadband 50 $\%$ interference beam splitter (Thorlabs), two identical SPAD detectors (MicroPhotonDevices, and QE $\sim$ 50 $\%$ at 500 nm, DCR ~ 10 Hz) and TCSPC electronics (Time Tagger Ultra by Swabian Instruments, 10  ps time resolution).

A highly sensitive cooled (-80°C) EMCCD-camera (Andor Ixon Ultra, QE $\sim$ 90 $\%$ at 500 nm, dark noise ~ 0.05 counts/pixel/s) was used to record single perovskite images in wide-field mode and to control the NC position during long-term measurements in confocal mode. An imaging spectrometer (SOL Instruments MS5204i), equipped with a cooled CCD camera (HS 101H, QE $\sim$ 90 $\%$), was used to measure the fluorescence spectra of single perovskite NCs (spectral resolution of ~ 0.2 nm for a diffraction grating of 300 grooves/mm).

\section{Results}
\subsection{Data processing}

\begin{figure}[H]
    \centering
    \includegraphics[width=1\linewidth]{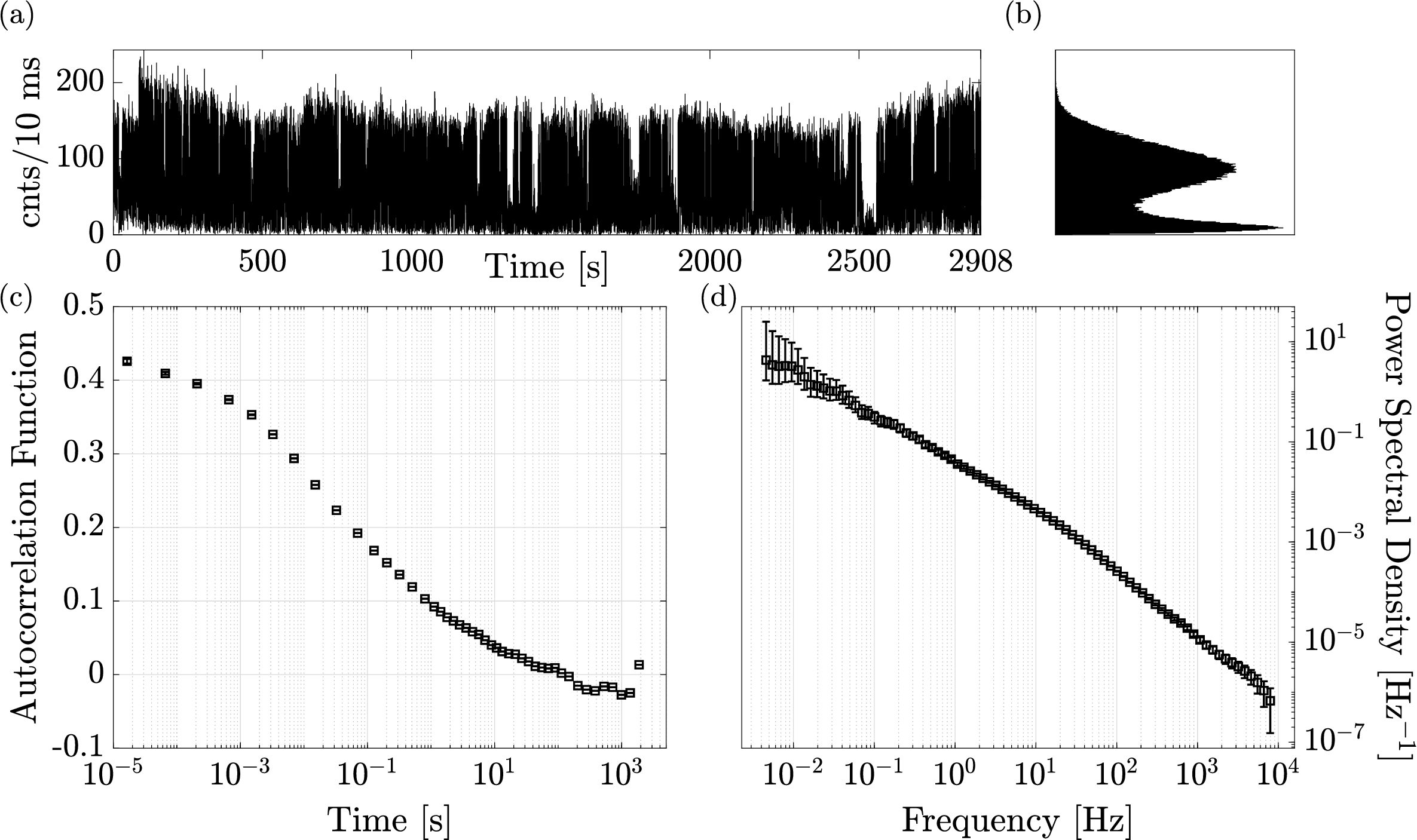}
    \caption{Data obtained from the experiments with single photon counting. (a) PL intensity time trajectory. (b) Photon distribution function (c) Autocorrelation function estimation and (d) its corresponding power spectral density.}
    \label{fig1}
\end{figure}

Time-correlated single photon counting (TCSPC) data were collected using the Hanbury-Brown and Twiss scheme to obtain the intensity trajectory and photon distribution function (PDF), as well as the long-term autocorrelation function (ACF) (ACF was calculated according to Ref.~\citenum{Laurence2006}). The power spectral density (PSD) of the luminescence blinking process was calculated using the recently proposed modified Blackman-Tukey method for the Fourier transform of the ACF.\cite{Podshivaylov2023} The results are shown in Fig.~\ref{fig1}. It is clearly seen that this method of estimating the PSD allows calculating the PSD in the range of 6-7 orders of magnitude in the frequency domain.   The PSD is a convenient tool for describing long-term fluctuations of emission of single nanocrystals. In the work \citenum{Podshivaylov2023} the PDF and PSD together with the FLID were proposed as basic quantities characterizing the blinking process.

\begin{figure}[H]
    \centering
    \includegraphics[width=.99\linewidth]{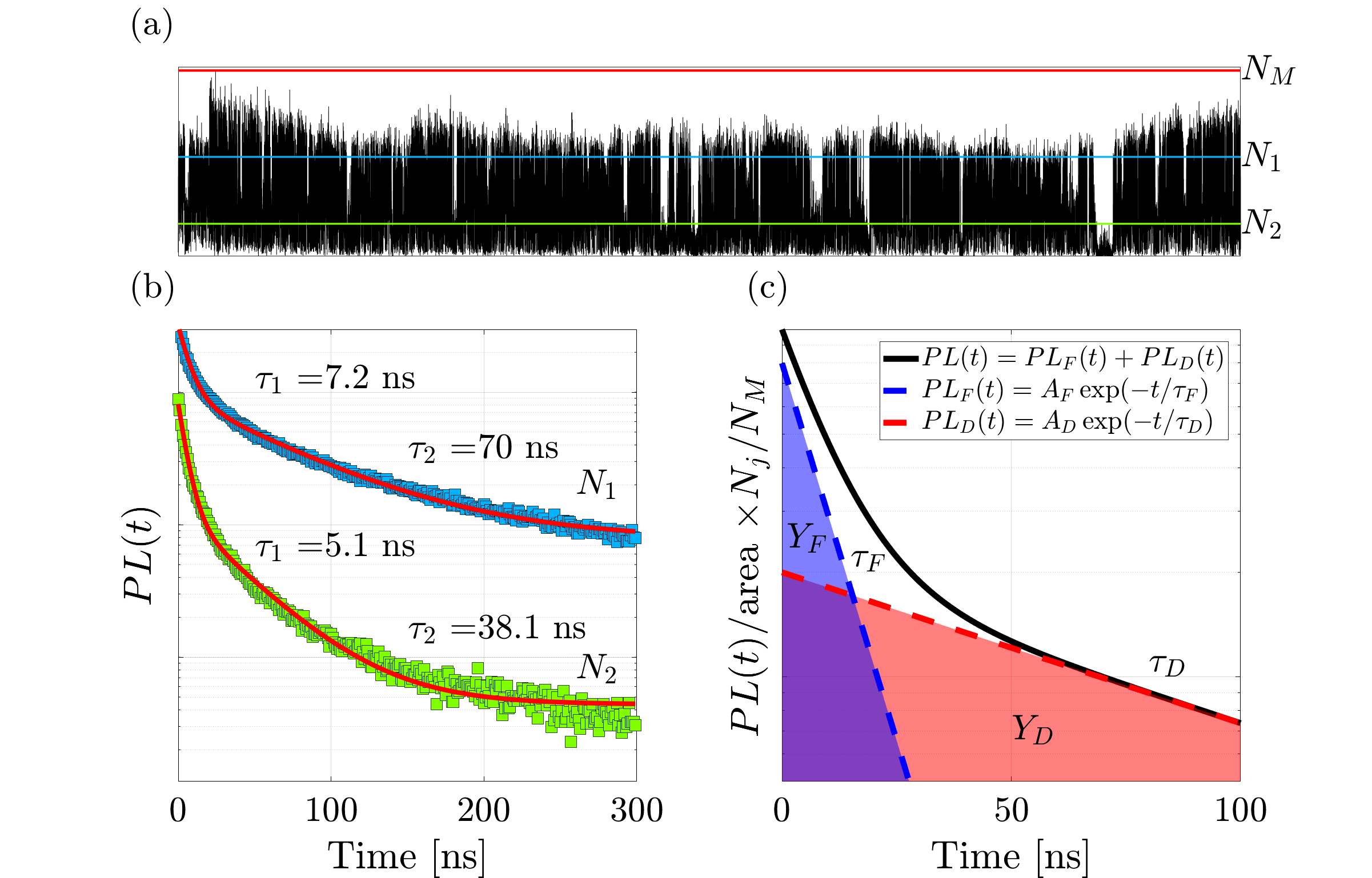}
    \caption{ (a) PL intensity time trajectory (black line) with 2 selected levels (blue and green lines). The red line marks the level that corresponds to the maximum intensity. (b) PL decay curves for selected levels (colored squares) and their fit with biexponential function using Eq.\ref{biexp} (red line). (c) Modeled biexponential PL decay (black line). The blue and red colors correspond to the fast and delayed PL components. The ordinate axis in graphs (b) and (c) is shown on a logarithmic scale. The parameters of the modeled curve are $A_F = 0.8$, $\tau_F =10$ ns, $A_D = 0.2$, $\tau_D =100$ ns. The slopes of the blue and red lines give the delay times of the fast and delayed components, respectively. The areas under the blue and red lines show the corresponding relative yields.}
    \label{fig2}
\end{figure}

To analyze the PL decay curves beyond the trajectory-averaging, we have selected two levels in the luminescence blinking trajectory (as shown in Fig.~\ref{fig2}a) with the number of counts per bin of $N_1$ and $N_2$. The characteristic width of the selected levels is 2-3 photons per level. The maximum number of counts per bin $N_M$ is marked with a red line. The PL decay curves can be well approximated using the biexponential decay in the following form:

\begin{equation}
    PL(t) = \frac{a_F}{\tau_F}e^{-t/\tau_F} + \frac{a_D}{\tau_D}e^{-t/\tau_D} + b,
    \label{biexp}
\end{equation}
where $a_F,a_D$ are the fast and delayed components amplitudes, $\tau_F,\tau_D$ are the fast and delayed components characteristic times, $b$ is the background level. The PL decay curves and their fit with Eq.\ref{biexp} are shown in Fig.~\ref{fig2}b. The PL decay curve shape is very different from the monoexponential law usually observed for cadmium selenide based quantum dots.

In general, four independent parameters are needed to describe biexponential PL decay curves: $a_F$, $a_D$, $\tau_F$ and $\tau_D$. When normalizing the PL decay curves to their area and to the total relative yield $N_j/N_M$, the area under the individual components can be interpreted as the relative yields of the fast and delayed components, as shown in Fig.~\ref{fig2}c. Thus, four other parameters can be obtained: $Y_F$, $Y_D$, $\tau_F$ and $\tau_D$, where $Y_F$, $Y_D$ are the fast and delayed components' relative yields, respectively. More details on the approximation of the biexponential decay and the estimation of relative yields can be found in the Supplementary Note 2 (ESI\dag).

\begin{figure}[H]
    \centering
    \includegraphics[width=1\linewidth]{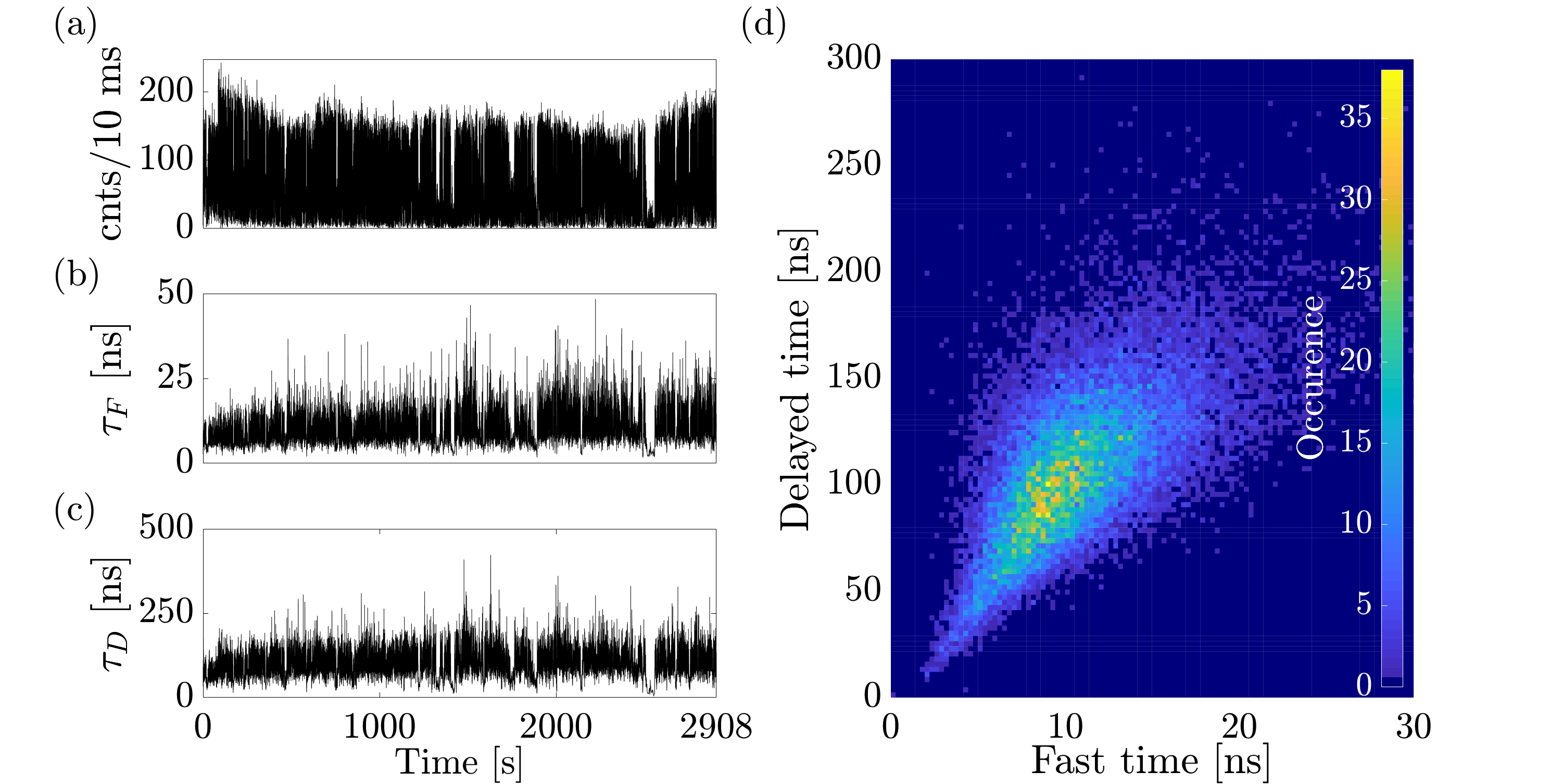}
    \caption{(a) PL intensity trajectory. (b) Fast component characteristic time trajectory. (c) Delayed component characteristic time trajectory. (d) Fast time -- delayed time two-dimensional distribution.}
    \label{fig3}
\end{figure}

The PL decay curves were then constructed and fitted with Eq.\ref{biexp} for each successive arrival of 1000 photons. Thus, we were able to construct trajectories of changes over time for the fast and delayed components PL decay time.   As can be concluded from  Fig.~\ref{fig3}(a-c), the luminescence intensity and the times of fast luminescence and delayed luminescence correlate with each other. The two-dimensional  fast time -- delayed time distribution  (Fig.~\ref{fig3}d) shows a strong correlation between the two characteristic times.
To date, there has been no evidence of such a correlation in the literature. This correlation for perovskite nanocrystals is very different from the observed connection between luminescence components for cadmium selenium based quantum dots. \cite{Rabouw2015, Hinterding2020}

In order to accurately study the dependencies of these characteristic times on the corresponding emission intensities, we carried out a procedure similar to that used in the FLID, where a biexponential fitting function (\ref{biexp}) is used instead of a monoexponential function. The processing details can be found in the Supplementary Note 2 (ESI\dag).

\begin{figure}[H]
    \centering
    \includegraphics[width=1\linewidth]{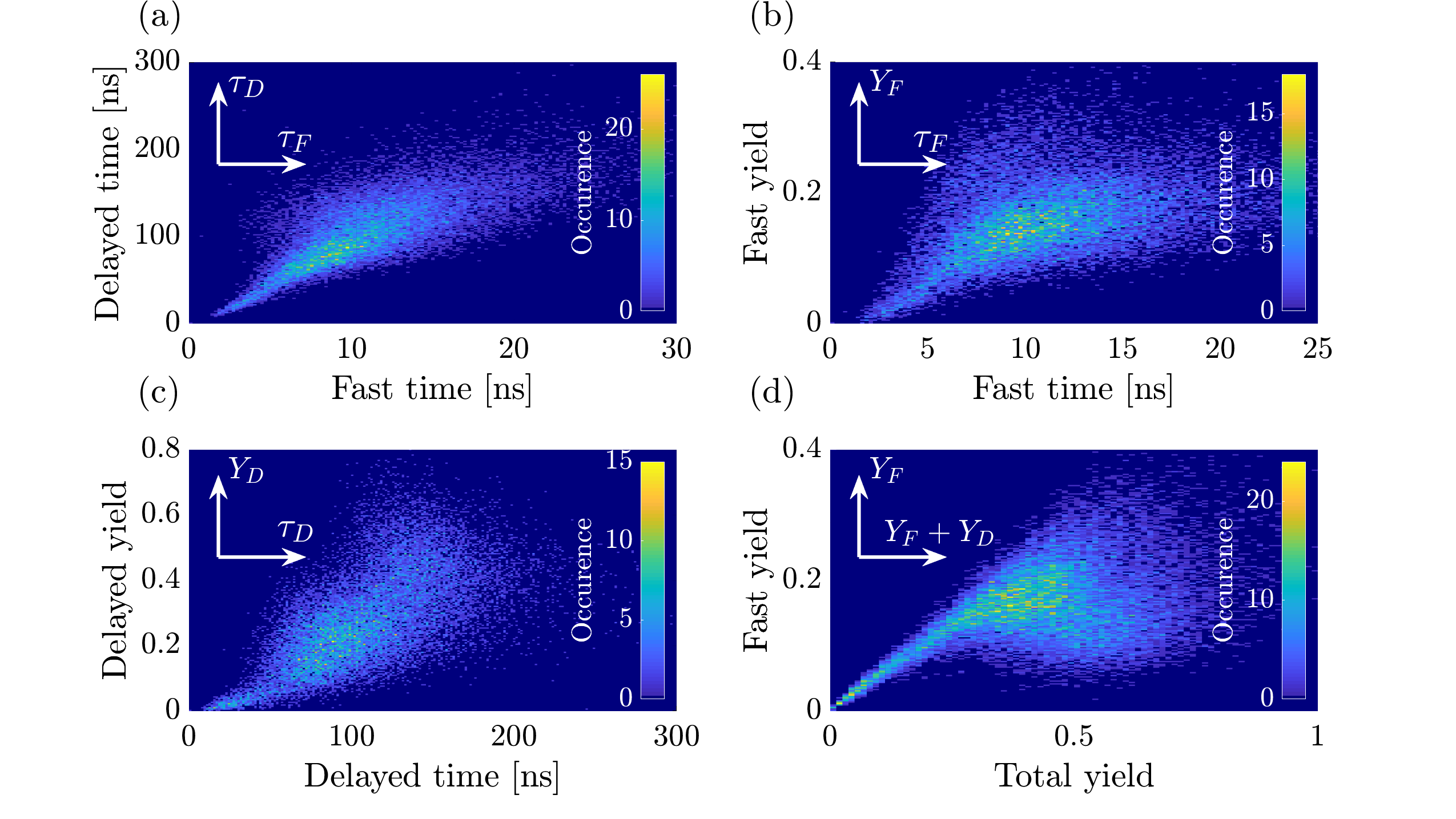}
    \caption{Two-dimensional distributions obtained by the FLID-like procedure. (a) Fast time -- delayed time distribution (b) Fast time --  fast component relative yield distribution(c) Delayed time -- delayed component relative yield distribution (d) Total relative yield -- short component relative yield distribution.}
    \label{fig4}
\end{figure}

Fig.~\ref{fig4} shows the two-dimensional distributions obtained by this procedure.  Fig.~\ref{fig4}a presents the fast time -- the delayed time distribution, which was previously shown in Fig.~\ref{fig3}d.  Fig.~\ref{fig4}b shows the fast time --  fast component relative yield distribution, which  corresponds to the standard FLID method. As expected for TM blinking, it shows an almost linear relationship between the fast time and fast component luminescence intensity. Fig.~\ref{fig4}c shows a strong non-linear relationship between the intensity and the characteristic time of delayed luminescence. Fig.~\ref{fig4}d shows the total yield - fast component relative yield distribution. It is evident that at values of the total yield above 0.3,  the  fast component relative yield reaches saturation. This can be attributed to a situation where the intensity of the delayed luminescence continues to fluctuate with time at a fixed intensity of the fast component contribution.

\begin{figure}[H]
    \centering  \includegraphics[width=1\linewidth]{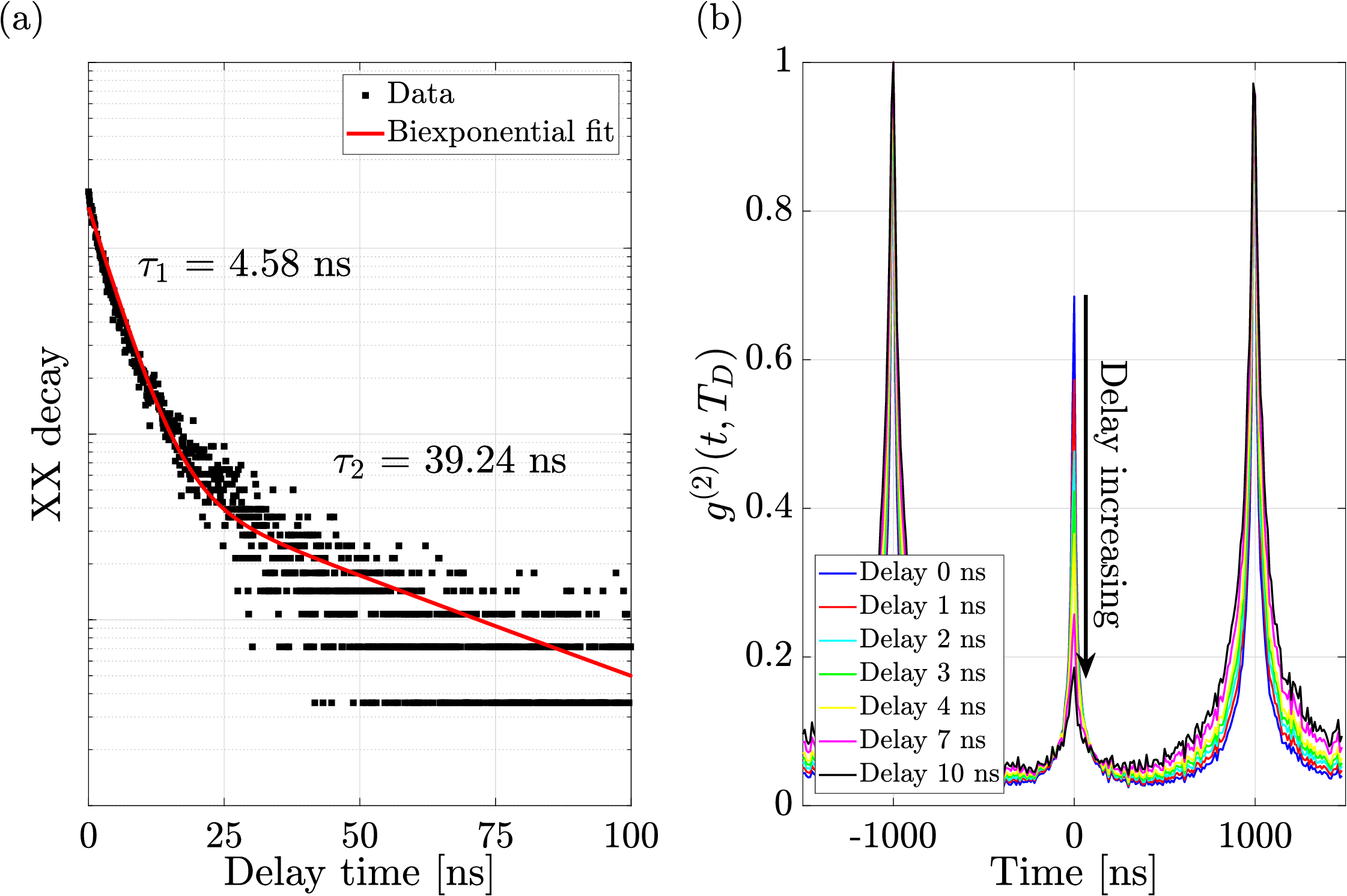}
    \caption{ (a) Distribution of arrival times of the first photon from a pair detected on two detectors (black squares and circles) and their biexponential fit (red line). (b) Normalized on maximum cross-correlation function at different minimal delay times $T_D$.}
    \label{fig5}
\end{figure}

To assess the influence of the biexciton on the resulting distributions, we also estimated the characteristic recombination time of the biexciton.For this purpose, we considered only events when both detectors in the HBT scheme were triggered as a result of a single excitation pulse. Then, the delay time distribution was constructed for the photon registered first.
Using this method, we almost exclude from consideration the part of the decay curve associated with exciton recombination.
 The details on this procedure can be found in the Supplementary Note 3 (ESI\dag). The calculated distribution is shown in Fig.~\ref{fig5}a.  We fitted the resulting distribution with the biexponential decay law.
 This fitting gives two characteristic times of 4.58 ns and 39.2 ns, as shown in Fig.~\ref{fig5}a. The first time can be attributed to the characteristic biexiton recombination time, while the second time refers to exciton recombination (fast and delayed) followed by the registration of an uncorrelated photon from the background noise. Thus, the estimated biexciton lifetime is shorter than the lifetime we observe in the two-dimensional distributions for the short component characteristic time (around 10 ns). We can conclude that the correlation is not related to the biexciton recombination.

The characteristic biexciton decay times described in the literature usually are of the order of hundreds of picoseconds or less.\cite{Castañeda2016,deJong2017,Eperon2018,Sonnichsen2021,Makarov2016}
However, in the work~\citenum{Li2018} higher values of characteristic times of the biexciton nonradiative recombination are given in the range from 250 ps to 1 ns.
A biexcitonic lifetime of about 4 ns was observed for CsPbBr nanocrystals\textsubscript{3} in Ref.~\citenum{Mi2023}.
 These long times were explained by the formation of a trapped exciton, in the presence of which the Auger recombination was strongly suppressed. Our observations suggest that we are most likely dealing with a similar condition.

 We have also tested  antibunching of the single PNC emission by the evaluation of the cross-correlation function $g^{(2)}(t,T_D)$, where $t$ is the time shift between the two photons and $T_D$ is the minimal allowed detection delay after the excitation pulse for the photons considered. This method was used earlier in Refs.~\citenum{Nair2011,Park2017,Eremchev2023}. Figure \ref{fig5}(b) shows the normalized on maximum cross-correlation functions of the luminescence intensity $g^{(2)}(t,T_D)$ with a sequential increase in the allowed delay of photon arrival. The $g^{(2)}(0,T_D)$ monotonically decreases with the increase of the minimal allowed photon detection delay. The obvious antibunching effect is achieved on approximately the same time scale that we obtained from the procedure for estimating the biexciton lifetime. The very presence of the antibunching effect indicates that we are dealing with a single nanocrystal.

\subsection{Model}

\begin{figure}[H]
    \centering
    \includegraphics[width=0.7\linewidth]{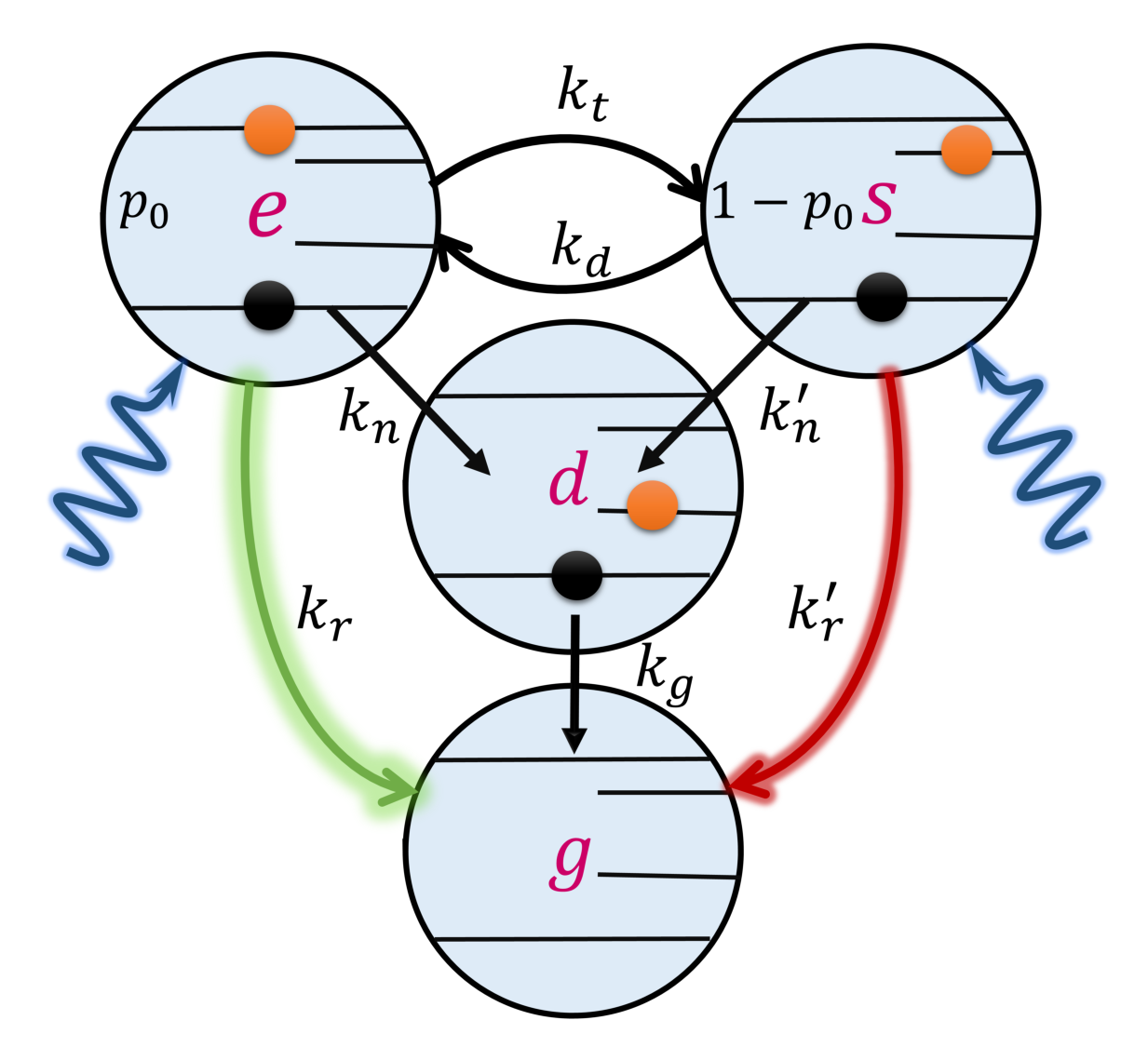}
    \caption{The proposed kinetic mechanism of the PNC excitation relaxation process in a single PNC. After the initial excitation (marked with blue arrows), the electron can be found in one of two possible states: excited (e) and shallow (s) with the probabilities $p_0$ and $1-p_0$, respectively. The electron can undergo a transition between these two states with rate constants of $k_t$ (e→s) and $k_d$ (s→e).  The electron  can be captured by a deep trap level (d) from both states  with rate constants of $k_n$ (e→d) and $k_n'$ (s→d) or recombine with the hole with photon emission, thus occurring in the ground (g) state with rate constants $k_r$ (e→g) and $k_r'$ (t→g). The transition from deep trap to a ground state with the rate constant $k_g$ (d→g) corresponds to a non-radiative electron-hole recombination.}
    \label{fig6}
\end{figure}

To explain the properties of single PNC blinking, we propose a kinetic mechanism for excitation recombination, shown in Fig.~\ref{fig6}. The mechanism involves the presence of two electronic states within the band gap, one of which is a shallow trap and the other is a deep trap. The deep trap likely occurs due to a defect on the surface of the PNC. The possible nature of the shallow trap will be discussed in the following section.
It is assumed that the PNC is excited by a laser pulse with a probability of $p_0$ to appear in the excited state (the electron is in the lower state in the conduction band and the hole is in the higher state in the valence band) or with the probability of $1-p_0$ to appear at the state when the electron is trapped in the shallow trap.

The electron in the excited state can recombine radiatively with a hole with a rate constant of $k_r$, be trapped in a shallow trap with a rate constant of $k_t$, or be captured in a deep trap  with a rate constant of $k_n$. An electron trapped in a shallow level trap can return to the upper state by a thermally activated process with a rate constant of $k_d$.
The rate constants are related by the detailed balance equation  $k_d=k_t e^{-\Delta E/k_bT}$, where $\Delta E$ is the energy difference between the excited and shallow states. Here we use the assumption that the levels' degeneracies are the same.
An electron from a shallow trap can radiatively recombine with a hole with a rate constant $k_r'$, and can also be captured in a deep trap with a rate constant $k_n'$. The trapped electron can than nonradiatively recombine with a hole with a rate constant $k_g$.

The kinetics of the excitation relaxation within this mechanism can be described by the following system of equations:

\begin{equation}
    \begin{cases}
        \frac{d p_e}{dt} = -(k_r + k_n +k_t) p_e +k_d p_t, \\
        \frac{d p_t}{dt} = -(k_r'+k_n' + k_d) p_t +k_t p_e,
    \end{cases}
    \label{sys}
\end{equation}
where $p_e$ and $p_t$ are the probabilities of finding the PNC in the excited and trapped states at time $t$, respectively. The initial conditions are as follows:

\begin{equation}
p_e(0) = p_0; \quad p_t(0)=1-p_0.
\end{equation}

The absolute quantum yield definition is as follows:

\begin{equation}
    Y =  k_r \int\limits_0^\infty p_e (t) dt + k_r' \int\limits_0^\infty p_t (t) dt
    \label{QY}
\end{equation}

The solution for the PL decay is the biexponential law:

\begin{equation}
    PL(t) = A_F e^{-t/\tau_F} + A_D e^{-t/\tau_D},
    \label{PL_th}
\end{equation}
where $A_F, A_D$ are the fast and delayed components amplitudes and $\tau_F,\tau_D$ are the fast and delayed components characteristic times. Introducing the rates:
$$ \Gamma_e \equiv k_r + k_n + k_t,$$
$$ \Gamma_t \equiv k_r'+ k_n'+ k_d,$$
one can express the characteristic decay times as follows:

\begin{equation}
    \frac{1}{\tau_{F,D}} = \frac{1}{2} \left\{ (\Gamma_e + \Gamma_t) \pm \left|\Gamma_e - \Gamma_t \right| \sqrt{1 + \frac{4 k_d k_t}{\left(\Gamma_e - \Gamma_t \right)^2}} \right\}.
\end{equation}

The total luminescence quantum yield consists of two parts:

\begin{equation}
Y=Y_F+Y_D,
\end{equation}
where the fast and delayed components' quantum yields are as follows:
\begin{equation}
    Y_F = (k_r X_1 + k_r') C_F \tau_F,
\end{equation}

\begin{equation}
    Y_D = (k_r X_2 + k_r') C_D \tau_D.
\end{equation}

The detailed derivation of the coefficients $X_{1,2}$ and $C_{F,D}$ can be found in the Supplementary Note 4 (ESI\dag).

To explain the PNC blinking, we assume that both trapping rate constants change with time, \textit{i.e.} $k_n=k_n(t)$ and $k_n'=k_n'(t)$. The change rate of these rate constants is much slower than the relaxation time of the excitation.

In the recent paper \cite{Podshivaylov2023}, we have proposed a model of colloidal QD blinking in which changes in the trapping constant are related to slow variations of the electron - LO phonon coupling value, \textit{i.e.} the Huang-Rhys parameter, as follows:

\begin{equation}
    k_n = k_0 S^{\alpha}(t),
    \label{Kn}
\end{equation}
where $\alpha=\delta E/\hbar\omega_{LO}$ is the number of phonons emitted during the non-radiative transition. $\delta E$ is the energy difference between the excited and deep trap levels. The typical value of $\alpha$ is close to 10 for a moderate deep trap. Here we assume that the trapping occurs via multiple-phonon mediated cascade trapping, while the value of the Huang-Rhys parameter is less than 1. In the model presented, we assume that both the trapping constants have a similar connection to the same Huang-Rhys parameter as follows:

\begin{equation}
    k_n(t)=k_0 S(t)^{\alpha}; \; k_n'(t)=k_0' S(t)^{\alpha'},
    \label{S}
\end{equation}
where $\alpha'=\alpha - \Delta E/\hbar \omega_{LO}$. The time dependence of the Huang-Rhys parameter is defined as:

\begin{equation}
    S(t)=s_0 + \sum\limits_i^N s_i \sigma_i(t),
\end{equation}
where $\sigma_i(t)$ is a random function of time that can take values of 0 and 1. These random functions represent the set of $N$ independent markovian two-level systems (TLS), which are assumed to switch exponentially. Such TLSs are assumed to correspond to a number of different nuclei configurations on the surface of a single PNC. $s_0$ is the Huang-Rhys parameter at the moment when all $\sigma_i(t)=0$, and $s_i$ is the corresponding change, when the $i$\textsuperscript{th} TLS is turned on. The switching rates for single TLS are $\gamma^+_i$ and $\gamma^-_i$ for the $0\rightarrow 1$ and $1\rightarrow 0$ switching, respectively.

Each configuration of the TLS' set can be described by a single binary number $\Sigma \equiv \left\{ \sigma_1, \dots, \sigma_N \right\}$. The quantities related to that  configuration are marked with the corresponding index, \textit{e.g.} $$S_{\Sigma}=s_0 + \sum\limits_i^N s_i \sigma_i$$ for the Huang-Rhys parameter.

Introducing the vector $\vec{P}(t)$ containing the probabilities of all possible  configurations $P_{\Sigma}(t)$  one can describe the dynamics of TLS's set by the master equation:

\begin{equation}
    \frac{d \vec{P}}{dt}= \hat{W}\vec{P}.
    \label{Master}
\end{equation}
where $\hat{W}$ is the matrix that contains the rates of the transitions between the  configurations. This equation was in details analyzed in Ref.~\citenum{Podshivaylov2023}. Introducing the Green's function $\hat{G}(t) \equiv \exp(\hat{W}t)$ one can obtain the ACF and PSD as follows:

\begin{equation}
    ACF(t) = \frac{1}{\langle Y \rangle^2}\left(\vec{Y}, \hat{G}(t) \hat{Y} \vec{P}_{st}\right); \quad \quad \: \: \: \langle Y \rangle = (\vec{Y}, \vec{P}_{st}),
\end{equation}

\begin{equation}
    PSD(f) = 4\int\limits_0^\infty ACF(t) \cos(2\pi f t) dt; \quad 0\le f < \infty,
    \label{PSD}
\end{equation}
where $(\vec{A},\vec{B})$ is the dot product, $\vec{P}_{st}$ is the steady-state solution of master equation (\ref{Master}).
$\vec{Y}$ is the vector of PL quantum yields. Each element of the vector corresponding to the configuration $\Sigma$ is defined as $Y_{\Sigma}=Y(S_{\Sigma})$, where $Y(S)$ is the PL quantum yield Eq. (\ref{QY}) for a given value of the Huang-Rhys parameter $S$. $\hat{Y}$ is the diagonal matrix with elements $Y_{\Sigma\Sigma} = Y_{\Sigma}$.  The PDF is calculated as follows:

\begin{equation}
     PDF(k) = (\vec{q}(k),\vec{P}_{st}); \quad q_\Sigma(k) = \frac{\bar{N}_\Sigma^k}{k!}e^{-\bar{N}_\Sigma}
    \label{PDF},
\end{equation}
where $\bar{N}_\Sigma$ is the mean value of photons detected in one bin with configuration $\Sigma$. The details on derivation can be found in Ref.~\citenum{Podshivaylov2023}. The integrated over all states PL decay curve can be calculated as follows:

\begin{equation}
    \overline{PL}(t) = \left(\vec{PL}(t),\vec{P}_{st}\right),
\end{equation}
 where $PL_{\Sigma}(t)$ is the PL decay curve calculated for the configuration $\Sigma$ according to Eq. (\ref{PL_th}).

\begin{figure}[H]
    \centering
    \includegraphics[width=1\linewidth]{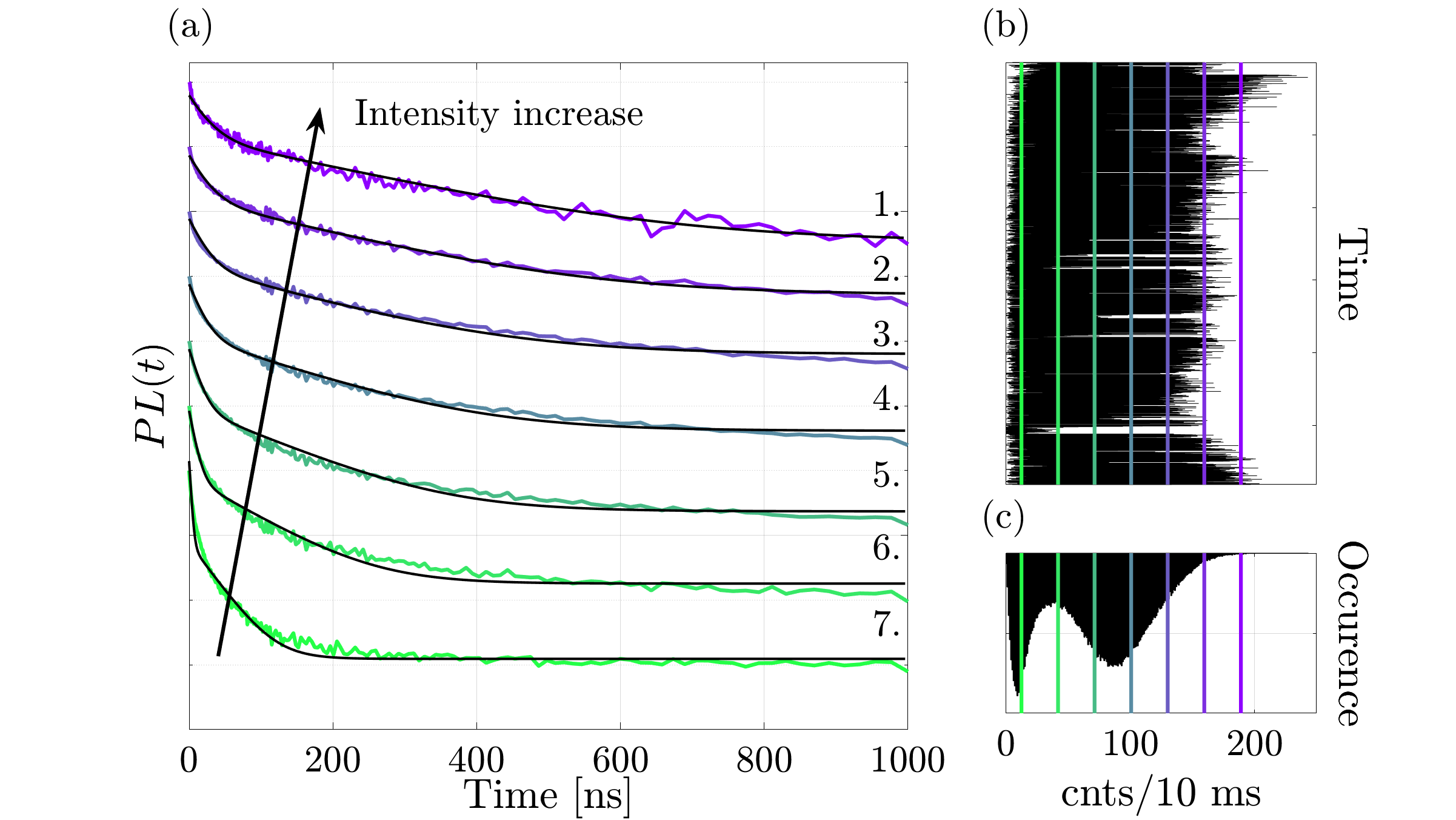}
    \caption{(a) PL decay curves for 7 different levels marked with the colors that correspond to selected intensity levels on (b) and (c) and their fit by model (black lines). Note that the PL decay curves were averaged using the logarithmic time bins on the scale beyond the 100 ns.}
    \label{fig7}
\end{figure}

To extract the kinetic parameters from our model, we propose the following method. Similarly to the procedure for obtaining FLID dependencies, we have selected 7 separate levels in the intensity trajectory with a spread of 2-3 photons for each level. The number of levels match the number of the fitting parameters. The PL decay at each level is then approximated by the solution of the kinetic model (\ref{PL_th}) with the same kinetic parameters for each level using the maximum likelihood method. The value that differs the levels was the value of the Huang-Rhys parameter, which determines the intensity level of the trajectory. We also assumed that the absolute quantum yield for the maximum intensity value in the trajectory corresponds to the value of 1. The details can be found in the Supplementary Note 5 (ESI\dag). The result is shown on Fig.~\ref{fig7}. It can be seen that the model successfully fits all PL decay curves. The values of the extracted parameters are as follows: $k_r = 0.027$ ns\textsuperscript{-1}, $k_t=0.55 \; k_r$, $\Delta E = 22$ meV, $k_0 =7.54 \cdot 10^{5}$ ns\textsuperscript{-1}, $k_0'= 8.13 \cdot 10^{3}$ ns\textsuperscript{-1}, $k_r' = 0.04 \; k_r$, $p_0= 0.5$.  Note, that the parameter $k_d$ is calculated using the detailed balance relation $k_d=k_t e^{-\Delta E/k_bT}$. The Huang-Rhys parameter is 0.23 for the lowest chosen level and 0.15 for the highest chosen level, respectively. The parameters $\alpha=10$ and $\hbar \omega_{LO} = 17$ meV were fixed during the fitting procedure. The value of the $\alpha$ is characteristic for moderate deep trap energies.  There are many reports in the literature on the maximum energy of LO-phonons. The reported values range from 16 to 20 meV\cite{Cho2022,Zhu2024,Zhou2020,Ramade2018,Iaru2017}, so the choice of 17 meV was optimal.

\begin{figure}[H]
    \centering    \includegraphics[width=1\linewidth]{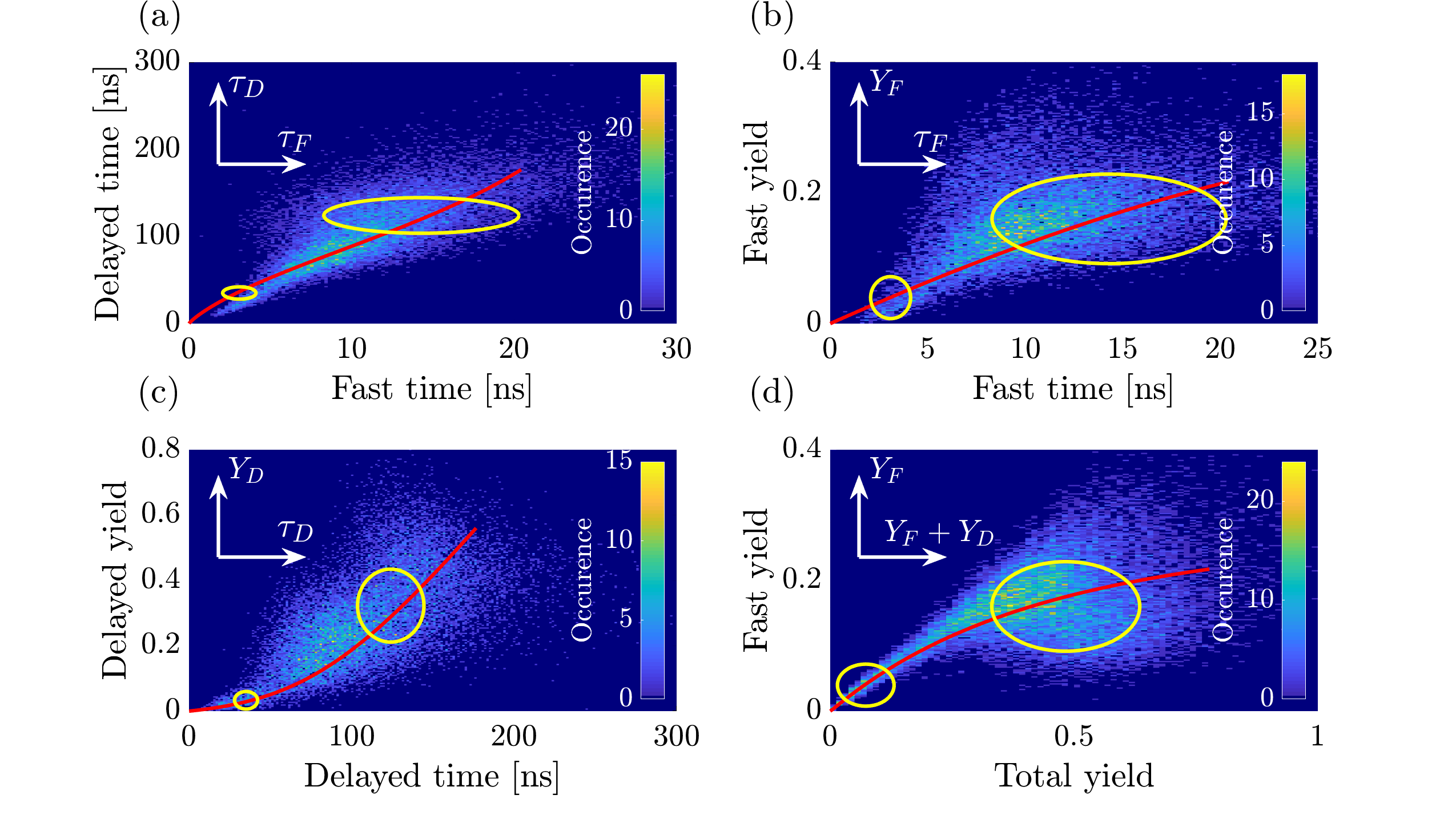}
    \caption{Two-dimensional distributions of the PL decay components and their theoretical prediction (red lines).  The yellow thin lines represent the 3$\sigma$ confidence regions for the two PL intensity levels.}
    \label{fig8}
\end{figure}

\begin{figure}[H]
    \centering    \includegraphics[width=1\linewidth]{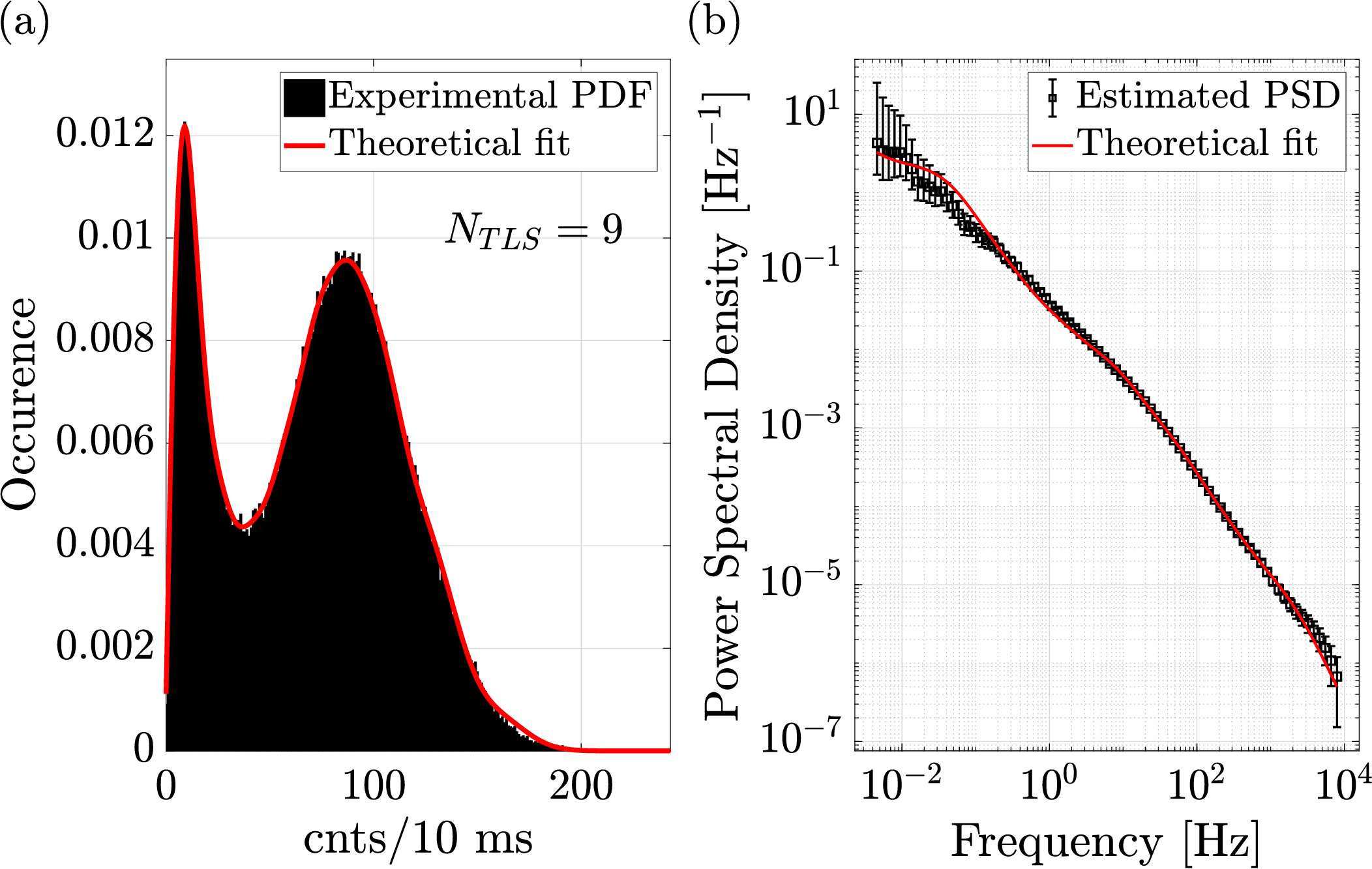}
    \caption{(a) The experimentally obtained PDF (black bars), (b) estimated PSD (black squares), and their model fit (red lines).}
    \label{fig9}
\end{figure}

Next, in order to calculate the dependencies of the PL decay individual components on each other, we varied the Huang-Rhys parameter  from the minimum value obtained from the approximation procedure to values several times larger than the maximum to cover the entire observed range. The calculation results are shown in Fig.~\ref{fig8}. The theoretical curves  successfully predict all four correlations measured in the experiment.
 The confidence regions were obtained using the procedure described in the Supplementary Note 5  (ESI\dag).

Using the fitted kinetics parameters, we also approximated the PDF and PSD estimating the parameters responsible for Huang-Rhys time dynamics, similar to what was done in Ref.~\citenum{Podshivaylov2023}. The result is presented in Fig.~\ref{fig9}. As in the case of the FLID, the model fit shows a successful agreement between the experimental results and the theoretical predictions. The mean value of the Huang-Rhys parameter extracted from this fit is $\langle S \rangle = 0.19$, which is consistent the parameter obtained from the previous procedure. The parameters obtained for each TLS can be found in the Supplementary Note 7  (ESI\dag).

We repeated all the previously described procedures for other nanocrystals from this set and, in all cases, were able to obtain agreement between the experimental data and the theory. Summary figures for each nanocrystal are shown in the Supplementary Note 6 (ESI\dag). The estimated parameters are presented in the Supplementary Note 7 (ESI\dag).

Fig.~\ref{fig10}a shows the trajectory-averaged decay of the luminescence intensity of a single perovskite nanocrystal. It can be clearly seen that the decay curve is a set of three power laws, the characteristic time scales of which can be attributed to biexciton luminescence (0-4 ns), the fast luminescence component (4-20 ns), and the delayed luminescence component (>20 ns).

We have calculated the PL decay curve using the parameters obtained from the previous fitting procedures. To account for the biexciton component, we calculate the PL decay using the following equation:

\begin{equation}
PL_{th} (t) = \frac{A_{xx}}{t^{\beta}} e^{-t/\tau_{XX}} + \overline{PL}(t),
\end{equation}
where $PL_{th}$ is the theoretical total PL decay curve, $\tau_{XX}$ is the biexciton component lifetime obtained from the biexciton analysis ($\tau_{XX} = $ 4.58 ns), $\beta$ is the power exponent equal to 0.35 (from Fig.~\ref{fig10}a), $A_{XX}$ is a coefficient calculated to improve the agreement between theory and experiment. The result is shown on Fig.~\ref{fig10}b. The model prediction corresponds well to the experimental PL decay curve. Therefore, it can be concluded that the power-law dependencies are the result of the averaging of exponential decays with the fluctuating characteristic times.

\begin{figure}[H]
    \centering    \includegraphics[width=1\linewidth]{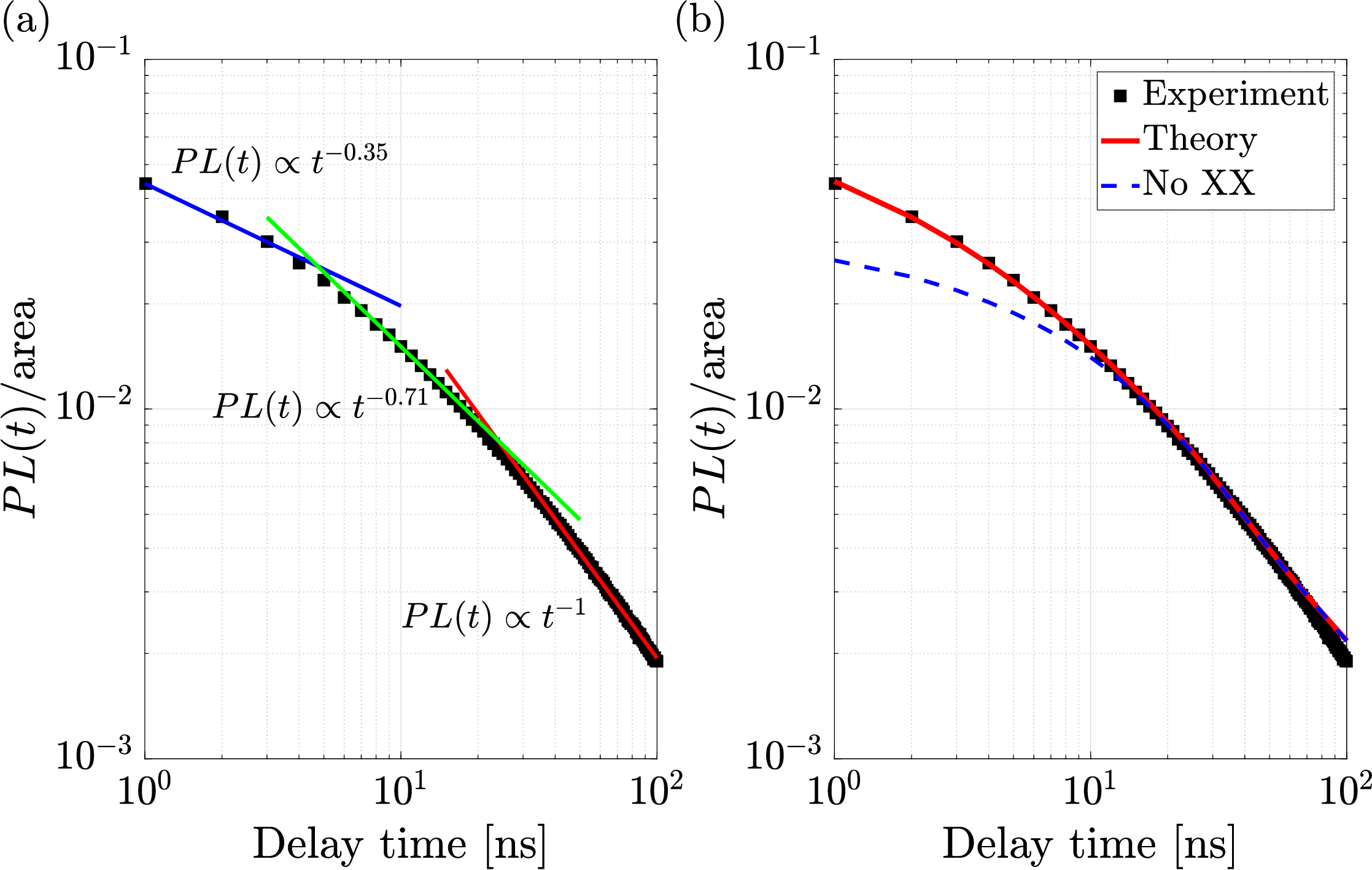}
    \caption{(a) Averaged PL decay (black squares) and guide-to-eye power law curves (colored lines) (b) Experimental PL decay curve (black squares) and its theoretical prediction (red line). The theoretical PL decay component without the biexciton contribution is marked with the blue line.}
    \label{fig10}
\end{figure}

\section{Discussion}

Delayed luminescence in CsPbBr\textsubscript{3} nanocrystals and quantum dots has already been observed. \cite{Chirvony2017,Ma2019,Dey2018,Becker2020,Shinde2017,He2016} In Ref.~\citenum{He2021} the delayed emission was observed for a PNC that was surface-functionalized with phenanthrene ligands. It is usually attributed to the  presence of a shallow energy level, which lies below the band gap at a distance of 20 to 100 meV below the excited state of the nanocrystal. Usually such shallow states are caused by the localized electronic states of the PNC. The observed characteristic times of delayed emission ranged from 10 to 500 ns. However,  the delayed luminescence time - quantum yield correlation, as well as  the delayed luminescence time -  fast time correlation, shown in Fig.~\ref{fig4}, have not yet been reported.

\begin{figure}[H]
    \centering    \includegraphics[width=1\linewidth]{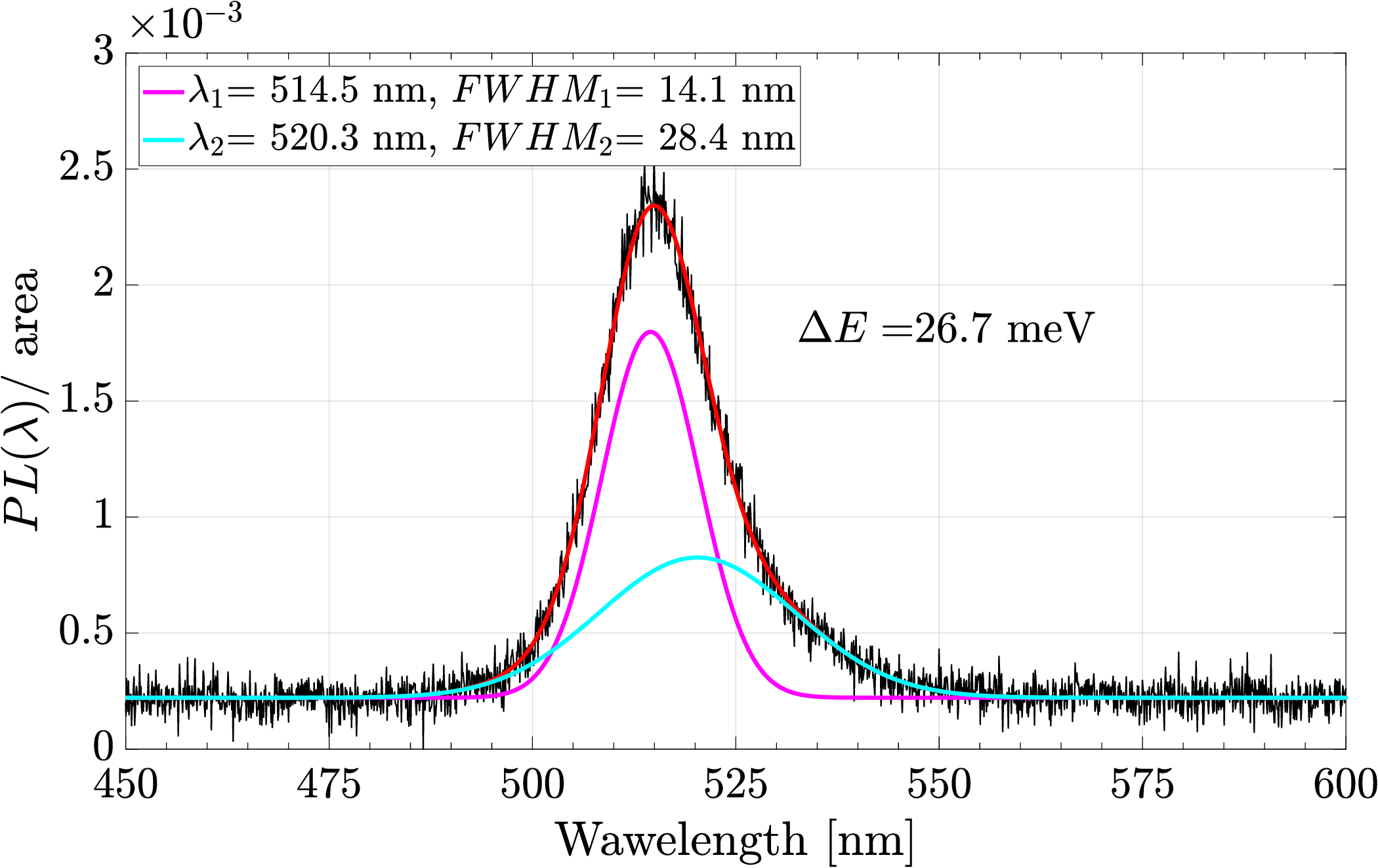}
    \caption{Normalized single PNC luminescence spectra (black line), its fit by two gaussian functions (red line), the short wavelength component (magenta line) and long wavelength component (cyan line).}
    \label{fig11}
\end{figure}

There is evidence in the literature of the presence of an additional peak in the luminescence spectra of single CsPbBr\textsubscript{3} nanocrystals. There are several explanations on the nature of this additional peak in the literature: the emission connected with localized states of PNC such as a bound or localized exciton, \cite{Shinde2017,Dey2018,He2016} bound state of an electron or hole, \cite{Santomauro2016,Neukirch2016} self-trapped exciton. \cite{Ma2019,Ma2020} In order to check the emission properties of the studied PNCs, we carried out the experiments measuring a single PNC's luminescence spectra. An example is presented in Fig.~\ref{fig11}. The spectra has an asymmetric form, that can be approximated by two gaussians. The energy difference between two peaks for the PNC presented is $\Delta E = 26.7$ meV.  The results for other PNCs are presented in the Supplementary Note 8 (ESI\dag). The energy difference  fluctuates in the range of about 20 to 35 meV. According to various data, the highest energies of an LO phonon are 16-20 meV, which is less than the observed values, which means the observed peak is not caused by a phonon replica. Thus, here is an energy level inside the band gap.

In Ref.~\citenum{Mi2023}, the involvement of a trapped exciton was used to explain the biexciton luminescence blinking of a single CsPbBr\textsubscript{3} PQDs, similar to the Auger mechanism. The characteristic short luminescence times observed by the authors correspond well to our data, and the charge recombination model is largely similar to the proposed one, although the authors considered the trapped exciton to be non-radiative. Considering that the rate of nonradiative transition in our model is correlated for the excited state and the shallow state, these states should be similar in nature. Thus, we consider both of these states to be excitonic, and the shallow state is a trapped exciton.

The Huang-Rhys parameter for the exciton-phonon coupling for an optical transition in CsPbBr\textsubscript{3} nanocrystals is reported to be less than 1. The value slightly changes with the size of the sample.\cite{Cho2022,Zhu2024,Iaru2017} The average Huang-Rhys parameter for nonradiative transitions should have the same order of magnitude, which allows us to use expression (\ref{Kn}) for the multiphonon cascade trapping rate constant.

\section{Conclusions}

In conclusion, experiments were carried out to register single photons emitted by single CsPbBr\textsubscript{3} nanocrystals in the process of luminescence blinking. A strong correlation was found between the fast and delayed luminescence components. To explain the observed phenomenon, we proposed a kinetic model of charge carrier recombination kinetics, as well as a luminescence blinking model corresponding to the trapping mechanism. The blinking is caused by slow fluctuations in the electron-phonon coupling.
The model assumes the existence of a correlation of the Huang-Rhys parameters for nonradiative transitions from excited and shallow levels in the process of multiphonon cascade capture into a deep trap.
 The proposed model successfully reproduces the two-dimensional distributions of the PL emission components, the photon distribution function, the power spectral density, and the integrated over trajectory PL decay.

\section{Acknowledgements}

E.A.P. acknowledges support from the Foundation for the Advancement of Theoretical Physics and Mathematics "BASIS" (22-1-5-36-1).
E.A.P., A.M.S., and P.A.F.  (Voevodsky Institute of Chemical Kinetics and Combustion SB RAS) acknowledge the core funding from the Russian Federal Ministry of Science and Higher Education (FWGF-2021-0002).
A.N. acknowledge the support from the Ministry of Science and Higher Education of the Russian Federation for the project “Physics of nanostructured materials and highly sensitive sensorics: Synthesis, fundamental research and applications in photonics, life sciences, quantum and nanotechnology” (theme no. 124031100005-5). I.E. and A.N. acknowledges support for the research project from the Ministry of  Science and Higher Education of the Russian Federation (Grant No. FFUU-2025-0004). A.T., M.K., A.N., and I.E. acknowledge support for the research project from the Ministry of Science and Higher Education of the Russian Federation (Grant No. FFMR-2024-0017).
E.V.S. and A.P.P. acknowledge support from the Russian Science Foundation (project no. 24-73-10072, synthesis of perovskite nanocrystals and TEM characterization).

\section{Conflicts of interest}

There are no conflicts to declare.

\section{Data availability}

The data supporting this article have been included as part of the Supplementary Information


\balance


\bibliography{Ref} 

\providecommand*{\mcitethebibliography}{\thebibliography}
\csname @ifundefined\endcsname{endmcitethebibliography}
{\let\endmcitethebibliography\endthebibliography}{}
\begin{mcitethebibliography}{65}
\providecommand*{\natexlab}[1]{#1}
\providecommand*{\mciteSetBstSublistMode}[1]{}
\providecommand*{\mciteSetBstMaxWidthForm}[2]{}
\providecommand*{\mciteBstWouldAddEndPuncttrue}
  {\def\EndOfBibitem{\unskip.}}
\providecommand*{\mciteBstWouldAddEndPunctfalse}
  {\let\EndOfBibitem\relax}
\providecommand*{\mciteSetBstMidEndSepPunct}[3]{}
\providecommand*{\mciteSetBstSublistLabelBeginEnd}[3]{}
\providecommand*{\EndOfBibitem}{}
\mciteSetBstSublistMode{f}
\mciteSetBstMaxWidthForm{subitem}
{(\emph{\alph{mcitesubitemcount}})}
\mciteSetBstSublistLabelBeginEnd{\mcitemaxwidthsubitemform\space}
{\relax}{\relax}

\bibitem[Swarnkar \emph{et~al.}(2015)Swarnkar, Chulliyil, Ravi, Irfanullah,
  Chowdhury, and Nag]{Swarnkar2015}
A.~Swarnkar, R.~Chulliyil, V.~K. Ravi, M.~Irfanullah, A.~Chowdhury and A.~Nag,
  \emph{Angewandte Chemie - International Edition}, 2015, \textbf{54},
  15424--15428\relax
\mciteBstWouldAddEndPuncttrue
\mciteSetBstMidEndSepPunct{\mcitedefaultmidpunct}
{\mcitedefaultendpunct}{\mcitedefaultseppunct}\relax
\EndOfBibitem
\bibitem[Rainò \emph{et~al.}(2022)Rainò, Yazdani, Boehme, Kober-Czerny, Zhu,
  Krieg, Rossell, Erni, Wood, Infante, and Kovalenko]{Raino2022}
G.~Rainò, N.~Yazdani, S.~C. Boehme, M.~Kober-Czerny, C.~Zhu, F.~Krieg, M.~D.
  Rossell, R.~Erni, V.~Wood, I.~Infante and M.~V. Kovalenko, \emph{Nature
  Communications}, 2022, \textbf{13}, 2587\relax
\mciteBstWouldAddEndPuncttrue
\mciteSetBstMidEndSepPunct{\mcitedefaultmidpunct}
{\mcitedefaultendpunct}{\mcitedefaultseppunct}\relax
\EndOfBibitem
\bibitem[Thakur \emph{et~al.}(2021)Thakur, Paul, Maiti, and
  Chattopadhyay]{THAKUR2021}
S.~Thakur, T.~Paul, S.~Maiti and K.~K. Chattopadhyay, \emph{Solid State
  Sciences}, 2021, \textbf{122}, 106769\relax
\mciteBstWouldAddEndPuncttrue
\mciteSetBstMidEndSepPunct{\mcitedefaultmidpunct}
{\mcitedefaultendpunct}{\mcitedefaultseppunct}\relax
\EndOfBibitem
\bibitem[Pang \emph{et~al.}(2023)Pang, Hoang, O'Mullane, and Wang]{Pang2023}
L.~Pang, M.~T. Hoang, A.~P. O'Mullane and H.~Wang, \emph{Energy Materials},
  2023, \textbf{3}, 300012\relax
\mciteBstWouldAddEndPuncttrue
\mciteSetBstMidEndSepPunct{\mcitedefaultmidpunct}
{\mcitedefaultendpunct}{\mcitedefaultseppunct}\relax
\EndOfBibitem
\bibitem[Yadav \emph{et~al.}(2023)Yadav, Shukla, Suhail, Kumar, and
  Bag]{Yadav2023}
A.~Yadav, P.~S. Shukla, A.~Suhail, J.~Kumar and M.~Bag, \emph{ACS Applied Nano
  Materials}, 2023, \textbf{6}, 16960--16969\relax
\mciteBstWouldAddEndPuncttrue
\mciteSetBstMidEndSepPunct{\mcitedefaultmidpunct}
{\mcitedefaultendpunct}{\mcitedefaultseppunct}\relax
\EndOfBibitem
\bibitem[Du \emph{et~al.}(2017)Du, Wu, Cheng, Dang, Ma, Zhang, Tan, and
  Chen]{Du2017}
X.~Du, G.~Wu, J.~Cheng, H.~Dang, K.~Ma, Y.~W. Zhang, P.~F. Tan and S.~Chen,
  \emph{RSC Advances}, 2017, \textbf{7}, 10391--10396\relax
\mciteBstWouldAddEndPuncttrue
\mciteSetBstMidEndSepPunct{\mcitedefaultmidpunct}
{\mcitedefaultendpunct}{\mcitedefaultseppunct}\relax
\EndOfBibitem
\bibitem[Yuan \emph{et~al.}(2018)Yuan, Wang, Zhuo, Tian, Jin, and
  Liao]{Yuan2018}
S.~Yuan, Z.-K. Wang, M.-P. Zhuo, Q.-S. Tian, Y.~Jin and L.-S. Liao, \emph{ACS
  Nano}, 2018, \textbf{12}, 9541--9548\relax
\mciteBstWouldAddEndPuncttrue
\mciteSetBstMidEndSepPunct{\mcitedefaultmidpunct}
{\mcitedefaultendpunct}{\mcitedefaultseppunct}\relax
\EndOfBibitem
\bibitem[Lan \emph{et~al.}(2019)Lan, Luo, Wang, Li, Wu, and Wang]{LAN2019}
J.~Lan, L.~Luo, M.~Wang, F.~Li, X.~Wu and F.~Wang, \emph{Journal of
  Luminescence}, 2019, \textbf{210}, 464--471\relax
\mciteBstWouldAddEndPuncttrue
\mciteSetBstMidEndSepPunct{\mcitedefaultmidpunct}
{\mcitedefaultendpunct}{\mcitedefaultseppunct}\relax
\EndOfBibitem
\bibitem[Krieg \emph{et~al.}(2021)Krieg, Sercel, Burian, Andrusiv, Bodnarchuk,
  Stöferle, Mahrt, Naumenko, Amenitsch, Rainò, and Kovalenko]{Krieg2021}
F.~Krieg, P.~C. Sercel, M.~Burian, H.~Andrusiv, M.~I. Bodnarchuk, T.~Stöferle,
  R.~F. Mahrt, D.~Naumenko, H.~Amenitsch, G.~Rainò and M.~V. Kovalenko,
  \emph{ACS Central Science}, 2021, \textbf{7}, 135--144\relax
\mciteBstWouldAddEndPuncttrue
\mciteSetBstMidEndSepPunct{\mcitedefaultmidpunct}
{\mcitedefaultendpunct}{\mcitedefaultseppunct}\relax
\EndOfBibitem
\bibitem[Rainò \emph{et~al.}(2018)Rainò, Becker, Bodnarchuk, Mahrt,
  Kovalenko, and Stöferle]{Rainò2018}
G.~Rainò, M.~A. Becker, M.~I. Bodnarchuk, R.~F. Mahrt, M.~V. Kovalenko and
  T.~Stöferle, \emph{Nature}, 2018, \textbf{563}, 671--675\relax
\mciteBstWouldAddEndPuncttrue
\mciteSetBstMidEndSepPunct{\mcitedefaultmidpunct}
{\mcitedefaultendpunct}{\mcitedefaultseppunct}\relax
\EndOfBibitem
\bibitem[Zai \emph{et~al.}(2018)Zai, Zhu, Xie, Zhao, Shi, Chen, Ke, Sui, Chen,
  Hu, Zhang, Gao, Zhou, Li, and Chen]{Zai2018}
H.~Zai, C.~Zhu, H.~Xie, Y.~Zhao, C.~Shi, Z.~Chen, X.~Ke, M.~Sui, C.~Chen,
  J.~Hu, Q.~Zhang, Y.~Gao, H.~Zhou, Y.~Li and Q.~Chen, \emph{ACS Energy
  Letters}, 2018, \textbf{3}, 30--38\relax
\mciteBstWouldAddEndPuncttrue
\mciteSetBstMidEndSepPunct{\mcitedefaultmidpunct}
{\mcitedefaultendpunct}{\mcitedefaultseppunct}\relax
\EndOfBibitem
\bibitem[Ullah \emph{et~al.}(2021)Ullah, Wang, Yang, Liu, Yang, Xia, Guo, and
  Chen]{Ullah2021}
S.~Ullah, J.~Wang, P.~Yang, L.~Liu, S.-E. Yang, T.~Xia, H.~Guo and Y.~Chen,
  \emph{Mater. Adv.}, 2021, \textbf{2}, 646--683\relax
\mciteBstWouldAddEndPuncttrue
\mciteSetBstMidEndSepPunct{\mcitedefaultmidpunct}
{\mcitedefaultendpunct}{\mcitedefaultseppunct}\relax
\EndOfBibitem
\bibitem[Gao \emph{et~al.}(2019)Gao, Wu, Lu, Chen, Liu, Bai, Yang, Yu, Dai, and
  Zhang]{Gao2019}
Y.~Gao, Y.~Wu, H.~Lu, C.~Chen, Y.~Liu, X.~Bai, L.~Yang, W.~W. Yu, Q.~Dai and
  Y.~Zhang, \emph{Nano Energy}, 2019, \textbf{59}, 517--526\relax
\mciteBstWouldAddEndPuncttrue
\mciteSetBstMidEndSepPunct{\mcitedefaultmidpunct}
{\mcitedefaultendpunct}{\mcitedefaultseppunct}\relax
\EndOfBibitem
\bibitem[Xie \emph{et~al.}(2021)Xie, Gong, Kong, Liu, Mi, Guo, and
  Luo]{Xie2022}
M.~Xie, W.~Gong, L.~Kong, Y.~Liu, Y.~Mi, H.~Guo and S.-N. Luo,
  \emph{Nanotechnology}, 2021, \textbf{33}, 115204\relax
\mciteBstWouldAddEndPuncttrue
\mciteSetBstMidEndSepPunct{\mcitedefaultmidpunct}
{\mcitedefaultendpunct}{\mcitedefaultseppunct}\relax
\EndOfBibitem
\bibitem[Wang \emph{et~al.}(2018)Wang, Wang, Xiao, and Song]{Wang2018}
K.~Wang, S.~Wang, S.~Xiao and Q.~Song, \emph{Advanced Optical Materials}, 2018,
  \textbf{6}, 1800278\relax
\mciteBstWouldAddEndPuncttrue
\mciteSetBstMidEndSepPunct{\mcitedefaultmidpunct}
{\mcitedefaultendpunct}{\mcitedefaultseppunct}\relax
\EndOfBibitem
\bibitem[Yu \emph{et~al.}(2022)Yu, Su, Pan, Gao, Wang, Chen, Zhang, Dou, Zhang,
  Ge, Shi, Zhai, and Wang]{Yu2022}
H.~Yu, X.~Su, Y.~Pan, D.~Gao, J.~Wang, R.~Chen, J.~Zhang, F.~Dou, X.~Zhang,
  K.~Ge, X.~Shi, T.~Zhai and L.~Wang, \emph{Optical Materials}, 2022,
  \textbf{133}, 112907\relax
\mciteBstWouldAddEndPuncttrue
\mciteSetBstMidEndSepPunct{\mcitedefaultmidpunct}
{\mcitedefaultendpunct}{\mcitedefaultseppunct}\relax
\EndOfBibitem
\bibitem[Zhu \emph{et~al.}(2022)Zhu, Marczak, Feld, Boehme, Bernasconi,
  Moskalenko, Cherniukh, Dirin, Bodnarchuk, Kovalenko, and Rainò]{Zhu2022}
C.~Zhu, M.~Marczak, L.~Feld, S.~C. Boehme, C.~Bernasconi, A.~Moskalenko,
  I.~Cherniukh, D.~Dirin, M.~I. Bodnarchuk, M.~V. Kovalenko and G.~Rainò,
  \emph{Nano Letters}, 2022, \textbf{22}, 3751--3760\relax
\mciteBstWouldAddEndPuncttrue
\mciteSetBstMidEndSepPunct{\mcitedefaultmidpunct}
{\mcitedefaultendpunct}{\mcitedefaultseppunct}\relax
\EndOfBibitem
\bibitem[Park \emph{et~al.}(2015)Park, Guo, Makarov, and Klimov]{Park2015}
Y.-S. Park, S.~Guo, N.~S. Makarov and V.~I. Klimov, \emph{ACS Nano}, 2015,
  \textbf{9}, 10386--10393\relax
\mciteBstWouldAddEndPuncttrue
\mciteSetBstMidEndSepPunct{\mcitedefaultmidpunct}
{\mcitedefaultendpunct}{\mcitedefaultseppunct}\relax
\EndOfBibitem
\bibitem[Hu \emph{et~al.}(2015)Hu, Zhang, Sun, Yin, Lv, Zhang, Yu, Wang, Zhang,
  and Xiao]{Hu2015}
F.~Hu, H.~Zhang, C.~Sun, C.~Yin, B.~Lv, C.~Zhang, W.~W. Yu, X.~Wang, Y.~Zhang
  and M.~Xiao, \emph{ACS Nano}, 2015, \textbf{9}, 12410--12416\relax
\mciteBstWouldAddEndPuncttrue
\mciteSetBstMidEndSepPunct{\mcitedefaultmidpunct}
{\mcitedefaultendpunct}{\mcitedefaultseppunct}\relax
\EndOfBibitem
\bibitem[Seth \emph{et~al.}(2016)Seth, Mondal, Patra, and Samanta]{Seth2016}
S.~Seth, N.~Mondal, S.~Patra and A.~Samanta, \emph{Journal of Physical
  Chemistry Letters}, 2016, \textbf{7}, 266--271\relax
\mciteBstWouldAddEndPuncttrue
\mciteSetBstMidEndSepPunct{\mcitedefaultmidpunct}
{\mcitedefaultendpunct}{\mcitedefaultseppunct}\relax
\EndOfBibitem
\bibitem[Seth \emph{et~al.}(2018)Seth, Ahmed, and Samanta]{Seth2018}
S.~Seth, T.~Ahmed and A.~Samanta, \emph{The Journal of Physical Chemistry
  Letters}, 2018, \textbf{9}, 7007--7014\relax
\mciteBstWouldAddEndPuncttrue
\mciteSetBstMidEndSepPunct{\mcitedefaultmidpunct}
{\mcitedefaultendpunct}{\mcitedefaultseppunct}\relax
\EndOfBibitem
\bibitem[Seth \emph{et~al.}(2019)Seth, Ahmed, De, and Samanta]{Seth2019}
S.~Seth, T.~Ahmed, A.~De and A.~Samanta, \emph{Tackling the Defects, Stability,
  and Photoluminescence of CsPbX3 Perovskite Nanocrystals}, 2019\relax
\mciteBstWouldAddEndPuncttrue
\mciteSetBstMidEndSepPunct{\mcitedefaultmidpunct}
{\mcitedefaultendpunct}{\mcitedefaultseppunct}\relax
\EndOfBibitem
\bibitem[Li \emph{et~al.}(2018)Li, Huang, Zhang, Yang, Guo, Chen, Qin, Gao,
  Biju, Rogach, Xiao, and Jia]{Li2018}
B.~Li, H.~Huang, G.~Zhang, C.~Yang, W.~Guo, R.~Chen, C.~Qin, Y.~Gao, V.~P.
  Biju, A.~L. Rogach, L.~Xiao and S.~Jia, \emph{Journal of Physical Chemistry
  Letters}, 2018, \textbf{9}, 6934--6940\relax
\mciteBstWouldAddEndPuncttrue
\mciteSetBstMidEndSepPunct{\mcitedefaultmidpunct}
{\mcitedefaultendpunct}{\mcitedefaultseppunct}\relax
\EndOfBibitem
\bibitem[Palstra \emph{et~al.}(2021)Palstra, de~Buy~Wenniger, Patra, Garnett,
  and Koenderink]{Palstra2021}
I.~M. Palstra, I.~M. de~Buy~Wenniger, B.~K. Patra, E.~C. Garnett and A.~F.
  Koenderink, \emph{The Journal of Physical Chemistry C}, 2021, \textbf{125},
  12061--12072\relax
\mciteBstWouldAddEndPuncttrue
\mciteSetBstMidEndSepPunct{\mcitedefaultmidpunct}
{\mcitedefaultendpunct}{\mcitedefaultseppunct}\relax
\EndOfBibitem
\bibitem[Gibson \emph{et~al.}(2018)Gibson, Koscher, Alivisatos, and
  Leone]{Gibson2018}
N.~A. Gibson, B.~A. Koscher, A.~P. Alivisatos and S.~R. Leone, \emph{Journal of
  Physical Chemistry C}, 2018, \textbf{122}, 12106--12113\relax
\mciteBstWouldAddEndPuncttrue
\mciteSetBstMidEndSepPunct{\mcitedefaultmidpunct}
{\mcitedefaultendpunct}{\mcitedefaultseppunct}\relax
\EndOfBibitem
\bibitem[Paul \emph{et~al.}(2023)Paul, Kishore, and Samanta]{Paul2023}
S.~Paul, G.~Kishore and A.~Samanta, \emph{The Journal of Physical Chemistry C},
  2023, \textbf{127}, 10207--10214\relax
\mciteBstWouldAddEndPuncttrue
\mciteSetBstMidEndSepPunct{\mcitedefaultmidpunct}
{\mcitedefaultendpunct}{\mcitedefaultseppunct}\relax
\EndOfBibitem
\bibitem[Efros and Rosen(1997)]{EfrosPRL1997}
A.~L. Efros and M.~Rosen, \emph{Phys. Rev. Lett.}, 1997, \textbf{78},
  1110--1113\relax
\mciteBstWouldAddEndPuncttrue
\mciteSetBstMidEndSepPunct{\mcitedefaultmidpunct}
{\mcitedefaultendpunct}{\mcitedefaultseppunct}\relax
\EndOfBibitem
\bibitem[Frantsuzov and Marcus(2005)]{FrantsuzovPRB2005}
P.~A. Frantsuzov and R.~A. Marcus, \emph{Phys. Rev. B.}, 2005, \textbf{72},
  155321\relax
\mciteBstWouldAddEndPuncttrue
\mciteSetBstMidEndSepPunct{\mcitedefaultmidpunct}
{\mcitedefaultendpunct}{\mcitedefaultseppunct}\relax
\EndOfBibitem
\bibitem[Galland \emph{et~al.}(2011)Galland, Ghosh, Steinbr\"uck, Sykora,
  Hollingsworth, Klimov, and Htoon]{KlimovNature2011}
C.~Galland, Y.~Ghosh, A.~Steinbr\"uck, M.~Sykora, J.~A. Hollingsworth, V.~I.
  Klimov and H.~Htoon, \emph{Nature}, 2011, \textbf{479}, 203--208\relax
\mciteBstWouldAddEndPuncttrue
\mciteSetBstMidEndSepPunct{\mcitedefaultmidpunct}
{\mcitedefaultendpunct}{\mcitedefaultseppunct}\relax
\EndOfBibitem
\bibitem[Ahmed \emph{et~al.}(2019)Ahmed, Seth, and Samanta]{Ahmed2019}
T.~Ahmed, S.~Seth and A.~Samanta, \emph{ACS Nano}, 2019, \textbf{13},
  13537--13544\relax
\mciteBstWouldAddEndPuncttrue
\mciteSetBstMidEndSepPunct{\mcitedefaultmidpunct}
{\mcitedefaultendpunct}{\mcitedefaultseppunct}\relax
\EndOfBibitem
\bibitem[Podshivaylov \emph{et~al.}(2023)Podshivaylov, Kniazeva, Tarasevich,
  Eremchev, Naumov, and Frantsuzov]{Podshivaylov2023}
E.~Podshivaylov, M.~Kniazeva, A.~Tarasevich, I.~Eremchev, A.~Naumov and
  P.~Frantsuzov, \emph{Journal of Materials Chemistry C}, 2023, \textbf{11},
  8570--8576\relax
\mciteBstWouldAddEndPuncttrue
\mciteSetBstMidEndSepPunct{\mcitedefaultmidpunct}
{\mcitedefaultendpunct}{\mcitedefaultseppunct}\relax
\EndOfBibitem
\bibitem[Frantsuzov \emph{et~al.}(2009)Frantsuzov, Volkán-Kacsó, and
  Jankó]{FrantsuzovPRL2009}
P.~A. Frantsuzov, S.~Volkán-Kacsó and B.~Jankó, \emph{Phys. Rev. Lett.},
  2009, \textbf{103}, 207402\relax
\mciteBstWouldAddEndPuncttrue
\mciteSetBstMidEndSepPunct{\mcitedefaultmidpunct}
{\mcitedefaultendpunct}{\mcitedefaultseppunct}\relax
\EndOfBibitem
\bibitem[Volk\'an-Kacs\'o \emph{et~al.}(2010)Volk\'an-Kacs\'o, Frantsuzov, and
  Jank\'o]{VolkanNL2010}
S.~Volk\'an-Kacs\'o, P.~A. Frantsuzov and B.~Jank\'o, \emph{Nano Lett.}, 2010,
  \textbf{10}, 2761--2765\relax
\mciteBstWouldAddEndPuncttrue
\mciteSetBstMidEndSepPunct{\mcitedefaultmidpunct}
{\mcitedefaultendpunct}{\mcitedefaultseppunct}\relax
\EndOfBibitem
\bibitem[Frantsuzov \emph{et~al.}(2013)Frantsuzov, Volkán-Kacsó, and
  Jankó]{FrantsuzovNL2013}
P.~A. Frantsuzov, S.~Volkán-Kacsó and B.~Jankó, \emph{Nano Letters}, 2013,
  \textbf{13}, 402--408\relax
\mciteBstWouldAddEndPuncttrue
\mciteSetBstMidEndSepPunct{\mcitedefaultmidpunct}
{\mcitedefaultendpunct}{\mcitedefaultseppunct}\relax
\EndOfBibitem
\bibitem[Busov and Frantsuzov(2019)]{BusovOS2019}
V.~K. Busov and P.~A. Frantsuzov, \emph{Opt. Spectr.}, 2019, \textbf{126}, 70
  -- 82\relax
\mciteBstWouldAddEndPuncttrue
\mciteSetBstMidEndSepPunct{\mcitedefaultmidpunct}
{\mcitedefaultendpunct}{\mcitedefaultseppunct}\relax
\EndOfBibitem
\bibitem[Hinterding \emph{et~al.}(2020)Hinterding, Hinterding, Vonk, Vonk,
  Harten, Rabouw, and Rabouw]{Hinterding2020}
S.~O. Hinterding, S.~O. Hinterding, S.~J. Vonk, S.~J. Vonk, E.~J.~V. Harten,
  F.~T. Rabouw and F.~T. Rabouw, \emph{Journal of Physical Chemistry Letters},
  2020, \textbf{11}, 4755--4761\relax
\mciteBstWouldAddEndPuncttrue
\mciteSetBstMidEndSepPunct{\mcitedefaultmidpunct}
{\mcitedefaultendpunct}{\mcitedefaultseppunct}\relax
\EndOfBibitem
\bibitem[Podshivaylov \emph{et~al.}(2019)Podshivaylov, Kniazeva, Gorshelev,
  Eremchev, Naumov, and Frantsuzov]{Podshivaylov2019}
E.~A. Podshivaylov, M.~A. Kniazeva, A.~A. Gorshelev, I.~Y. Eremchev, A.~V.
  Naumov and P.~A. Frantsuzov, \emph{The Journal of chemical physics}, 2019,
  \textbf{151 17}, 174710\relax
\mciteBstWouldAddEndPuncttrue
\mciteSetBstMidEndSepPunct{\mcitedefaultmidpunct}
{\mcitedefaultendpunct}{\mcitedefaultseppunct}\relax
\EndOfBibitem
\bibitem[Protesescu \emph{et~al.}(2018)Protesescu, Yakunin, Nazarenko, Dirin,
  and Kovalenko]{Protesescu2018}
L.~Protesescu, S.~Yakunin, O.~Nazarenko, D.~N. Dirin and M.~V. Kovalenko,
  \emph{ACS Applied Nano Materials}, 2018, \textbf{1}, 1300--1308\relax
\mciteBstWouldAddEndPuncttrue
\mciteSetBstMidEndSepPunct{\mcitedefaultmidpunct}
{\mcitedefaultendpunct}{\mcitedefaultseppunct}\relax
\EndOfBibitem
\bibitem[Baitova \emph{et~al.}(2023)Baitova, Knyazeva, Mukanov, Tarasevich,
  Naumov, Son, Kozyukhin, and Eremchev]{Baitova2023}
V.~A. Baitova, M.~A. Knyazeva, I.~A. Mukanov, A.~O. Tarasevich, A.~V. Naumov,
  A.~G. Son, S.~A. Kozyukhin and I.~Y. Eremchev, \emph{JETP Letters}, 2023,
  \textbf{118}, 560--567\relax
\mciteBstWouldAddEndPuncttrue
\mciteSetBstMidEndSepPunct{\mcitedefaultmidpunct}
{\mcitedefaultendpunct}{\mcitedefaultseppunct}\relax
\EndOfBibitem
\bibitem[Laurence \emph{et~al.}(2006)Laurence, Fore, and Huser]{Laurence2006}
T.~A. Laurence, S.~Fore and T.~Huser, \emph{Fast, flexible algorithm for
  calculating photon correlations}, 2006\relax
\mciteBstWouldAddEndPuncttrue
\mciteSetBstMidEndSepPunct{\mcitedefaultmidpunct}
{\mcitedefaultendpunct}{\mcitedefaultseppunct}\relax
\EndOfBibitem
\bibitem[Rabouw \emph{et~al.}(2015)Rabouw, Kamp, Dijk-Moes, Gamelin,
  Koenderink, Meijerink, and Vanmaekelbergh]{Rabouw2015}
F.~T. Rabouw, M.~Kamp, R.~J.~V. Dijk-Moes, D.~R. Gamelin, A.~F. Koenderink,
  A.~Meijerink and D.~Vanmaekelbergh, \emph{Nano Letters}, 2015, \textbf{15},
  7718--7725\relax
\mciteBstWouldAddEndPuncttrue
\mciteSetBstMidEndSepPunct{\mcitedefaultmidpunct}
{\mcitedefaultendpunct}{\mcitedefaultseppunct}\relax
\EndOfBibitem
\bibitem[Castañeda \emph{et~al.}(2016)Castañeda, Nagamine, Yassitepe, Bonato,
  Voznyy, Hoogland, Nogueira, Sargent, Cruz, and Padilha]{Castañeda2016}
J.~A. Castañeda, G.~Nagamine, E.~Yassitepe, L.~G. Bonato, O.~Voznyy,
  S.~Hoogland, A.~F. Nogueira, E.~H. Sargent, C.~H.~B. Cruz and L.~A. Padilha,
  \emph{ACS Nano}, 2016, \textbf{10}, 8603--8609\relax
\mciteBstWouldAddEndPuncttrue
\mciteSetBstMidEndSepPunct{\mcitedefaultmidpunct}
{\mcitedefaultendpunct}{\mcitedefaultseppunct}\relax
\EndOfBibitem
\bibitem[de~Jong \emph{et~al.}(2017)de~Jong, Yamashita, Gomez, Ashida,
  Fujiwara, and Gregorkiewicz]{deJong2017}
E.~M. L.~D. de~Jong, G.~Yamashita, L.~Gomez, M.~Ashida, Y.~Fujiwara and
  T.~Gregorkiewicz, \emph{The Journal of Physical Chemistry C}, 2017,
  \textbf{121}, 1941--1947\relax
\mciteBstWouldAddEndPuncttrue
\mciteSetBstMidEndSepPunct{\mcitedefaultmidpunct}
{\mcitedefaultendpunct}{\mcitedefaultseppunct}\relax
\EndOfBibitem
\bibitem[Eperon \emph{et~al.}(2018)Eperon, Jedlicka, and Ginger]{Eperon2018}
G.~E. Eperon, E.~Jedlicka and D.~S. Ginger, \emph{The Journal of Physical
  Chemistry Letters}, 2018, \textbf{9}, 104--109\relax
\mciteBstWouldAddEndPuncttrue
\mciteSetBstMidEndSepPunct{\mcitedefaultmidpunct}
{\mcitedefaultendpunct}{\mcitedefaultseppunct}\relax
\EndOfBibitem
\bibitem[Sonnichsen \emph{et~al.}(2021)Sonnichsen, Strandell, Brosseau, and
  Kambhampati]{Sonnichsen2021}
C.~D. Sonnichsen, D.~P. Strandell, P.~J. Brosseau and P.~Kambhampati,
  \emph{Phys. Rev. Res.}, 2021, \textbf{3}, 023147\relax
\mciteBstWouldAddEndPuncttrue
\mciteSetBstMidEndSepPunct{\mcitedefaultmidpunct}
{\mcitedefaultendpunct}{\mcitedefaultseppunct}\relax
\EndOfBibitem
\bibitem[Makarov \emph{et~al.}(2016)Makarov, Guo, Isaienko, Liu, Robel, and
  Klimov]{Makarov2016}
N.~S. Makarov, S.~Guo, O.~Isaienko, W.~Liu, I.~Robel and V.~I. Klimov,
  \emph{Nano Letters}, 2016, \textbf{16}, 2349--2362\relax
\mciteBstWouldAddEndPuncttrue
\mciteSetBstMidEndSepPunct{\mcitedefaultmidpunct}
{\mcitedefaultendpunct}{\mcitedefaultseppunct}\relax
\EndOfBibitem
\bibitem[Mi \emph{et~al.}(2023)Mi, Atteberry, Mapara, Hidayatova, Gee, Furis,
  Yip, Weng, and Dong]{Mi2023}
C.~Mi, M.~L. Atteberry, V.~Mapara, L.~Hidayatova, G.~C. Gee, M.~Furis, W.~T.
  Yip, B.~Weng and Y.~Dong, \emph{The Journal of Physical Chemistry Letters},
  2023, \textbf{14}, 5466--5474\relax
\mciteBstWouldAddEndPuncttrue
\mciteSetBstMidEndSepPunct{\mcitedefaultmidpunct}
{\mcitedefaultendpunct}{\mcitedefaultseppunct}\relax
\EndOfBibitem
\bibitem[Nair \emph{et~al.}(2011)Nair, Zhao, and Bawendi]{Nair2011}
G.~Nair, J.~Zhao and M.~G. Bawendi, \emph{Nano Letters}, 2011, \textbf{11},
  1136--1140\relax
\mciteBstWouldAddEndPuncttrue
\mciteSetBstMidEndSepPunct{\mcitedefaultmidpunct}
{\mcitedefaultendpunct}{\mcitedefaultseppunct}\relax
\EndOfBibitem
\bibitem[Park \emph{et~al.}(2017)Park, Lim, Makarov, and Klimov]{Park2017}
Y.-S. Park, J.~Lim, N.~S. Makarov and V.~I. Klimov, \emph{Nano Letters}, 2017,
  \textbf{17}, 5607--5613\relax
\mciteBstWouldAddEndPuncttrue
\mciteSetBstMidEndSepPunct{\mcitedefaultmidpunct}
{\mcitedefaultendpunct}{\mcitedefaultseppunct}\relax
\EndOfBibitem
\bibitem[Eremchev \emph{et~al.}(2023)Eremchev, Tarasevich, Kniazeva, Li,
  Naumov, and Scheblykin]{Eremchev2023}
I.~Y. Eremchev, A.~O. Tarasevich, M.~A. Kniazeva, J.~Li, A.~V. Naumov and I.~G.
  Scheblykin, \emph{Nano Letters}, 2023, \textbf{23}, 2087--2093\relax
\mciteBstWouldAddEndPuncttrue
\mciteSetBstMidEndSepPunct{\mcitedefaultmidpunct}
{\mcitedefaultendpunct}{\mcitedefaultseppunct}\relax
\EndOfBibitem
\bibitem[Cho \emph{et~al.}(2022)Cho, Tahara, Yamada, Suzuura, Tadano, Sato,
  Saruyama, Hirori, Teranishi, and Kanemitsu]{Cho2022}
K.~Cho, H.~Tahara, T.~Yamada, H.~Suzuura, T.~Tadano, R.~Sato, M.~Saruyama,
  H.~Hirori, T.~Teranishi and Y.~Kanemitsu, \emph{Nano Letters}, 2022,
  \textbf{22}, 7674--7681\relax
\mciteBstWouldAddEndPuncttrue
\mciteSetBstMidEndSepPunct{\mcitedefaultmidpunct}
{\mcitedefaultendpunct}{\mcitedefaultseppunct}\relax
\EndOfBibitem
\bibitem[Zhu \emph{et~al.}(2024)Zhu, Feld, Svyrydenko, Cherniukh, Dirin,
  Bodnarchuk, Wood, Yazdani, Boehme, Kovalenko, and Rainò]{Zhu2024}
C.~Zhu, L.~G. Feld, M.~Svyrydenko, I.~Cherniukh, D.~N. Dirin, M.~I. Bodnarchuk,
  V.~Wood, N.~Yazdani, S.~C. Boehme, M.~V. Kovalenko and G.~Rainò,
  \emph{Advanced Optical Materials}, 2024, \textbf{12}, 2301534\relax
\mciteBstWouldAddEndPuncttrue
\mciteSetBstMidEndSepPunct{\mcitedefaultmidpunct}
{\mcitedefaultendpunct}{\mcitedefaultseppunct}\relax
\EndOfBibitem
\bibitem[Zhou and Zhang(2020)]{Zhou2020}
X.~Zhou and Z.~Zhang, \emph{AIP Advances}, 2020, \textbf{10}, 125015\relax
\mciteBstWouldAddEndPuncttrue
\mciteSetBstMidEndSepPunct{\mcitedefaultmidpunct}
{\mcitedefaultendpunct}{\mcitedefaultseppunct}\relax
\EndOfBibitem
\bibitem[Ramade \emph{et~al.}(2018)Ramade, Andriambariarijaona, Steinmetz,
  Goubet, Legrand, Barisien, Bernardot, Testelin, Lhuillier, Bramati, and
  Chamarro]{Ramade2018}
J.~Ramade, L.~M. Andriambariarijaona, V.~Steinmetz, N.~Goubet, L.~Legrand,
  T.~Barisien, F.~Bernardot, C.~Testelin, E.~Lhuillier, A.~Bramati and
  M.~Chamarro, \emph{Applied Physics Letters}, 2018, \textbf{112}, 072104\relax
\mciteBstWouldAddEndPuncttrue
\mciteSetBstMidEndSepPunct{\mcitedefaultmidpunct}
{\mcitedefaultendpunct}{\mcitedefaultseppunct}\relax
\EndOfBibitem
\bibitem[Iaru \emph{et~al.}(2017)Iaru, Geuchies, Koenraad, Vanmaekelbergh, and
  Silov]{Iaru2017}
C.~M. Iaru, J.~J. Geuchies, P.~M. Koenraad, D.~Vanmaekelbergh and A.~Y. Silov,
  \emph{ACS Nano}, 2017, \textbf{11}, 11024--11030\relax
\mciteBstWouldAddEndPuncttrue
\mciteSetBstMidEndSepPunct{\mcitedefaultmidpunct}
{\mcitedefaultendpunct}{\mcitedefaultseppunct}\relax
\EndOfBibitem
\bibitem[Chirvony \emph{et~al.}(2017)Chirvony, González-Carrero, Suárez,
  Galian, Sessolo, Bolink, Martínez-Pastor, and Pérez-Prieto]{Chirvony2017}
V.~S. Chirvony, S.~González-Carrero, I.~Suárez, R.~E. Galian, M.~Sessolo,
  H.~J. Bolink, J.~P. Martínez-Pastor and J.~Pérez-Prieto, \emph{Journal of
  Physical Chemistry C}, 2017, \textbf{121}, 13381--13390\relax
\mciteBstWouldAddEndPuncttrue
\mciteSetBstMidEndSepPunct{\mcitedefaultmidpunct}
{\mcitedefaultendpunct}{\mcitedefaultseppunct}\relax
\EndOfBibitem
\bibitem[Ma \emph{et~al.}(2019)Ma, Pan, Li, Shen, Ma, Zhang, Niu, Zhu, Xu, and
  Ye]{Ma2019}
X.~Ma, F.~Pan, H.~Li, P.~Shen, C.~Ma, L.~Zhang, H.~Niu, Y.~Zhu, S.~Xu and
  H.~Ye, \emph{Journal of Physical Chemistry Letters}, 2019, \textbf{10},
  5989--5996\relax
\mciteBstWouldAddEndPuncttrue
\mciteSetBstMidEndSepPunct{\mcitedefaultmidpunct}
{\mcitedefaultendpunct}{\mcitedefaultseppunct}\relax
\EndOfBibitem
\bibitem[Dey \emph{et~al.}(2018)Dey, Rathod, and Kabra]{Dey2018}
A.~Dey, P.~Rathod and D.~Kabra, \emph{Advanced Optical Materials}, 2018,
  \textbf{6}, 1800109\relax
\mciteBstWouldAddEndPuncttrue
\mciteSetBstMidEndSepPunct{\mcitedefaultmidpunct}
{\mcitedefaultendpunct}{\mcitedefaultseppunct}\relax
\EndOfBibitem
\bibitem[Becker \emph{et~al.}(2020)Becker, Bernasconi, Bodnarchuk, Rainò,
  Kovalenko, Norris, Mahrt, and Stöferle]{Becker2020}
M.~A. Becker, C.~Bernasconi, M.~I. Bodnarchuk, G.~Rainò, M.~V. Kovalenko,
  D.~J. Norris, R.~F. Mahrt and T.~Stöferle, \emph{ACS Nano}, 2020,
  \textbf{14}, 14939--14946\relax
\mciteBstWouldAddEndPuncttrue
\mciteSetBstMidEndSepPunct{\mcitedefaultmidpunct}
{\mcitedefaultendpunct}{\mcitedefaultseppunct}\relax
\EndOfBibitem
\bibitem[Shinde \emph{et~al.}(2017)Shinde, Gahlaut, and Mahamuni]{Shinde2017}
A.~Shinde, R.~Gahlaut and S.~Mahamuni, \emph{The Journal of Physical Chemistry
  C}, 2017, \textbf{121}, 14872--14878\relax
\mciteBstWouldAddEndPuncttrue
\mciteSetBstMidEndSepPunct{\mcitedefaultmidpunct}
{\mcitedefaultendpunct}{\mcitedefaultseppunct}\relax
\EndOfBibitem
\bibitem[He \emph{et~al.}(2016)He, Yu, Li, Li, Si, Jin, Wang, Wang, He, Wang,
  Zhang, and Ye]{He2016}
H.~He, Q.~Yu, H.~Li, J.~Li, J.~Si, Y.~Jin, N.~Wang, J.~Wang, J.~He, X.~Wang,
  Y.~Zhang and Z.~Ye, \emph{Nature Communications}, 2016, \textbf{7},
  10896\relax
\mciteBstWouldAddEndPuncttrue
\mciteSetBstMidEndSepPunct{\mcitedefaultmidpunct}
{\mcitedefaultendpunct}{\mcitedefaultseppunct}\relax
\EndOfBibitem
\bibitem[He \emph{et~al.}(2021)He, Han, Guo, and Wu]{He2021}
S.~He, Y.~Han, J.~Guo and K.~Wu, \emph{ACS Energy Letters}, 2021, \textbf{6},
  2786--2791\relax
\mciteBstWouldAddEndPuncttrue
\mciteSetBstMidEndSepPunct{\mcitedefaultmidpunct}
{\mcitedefaultendpunct}{\mcitedefaultseppunct}\relax
\EndOfBibitem
\bibitem[Santomauro \emph{et~al.}(2016)Santomauro, Grilj, Mewes, Nedelcu,
  Yakunin, Rossi, Capano, Al~Haddad, Budarz, Kinschel, Ferreira, Rossi,
  Gutierrez~Tovar, Grolimund, Samson, Nachtegaal, Smolentsev, Kovalenko, and
  Chergui]{Santomauro2016}
F.~G. Santomauro, J.~Grilj, L.~Mewes, G.~Nedelcu, S.~Yakunin, T.~Rossi,
  G.~Capano, A.~Al~Haddad, J.~Budarz, D.~Kinschel, D.~S. Ferreira, G.~Rossi,
  M.~Gutierrez~Tovar, D.~Grolimund, V.~Samson, M.~Nachtegaal, G.~Smolentsev,
  M.~V. Kovalenko and M.~Chergui, \emph{Structural Dynamics}, 2016, \textbf{4},
  044002\relax
\mciteBstWouldAddEndPuncttrue
\mciteSetBstMidEndSepPunct{\mcitedefaultmidpunct}
{\mcitedefaultendpunct}{\mcitedefaultseppunct}\relax
\EndOfBibitem
\bibitem[Neukirch \emph{et~al.}(2016)Neukirch, Nie, Blancon, Appavoo, Tsai,
  Sfeir, Katan, Pedesseau, Even, Crochet, Gupta, Mohite, and
  Tretiak]{Neukirch2016}
A.~J. Neukirch, W.~Nie, J.-C. Blancon, K.~Appavoo, H.~Tsai, M.~Y. Sfeir,
  C.~Katan, L.~Pedesseau, J.~Even, J.~J. Crochet, G.~Gupta, A.~D. Mohite and
  S.~Tretiak, \emph{Nano Letters}, 2016, \textbf{16}, 3809--3816\relax
\mciteBstWouldAddEndPuncttrue
\mciteSetBstMidEndSepPunct{\mcitedefaultmidpunct}
{\mcitedefaultendpunct}{\mcitedefaultseppunct}\relax
\EndOfBibitem
\bibitem[Ma \emph{et~al.}(2020)Ma, Shen, Wang, Pan, Chen, Xu, and Ye]{Ma2020}
X.~Ma, P.~Shen, Y.~N. Wang, F.~Pan, G.~Chen, S.~Xu and H.~Ye, \emph{Journal of
  Physical Chemistry C}, 2020, \textbf{124}, 27130--27135\relax
\mciteBstWouldAddEndPuncttrue
\mciteSetBstMidEndSepPunct{\mcitedefaultmidpunct}
{\mcitedefaultendpunct}{\mcitedefaultseppunct}\relax
\EndOfBibitem
\end{mcitethebibliography}


\providecommand*{\mcitethebibliography}{\thebibliography}
\csname @ifundefined\endcsname{endmcitethebibliography}
{\let\endmcitethebibliography\endthebibliography}{}
\begin{mcitethebibliography}{2}
\providecommand*{\natexlab}[1]{#1}
\providecommand*{\mciteSetBstSublistMode}[1]{}
\providecommand*{\mciteSetBstMaxWidthForm}[2]{}
\providecommand*{\mciteBstWouldAddEndPuncttrue}
  {\def\EndOfBibitem{\unskip.}}
\providecommand*{\mciteBstWouldAddEndPunctfalse}
  {\let\EndOfBibitem\relax}
\providecommand*{\mciteSetBstMidEndSepPunct}[3]{}
\providecommand*{\mciteSetBstSublistLabelBeginEnd}[3]{}
\providecommand*{\EndOfBibitem}{}
\mciteSetBstSublistMode{f}
\mciteSetBstMaxWidthForm{subitem}
{(\emph{\alph{mcitesubitemcount}})}
\mciteSetBstSublistLabelBeginEnd{\mcitemaxwidthsubitemform\space}
{\relax}{\relax}

\bibitem[Podshivaylov \emph{et~al.}(2023)Podshivaylov, Kniazeva, Tarasevich,
  Eremchev, Naumov, and Frantsuzov]{Podshivaylov2023}
E.~Podshivaylov, M.~Kniazeva, A.~Tarasevich, I.~Eremchev, A.~Naumov and
  P.~Frantsuzov, \emph{Journal of Materials Chemistry C}, 2023, \textbf{11},
  8570--8576\relax
\mciteBstWouldAddEndPuncttrue
\mciteSetBstMidEndSepPunct{\mcitedefaultmidpunct}
{\mcitedefaultendpunct}{\mcitedefaultseppunct}\relax
\EndOfBibitem
\bibitem[Bouchet \emph{et~al.}(2019)Bouchet, Krachmalnicoff, and
  Izeddin]{Bouchet2019}
D.~Bouchet, V.~Krachmalnicoff and I.~Izeddin, \emph{Opt. Express}, 2019,
  \textbf{27}, 21239--21252\relax
\mciteBstWouldAddEndPuncttrue
\mciteSetBstMidEndSepPunct{\mcitedefaultmidpunct}
{\mcitedefaultendpunct}{\mcitedefaultseppunct}\relax
\EndOfBibitem
\end{mcitethebibliography}
\bibliographystyle{rsc} 
\end{document}


\maketitle

\newpage
\section*{Supplementary Note 1: STEM images of the studied PNCs}

\begin{figure}[h!]
\center{\includegraphics[width=1\linewidth]{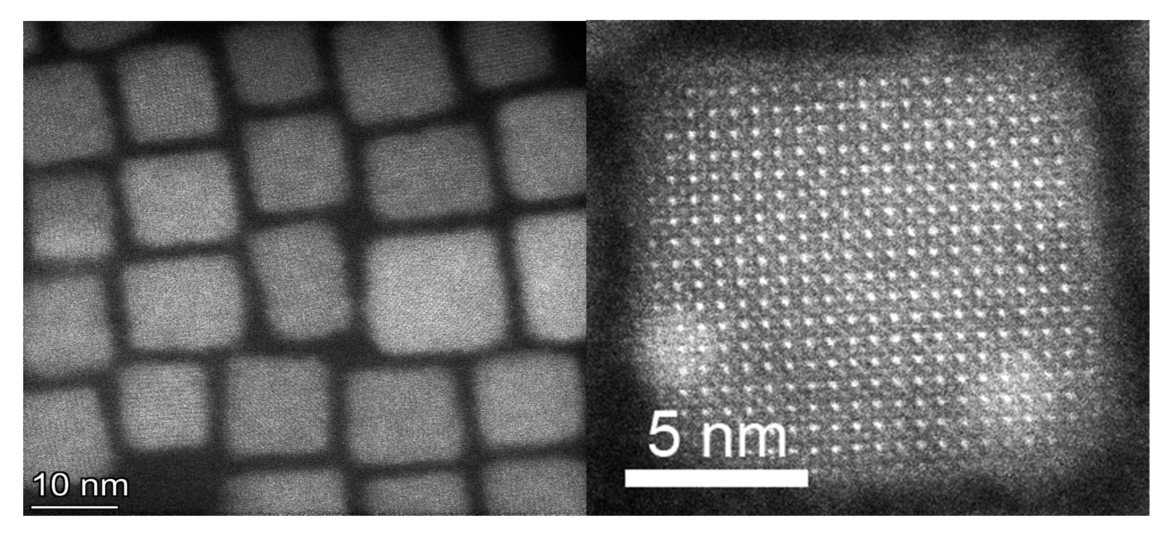}}
\caption{Left panel: STEM image of the ensemble of CsPbBr\textsubscript{3} PNCs. Right panel: STEM image of a single PNC.}
\label{STEM}
\end{figure}

\newpage
\section*{Supplementary Note 2: Obtaining two-dimensional distributions}

To obtain two-dimensional distributions, we divided the range of PL intensities (number of photon counts per bin) into levels numbered with index $m$.
The PL intensity of the $m$-th level is $N_m$.
 The number of levels depends on the maximum PL intensity. Typically one level has a width of two to three photons per bin.
The detected photons corresponding to each level were then divided into groups containing 1000 photons each.
The set of photon delay times of each group $(m,j)$ is used to calculate the PL decay curve.
In this way, one can obtain a set of PL decay curves, each of them is a numerical sequence $N_i^{(m,j)}$, where $N_i^{(m,j)}$ is the number of photon counts with delay time within the $i$-th time bin.
 The $i$-th time bin is defined as time interval between $t_i \equiv i\delta t$ and $t_{i+1}$, where $\delta t$ is the width of the time bin.
We assume the PL decay to have a biexponential form with background noise for any given set $(m,j)$:
\begin{equation}
    w_i(a,\tau_F,\tau_D,b) = A\left(\frac{1}{\tau_F}e^{-t_i/\tau_F} + \frac{a}{\tau_D}e^{-t_1/\tau_D} + b\right)
    \label{Biexp}
\end{equation}
where $w_i$ is the probability of detecting a photon within the $i$-th bin,  $a$ is a relative intensity of the delayed component, and $b$ is the background level.
The coefficient $A$ is determined as:
$$A^{-1}=\sum\limits_i \left(\frac{1}{\tau_F}e^{-t_i/\tau_F} + \frac{a}{\tau_D}e^{-t_1/\tau_D} + b\right)$$
Thus, $w_i$  satisfies the normalization condition:
\begin{equation}
    \sum_i w_i(a,\tau_F,\tau_D,b) = 1.
\end{equation}

 To estimate the parameters $a$, $\tau_F$, $\tau_D$, and $b$, we applied  the likelihood function for each group $(m,j)$ using the multinomial distribution as follows:
\begin{equation}
    \ln L^{(m,j)}(a,\tau_F,\tau_D,b) = \sum N_i^{(m,j)} \ln w_i(a,\tau_F,\tau_D,b) + const
    \label{MLflid}
\end{equation}
 We have found the minima of the negative likelihood function logarithm Eq.(\ref{MLflid}) using the MATLAB fminsearch function.
The resulting estimated parameters are denoted as $a^{(m,j)}$, $\tau_F^{(m,j)}$, $\tau_D^{(m,j)}$, and $b^{(m,j)}$.
Estimated PL intensities of the fast and delayed component are
$$N_F^{(m,j)}=N_m Q_F\left(a^{(m,j)},\tau_F^{(m,j)},\tau_D^{(m,j)},b^{(m,j)}\right)$$
$$N_D^{(m,j)}=N_m Q_D\left(a^{(m,j}),\tau_F^{(m,j)},\tau_D^{(m,j)},b^{(m,j)}\right)$$
 where
 $$Q_F(a,\tau_F,\tau_D,b)=\sum_i  A\frac{1}{\tau_F}e^{-t_i/\tau_F}$$
 $$Q_D(a,\tau_F,\tau_D,b)=\sum_i  A\frac{a}{\tau_D}e^{-t_i/\tau_D}$$

We assume that the maximum value of $N_F^{(m,j)} + N_D^{(m,j)}$ for all groups
$$N_{\max}=\max_{m,j}\,(N_F^{(m,j)} + N_D^{(m,j)})$$
corresponds to a PL quantum yield of unity.
Then, the PL quantum yields of the  fast and slow components can be estimated as
\begin{equation}
    Y_F^{(m,j)} = \frac{N_F^{(m,j)}}{N_{\max}}, \: Y_D^{(m,j)} = \frac{N_D^{(m,j)}}{N_{\max}}
\end{equation}

Finally, we get four parameter estimations $\tau_F^{(m,j)}$, $\tau_D^{(m,j)}$, $Y_F^{(m,j)}$, and $Y_D^{(m,j)}$ for each group of photons. This allows us to obtain four different two-dimensional distributions $\rho_1(\tau_F,\tau_D)$, $\rho_2(\tau_F,Y_F)$, $\rho_3(\tau_D,Y_D)$ and $\rho_4(Y_D + Y_F,Y_F)$, which are presented in Fig.~4 in the main text and in the Supplementary Note 6.

\newpage
\section*{Supplementary Note 3: Estimation of the biexciton lifetime}

\begin{figure}[h]
    \centering
    \includegraphics[width=1\textwidth]{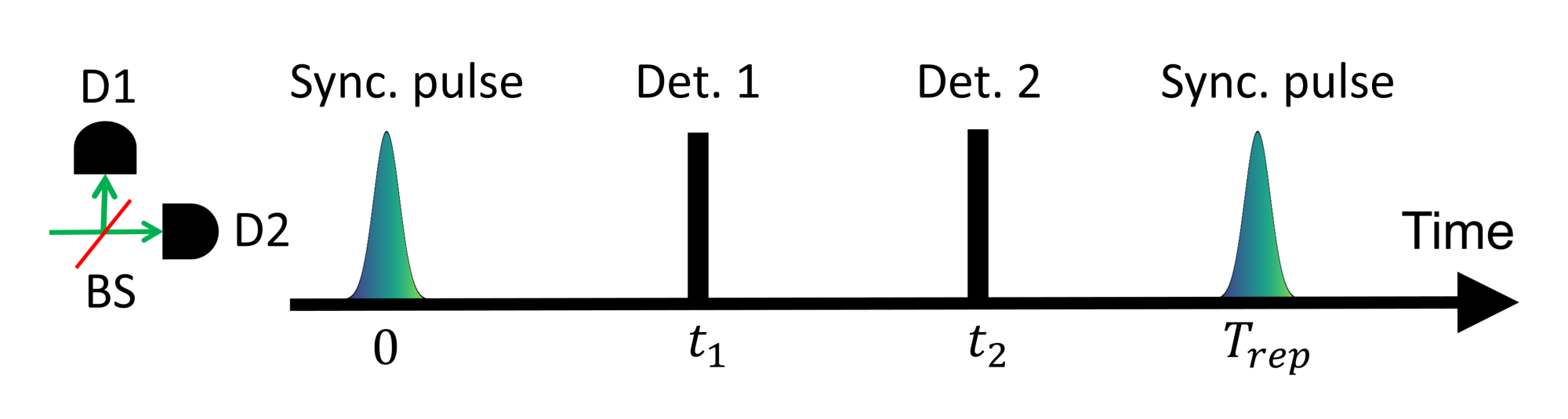}
    \caption{Schematic picture of pair detection after the one pulse excitation}
    \label{Pair}
\end{figure}

To estimate the biexciton lifetime, we only collected   events when both detectors in the Hanbury Brown and Twiss scheme were triggered as a result of a single excitation pulse.
The schematic picture of such an event is presented in Fig.~\ref{Pair}: two photons are registered on  different detectors.
  Denoting the smallest of time delays (the first photon delay time) as $t_1$, one can obtain the decay curve as a numerical sequence $N^{XX}_i$,
  where $N^{XX}_i$ is the number of first photons delay times within the $i$-th time bin.
 This distribution can be fitted with the biexponential decay law Eq.(\ref{Biexp}) in the same manner as was done in the Supplementary Note 2.
 An example of the $t_1$ normalized distribution
 \begin{equation}
 p^{XX}_i=\frac{N^{XX}_i}{\sum\limits_k N^{XX}_i}
 \label{pXX}
 \end{equation}
 and its fit is presented in Fig.~5a in the main text.

\newpage
\section*{Supplementary Note 4: Analytical solution for excitation relaxation kinetics}

The kinetic mechanism of the excitation relaxation proposed in the main article (Fig.~6) is described by the following set of equations:

\begin{equation}
 \begin{cases}
\frac{d p_e}{dt}=-(k_r+k_n+k_t)p_e+k_d p_t\equiv -\Gamma_e p_e + k_d p_t
\\
\frac{d p_t}{dt}=-(k_r'+k_n'+k_d)p_t
 + k_t p_e\equiv -\Gamma_t p_e + k_t p_e
\end{cases}
\label{sys_sup}
\end{equation}
with the initial conditions:

\begin{equation}
    p_e(0)=p_0, \quad p_t(0)=1-p_0
    \label{cond_sup}
\end{equation}
The system Eq.(\ref{sys_sup}) can be written in matrix form:
\begin{equation}
\frac{d}{dt}
    \begin{pmatrix}
        p_e(t)\\
        p_t(t)
    \end{pmatrix} = \boldsymbol{\hat{M}}     \begin{pmatrix}
        p_e(t)\\
        p_t(t)
    \end{pmatrix}
    \label{mat}
\end{equation}
where:
$$\boldsymbol{\hat{M}}=\begin{pmatrix}
-\Gamma_e&k_d\\
k_t&-\Gamma_t
\end{pmatrix}$$
The eigenvalues of the matrix $\boldsymbol{\hat{M}}$ are solutions of the following equation:

\begin{equation}
    \det\left(\boldsymbol{\hat{M}} - \lambda \boldsymbol{\hat{E}}\right) = \lambda^2+\lambda (\Gamma_e+\Gamma_t)+(\Gamma_e \Gamma_t- k_t k_d)=0
    \label{lam}
\end{equation}
where $\boldsymbol{\hat{E}}$ is the unity matrix. Solving Eq.(\ref{lam}), we obtain two eigenvalues:
\begin{equation}
    \lambda_{F,D}\equiv - \frac{1}{\tau_{F,D}}=-\frac{1}{2}\left[(\Gamma_e+\Gamma_t) \pm \left.|\Gamma_e-\Gamma_t\right.|\sqrt{1+\frac{4 k_t k_d}{(\Gamma_e-\Gamma_t)^2}}\right].
\end{equation}
That corresponds to Eq. (6) in the main text for fast and delayed component times. The equation for eigenvectors is as follows:

\begin{equation}
    \left(\boldsymbol{\hat{M}} - \lambda_{F,D} \boldsymbol{\hat{E}}\right) \boldsymbol{\vec{X}}_{1,2} = 0.
\end{equation}
The solutions are as follows:
\begin{equation}
    \boldsymbol{\vec{X}}_{1} =
    \begin{pmatrix}
    X_1
  \\
    1
    \end{pmatrix};
    \quad \lambda = \lambda_F,
\end{equation}

\begin{equation}
    \boldsymbol{\vec{X}}_{2} =     \begin{pmatrix}
    X_2
  \\
    1
    \end{pmatrix};
    \quad \lambda = \lambda_D,
\end{equation}
where:

$$ X_1 =2 k_d \left[(\Gamma_e - \Gamma_t) - \left.|\Gamma_e-\Gamma_t\right.|\sqrt{1+\frac{4 k_t k_d}{(\Gamma_e-\Gamma_t)^2}}\right]^{-1},$$
$$ X_2 = 2 k_d \left[(\Gamma_e - \Gamma_t) + \left.|\Gamma_e-\Gamma_t\right.|\sqrt{1+\frac{4 k_t k_d}{(\Gamma_e-\Gamma_t)^2}}\right]^{-1}.$$

A general solution of Eq.(\ref{mat}) is a sum of two vectors:
\begin{equation}
\begin{pmatrix}p_e(t)\\p_t(t)\end{pmatrix}=C_F\begin{pmatrix}
X_{1}\\1\end{pmatrix}e^{-t/\tau_F}+C_D\begin{pmatrix}
X_{2}\\1\end{pmatrix}e^{-t/\tau_D},
\end{equation}
where $C_1$ and $C_2$ are unknown constants that can be found using the initial conditions (\ref{cond_sup}). The solution for the coefficients is as follows:

\begin{equation}
C_F=\frac{X_2(1-p_0)-p_0}{X_2-X_1},
\end{equation}

\begin{equation}
    C_D=\frac{p_0-X_1(1-p_0)}{X_2-X_1}.
\end{equation}
PL decay curve is given by the following expression:
\begin{equation}
p(t)=k_r p_e(t)+k_r' p_t(t)
\label{PL}
\end{equation}
 Finally we obtain expressions for the quantum yields:
\begin{equation}
    Y_F = C_F \tau_F (k_r X_1+ k_r') , \quad Y_D = C_D \tau_D (k_r X_2+ k_r')
    \label{yield_th}
\end{equation}
The total quantum yield of PL emission is defined as
\begin{equation}
  Y=Y_F+Y_D
    \label{yield}
\end{equation}

The function $p(t)$, characteristic times $\tau_F$,  $\tau_D$  and quantum yields $Y_F$, $Y_D$, and $Y$ are determined by the values of the kinetic parameters
$$k_r, k_r',k_n, k_n', k_t, k_d, p_0$$

\newpage
\section*{Supplementary Note 5. Fitting procedure}

\subsection*{Fitting the PL decay kinetics}

From the PL intensity levels defined in the Supplementary Note 2, we selected seven equally spaced levels with indices $m_k$, where $k=1\dots 7$.
We will denote $ (k) \equiv m_k$.
We collected photon delay times within each level and obtained seven PL decay curves $N^{(k)}_i$.
Then we defined the likelihood function for each level according to the multinomial distribution:

\begin{equation}
    \ln l^{(k)} \left(\boldsymbol{\vec{\nu}},S_k\right) = \ln C^{(k)} + \sum_i N^{(k)}_i \ln w^{(k)}_i \left(\boldsymbol{\vec{\nu}},S_k\right),
\end{equation}
where $C^{(k)}$ is the multinomial coefficient:
$$ C^{(k)} \equiv \begin{pmatrix} \sum_i N^{(k)}_i \\
N^{(k)}_1 N^{(k)}_2  \ldots \; N^{(k)}_{i_m}
\end{pmatrix} = \frac{\left(\sum_i N^{(k)}_i\right)!}{N^{(k)}_1!  \ldots N^{(k)}_{i_m}!} $$
The $ b_k $ parameter describes the level of background emission, $ w_i^{(k)} $ is the  theoretical distribution of probabilities
\begin{equation}
    w_i^{(k)} \left(\boldsymbol{\vec{\nu}},b_k\right) = \frac{p\left[\boldsymbol{\vec{\nu}},S_k\right](t_i) +b_k}{\sum_i \left(p\left[\boldsymbol{\vec{\nu}},S_k\right](t_i) +b_k\right)}
\end{equation}
$\boldsymbol{\vec{\nu}}$ is the vector containing the fitting parameters
$$\boldsymbol{\vec{\nu}}=\left(k_r, k_r',k_0, k_0', k_t, p_0, \Delta E\right)$$

The Huang-Rhys parameter for the $(k)$-th level $S_k$ is found by solving the following equation:
\begin{equation}
Y\left[\boldsymbol{\vec{\nu}},S_k\right] = \frac {N_{(k)}}{N_M}
\label{Y}
\end{equation}
$p\left[\boldsymbol{\vec{\nu}},S\right](t)$ and  $Y\left[\boldsymbol{\vec{\nu}},S\right]$ are  the PL decay function Eq.(\ref{PL}) and the total quantum yield Eq.(\ref{yield}), respectively,  with the following parameters:
$$k_r, k_r',k_n=k_0 S^\alpha, k_n'=k_0'S^{\alpha'}, k_t, k_d= k_t \exp(-\Delta E/k_BT), p_0$$
The total likelihood function is as follows:
\begin{equation}
    \ln L \left(\boldsymbol{\vec{\nu}},\boldsymbol{\vec{b}} \right) = \sum_{k=1}^{7} \ln l^{(k)} = \sum_{k=1}^{7}  \left( \ln C^{(k)} + \sum_i N^{(k)}_i \ln w_i^{(k)} \left(\boldsymbol{\vec{\nu}},b_k\right) \right)
\end{equation}
where $\boldsymbol{\vec{b}}$ is the vector containing $b_k$ values.
The fitting parameters $\boldsymbol{\vec{\nu}}$ and $\boldsymbol{\vec{b}}$ estimations were found by searching for the minimum negative value of the likelihood function logarithm using the MATLAB fminsearch function.
Eq.(\ref{Y}) was solved numerically  for every step of the fitting procedure using the MATLAB fminserch function.

Using the estimated parameters, the theoretical Huang-Rhys parameter dependencies of $\tau_F(S)$, $\tau_D(S)$, $Y_F(S)$, $Y_D(S)$ and $Y(S)$ were calculated.
As a result, we obtain theoretical dependencies $\tau_F\,-\,\tau_D$, $\tau_F\,-\,Y_F$, $\tau_D\,-\,D$ and $Y\,-\,Y_F$.
These dependencies are shown in Fig.~8 of the main article as red lines.

\subsection*{Fitting PSD and PDF }

The theoretical calculations of the power spectral density (Eq. (15) in the main article) and the photon distribution function  (Eq. (16) in the main article) within the proposed model  were performed according to the procedure described in detail in the Ref.~\citenum{Podshivaylov2023} Supplementary  notes 8 and 9.
The procedure used the theoretical dependence of the quantum yield on the Huang-Rhys parameter  $Y\left[\boldsymbol{\vec{\nu}},S\right]$, given by Eq.(\ref{yield}), which differs from the dependence used in Ref.~\citenum{Podshivaylov2023}.
Fitting of the estimated PSD and PDF  was  performed according to the procedure described in detail in the Ref.~\citenum{Podshivaylov2023} Supplementary  note 10.

\subsection*{Error estimation for two-dimensional distributions}

The mean values of $\tau_{F,D}$, $Y_{F,D}$, and $Y$ over a time bin, which had a starting system configuration of $\Sigma$, can be found as (see the Ref.~\citenum{Podshivaylov2023} Supplementary  note 11):

\begin{equation}
    \overline{\tau}_{F,D}(\Sigma) =  \sum_{\Sigma'} \tau_{F,D}(\Sigma') \overline{G}_{\Sigma'\Sigma}
\end{equation}

\begin{equation}
    \overline{Y}(\Sigma) =  \sum_{\Sigma'} Y_{\Sigma'} \overline{G}_{\Sigma'\Sigma}; \quad  \overline{Y}_{F,D}(\Sigma) =  \sum_{\Sigma'} Y_{F,D}(\Sigma') \overline{G}_{\Sigma'\Sigma}
\end{equation}
where
$$\overline{G}_{\Sigma'\Sigma} = \frac 1 {\Delta} \int\limits_0^{\Delta t} {G}_{\Sigma'\Sigma} (t)\,dt $$
and
$$\tau_{F,D}(\Sigma)\equiv \tau_{F,D}(S(\Sigma)); \quad Y_{F,D}(\Sigma)\equiv Y_{F,D}(S(\Sigma))$$
The mean number of detected photons per time bin is
$$\overline{N}_{\Sigma} = N_{\max} \overline{Y} (\Sigma)$$
The probability to detect $k$ photons within the bin is
$$P_{\Sigma}(k) = \frac {\overline{N}_{\Sigma}^k}{k!}\exp\left(-\overline{N}_{\Sigma}\right)$$
We chose bins that had a number of detected photons $k$ within level $m$, so $N_m \le k < N_{m+1}$.
The prior probability of the configuration $\Sigma$ is $P_{st}(\Sigma)$.
The posterior Bayes probability when $k$ is within the $m$-th  level is
 \begin{equation}
 P_m(\Sigma) = \frac{ \sum\limits_{N_m \le k < N_{m+1}} P_{\Sigma}(k) P_{st}(\Sigma)}{ \sum\limits_{N_m \le k < N_{m+1}} PDF(k)}
\end{equation}
The mean value of the PL quantum yield for the $m$-th level bin  is
$$ {\overline{Y}}_{m}=  \sum_{\Sigma} \overline{Y}_{\Sigma} P_m(\Sigma) $$
and the variance of ${Y}_{m}$ is

\begin{equation}
\mathrm{Var} \left[ {\overline{Y}_m}\right]=  \sum_{\Sigma} \overline{Y}_{\Sigma}^2 P_m(\Sigma) - {\overline{Y}}_{m}^2
\end{equation}

Recall that the estimations of the $\tau_{F,D}$ and $Y_{F,D}$ are calculated using the maximum likelihood method for multinomial distribution. We will consider the special case of $\delta t \ll \tau_F$ and $T_{rep} \gg \tau_D$. The probability density is as follows:

\begin{equation}
    w_{\theta}(t) = \frac{Q_F}{\tau_F} e^{-t/\tau_F} + \frac{Q_D}{\tau_D} e^{-t/\tau_D}
\end{equation}
where $Q_D = 1 - Q_F$ and $ \boldsymbol {\vec \theta}$ is the vector parameter
$$ { \boldsymbol{ \vec \theta}}= \left( \tau_F,  \tau_D,  Q_F\right)$$
We also neglect background noise.
The variance of an  unbiased estimator of any parameter $\theta_i$ obtained from the maximum likelihood method obeys the Cramer–Rao bound (see Ref.~\citenum{Bouchet2019} for example):

\begin{equation}
    \sigma^2_{\theta_i} \ge \left[ \boldsymbol {\hat F^{-1}} \right]_{ii}
    \label{variance}
\end{equation}
where $ \boldsymbol{\hat F}$ is the Fisher's information matrix with elements:

\begin{equation}
    F_{ij}  = N_p \int\limits_0^\infty \frac{1}{w_{\theta}(t)} \left(\frac{\partial w_{\theta}(t)}{\partial \theta_i}\right) \left(\frac{\partial w_{\theta}(t)}{\partial \theta_j}\right)\, dt
    \label{fisher}
\end{equation}
  where $N_p$ is the total number of counts in a single PL decay curve. The partial derivatives of the probability density are given by:

$$\frac{\partial w_{\theta}(t)}{\partial Q_F} = \frac{1}{\tau_F} e^{-t/\tau_F} - \frac{1}{\tau_D} e^{-t/\tau_D}$$

$$\frac{\partial w_{\theta}(t)}{\partial \tau_{F,D}} = \frac{Q_{F,D}}{\tau_{F,D}^2} e^{-t/\tau_{F,D}} \left(\frac{t}{\tau_{F,D}}-1\right)$$

Substituting these expressions into Eqs. (\ref{variance}-\ref{fisher}), one can obtain the variances of the parameters' estimators $\sigma^2_{\tau_F}$, $\sigma^2_{\tau_D}$, and $\sigma^2_{Q_F}$.
We have calculated these values  using the numerical integration for each configuration $\Sigma$  by setting
$$\tau_{F,D}=\tau_{F,D}(\Sigma); \quad Q_F (\Sigma) = Y_F(\Sigma)/Y(\Sigma)$$
 The obtained values are denoted as $\sigma^2_{\theta_i}(\Sigma)$.
After averaging the variances of the parameters over the configurations at the beginning of the bin and over the fluctuations within the bin, we obtain the total variances at the level $m$ :

\begin{equation}
    \overline{\sigma_{\theta_i}^2} =\sum_\Sigma \sum_{\Sigma'} \sigma^2_{\theta_i} (\Sigma') \overline{G}_{\Sigma' \Sigma}P_m(\Sigma)
\end{equation}

The variance for characteristic times at level $m$  can be then found as:

\begin{equation}
    \mathrm{Var}_m[\tau_{F,D}] =   \overline{\sigma_{\tau_{F,D}}^2}
\end{equation}

The estimators of $Y_{F,D}$ are the products of two independent estimators:
\begin{equation}
    Y_{F,D} = Q_{F,D}  \overline{Y}_m
\end{equation}

Thus, the total variance for $Y_{F,D}$ at level $m$ can be calculated as:
\begin{equation}
    \mathrm{Var}_m[Y_{F,D}] =  \bar{Y}_m^2 \overline{\sigma_{Q_F}^2} + Q_{F,D}^2 \mathrm{Var} \left[ {\overline{Y}_m}\right]
\end{equation}

The calculated $3\sigma$ confidence regions for the two levels are shown in Fig.~8 of the main article as yellow ellipses.

\newpage
\section*{Supplementary Note 6. Processing results for the studied PNCs}

\begin{figure}[h]
    \centering
    \includegraphics[width=1\textwidth]{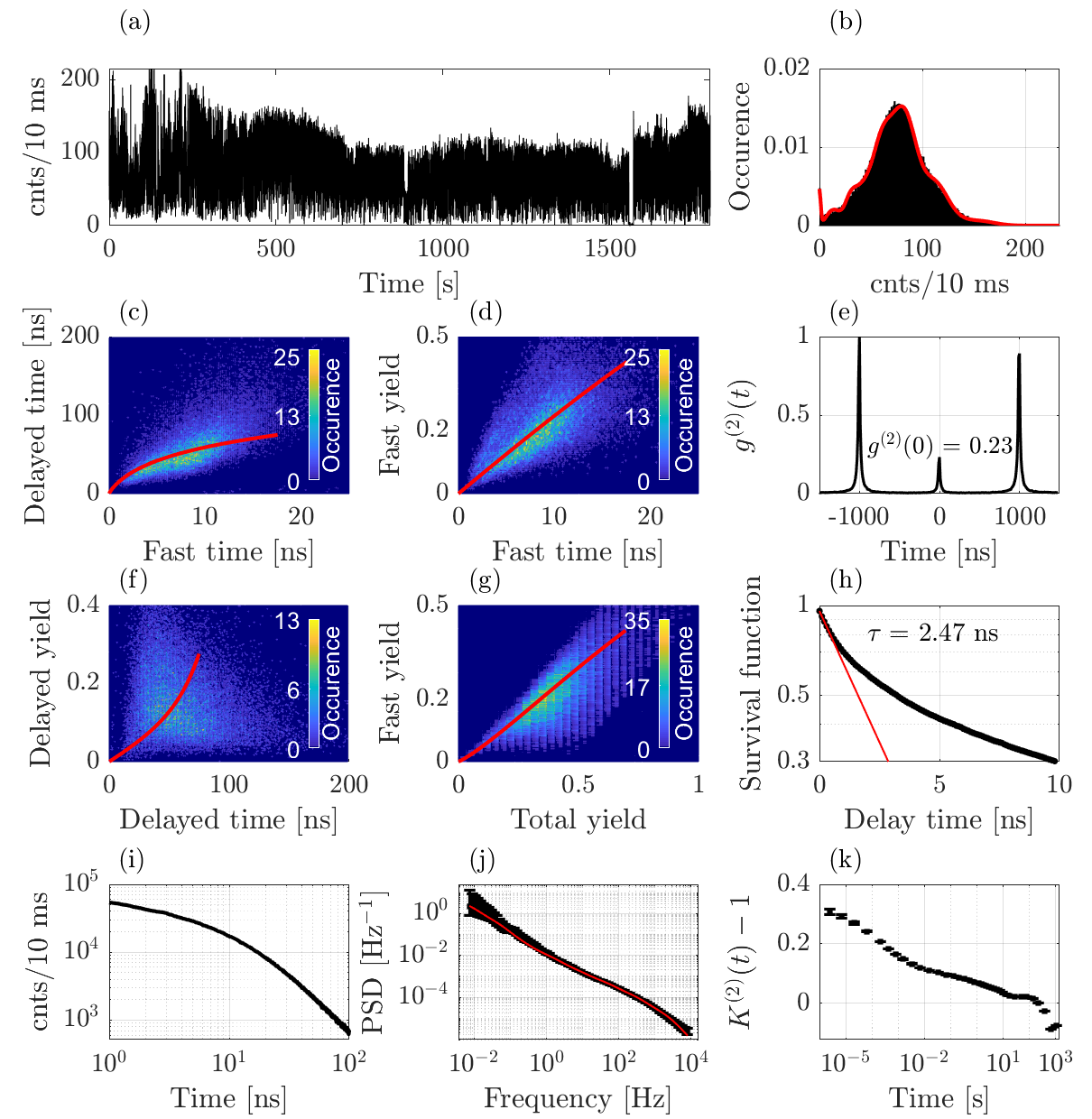}
    \caption{PNC \#1}
    \label{figS1}
\end{figure}

\begin{figure}[h]
    \centering
    \includegraphics[width=1\textwidth]{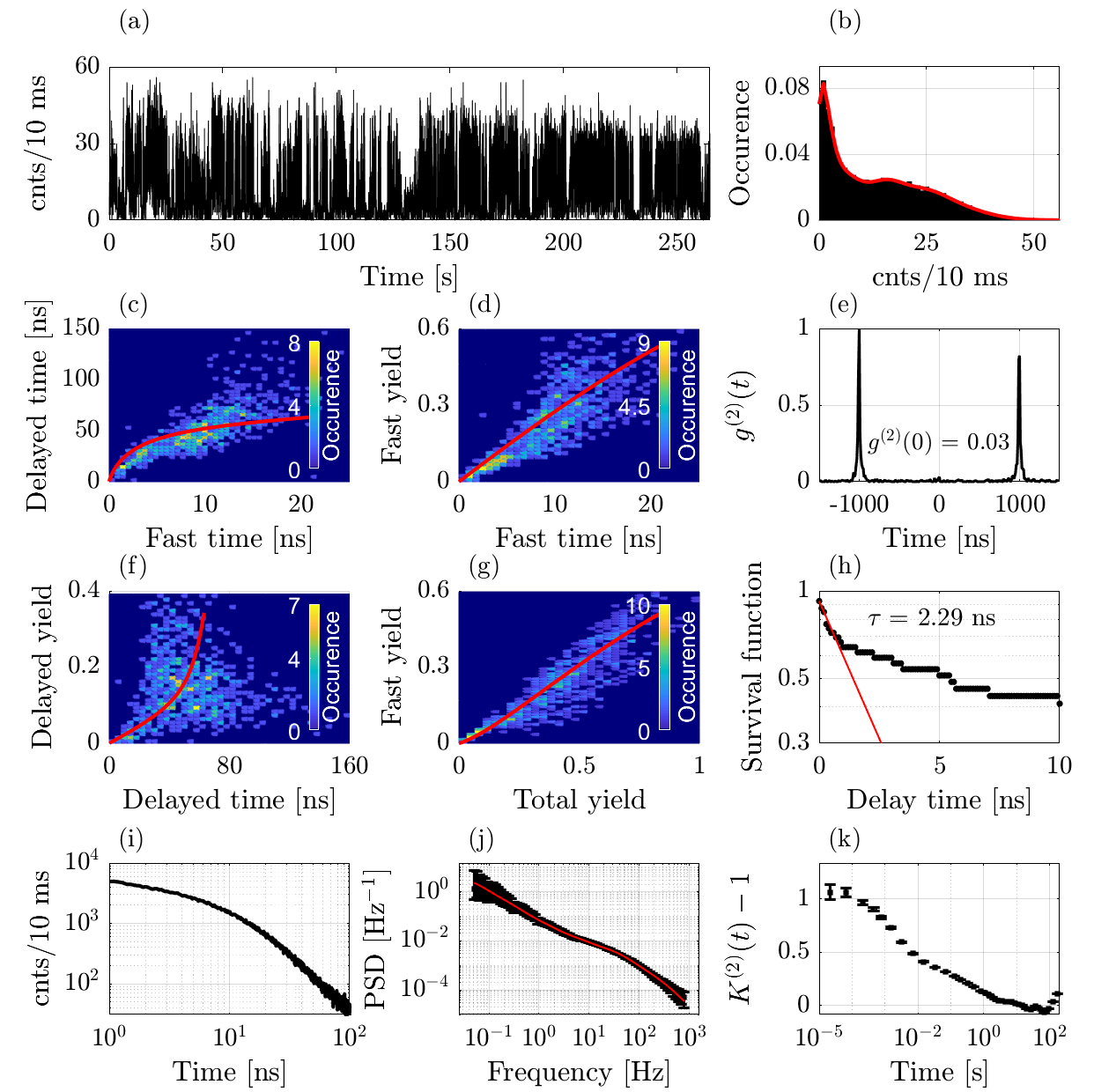}
    \caption{PNC \#2}
    \label{figS2}
\end{figure}

\begin{figure}[h]
    \centering
    \includegraphics[width=1\textwidth]{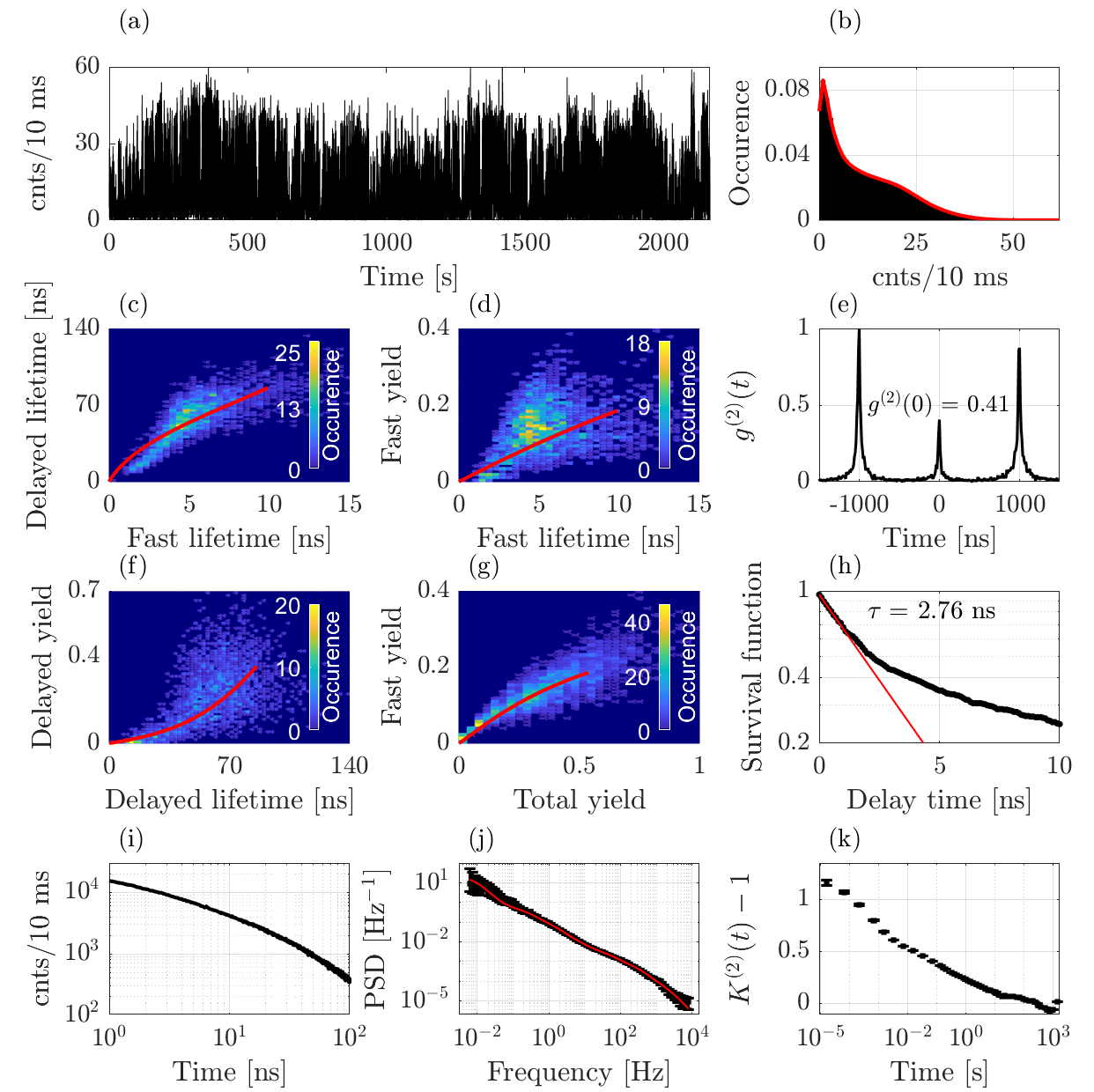}
    \caption{PNC \#3}
    \label{figS3}
\end{figure}

\begin{figure}[h]
    \centering
    \includegraphics[width=1\textwidth]{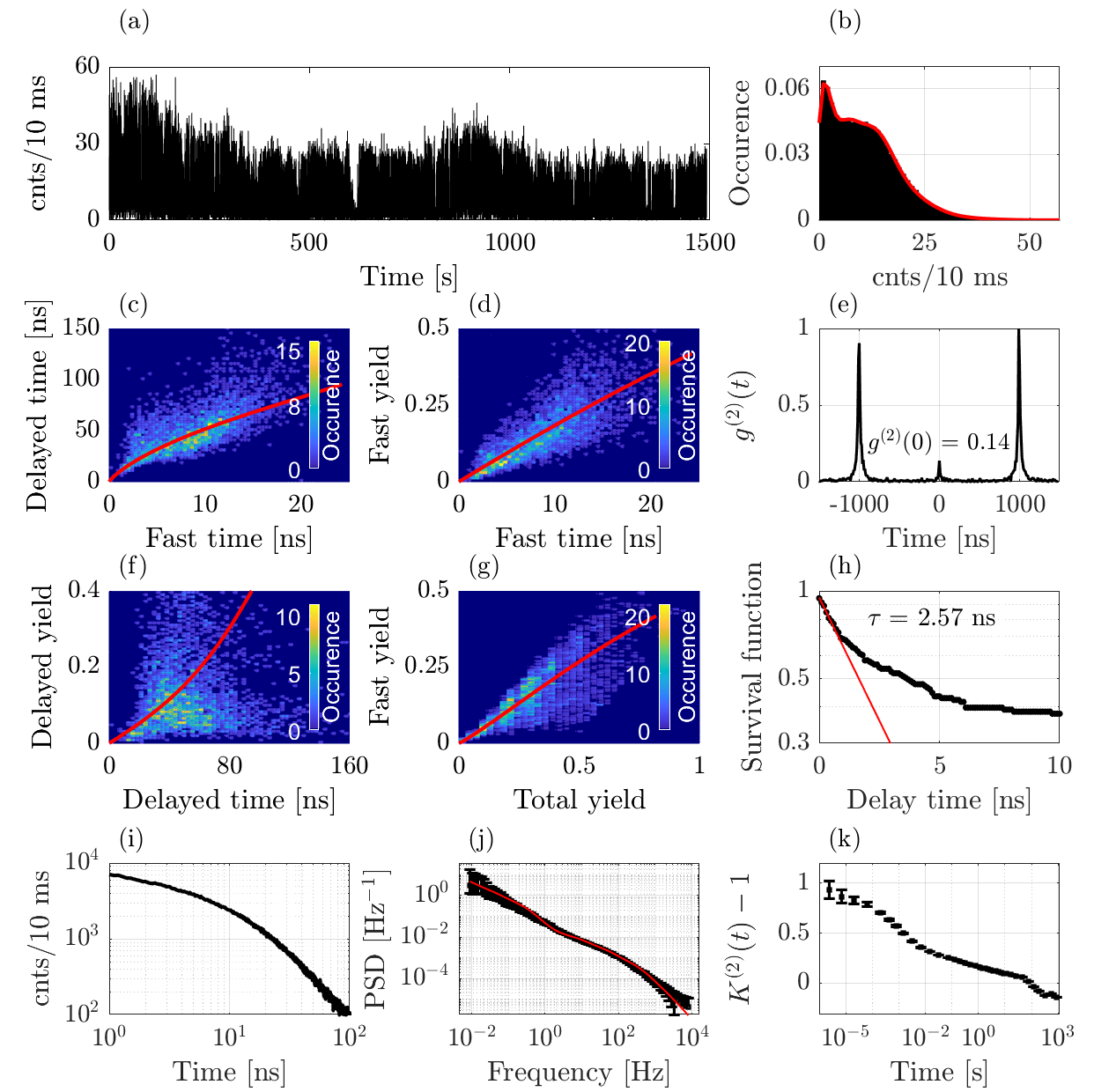}
    \caption{PNC \#4}
    \label{figS4}
\end{figure}

\begin{figure}[h]
    \centering
    \includegraphics[width=1\textwidth]{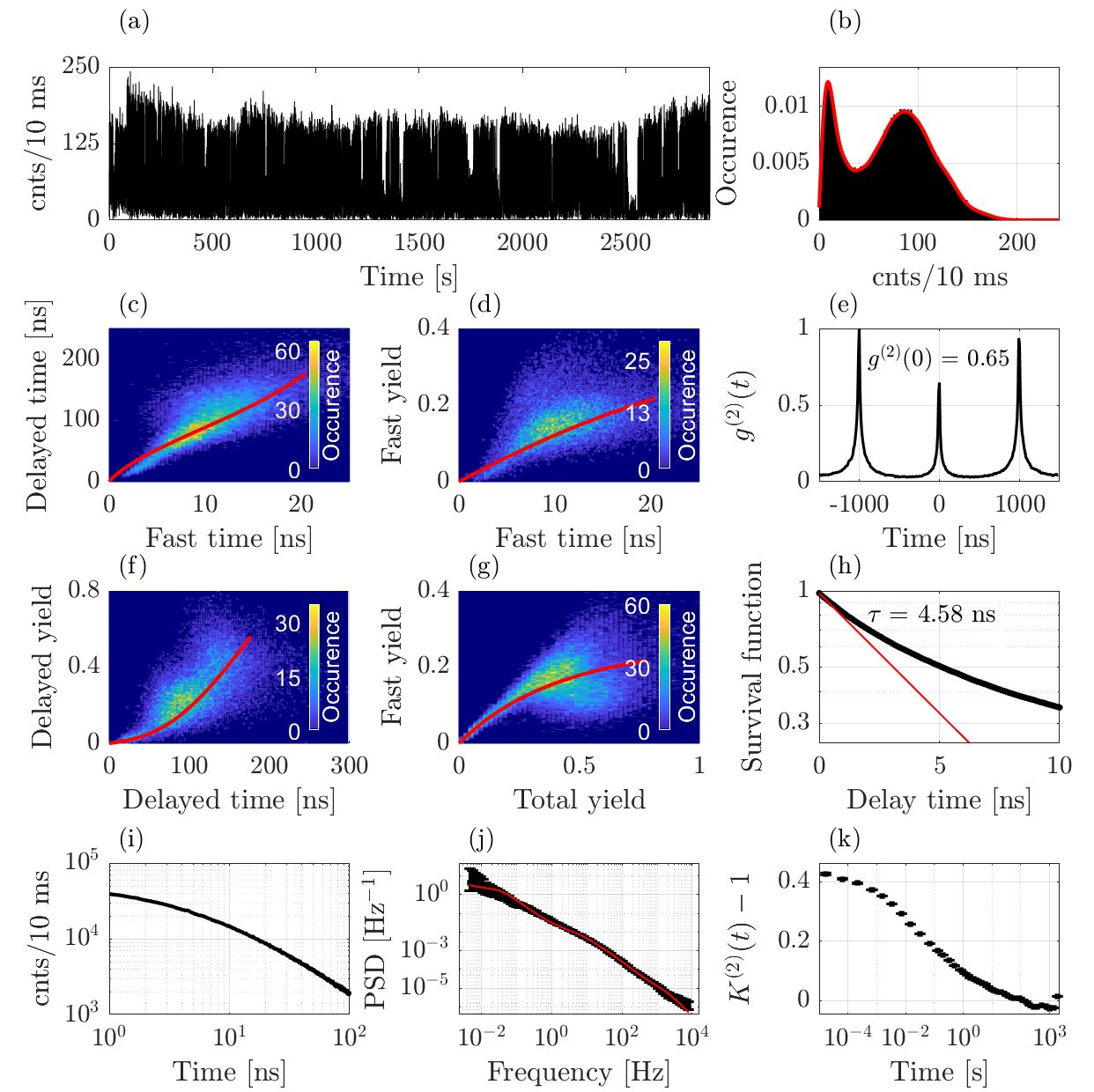}
    \caption{PNC \#5. The results for this PNC are presented in the main article.}
    \label{figS5}
\end{figure}

\begin{figure}[h]
    \centering
    \includegraphics[width=1\textwidth]{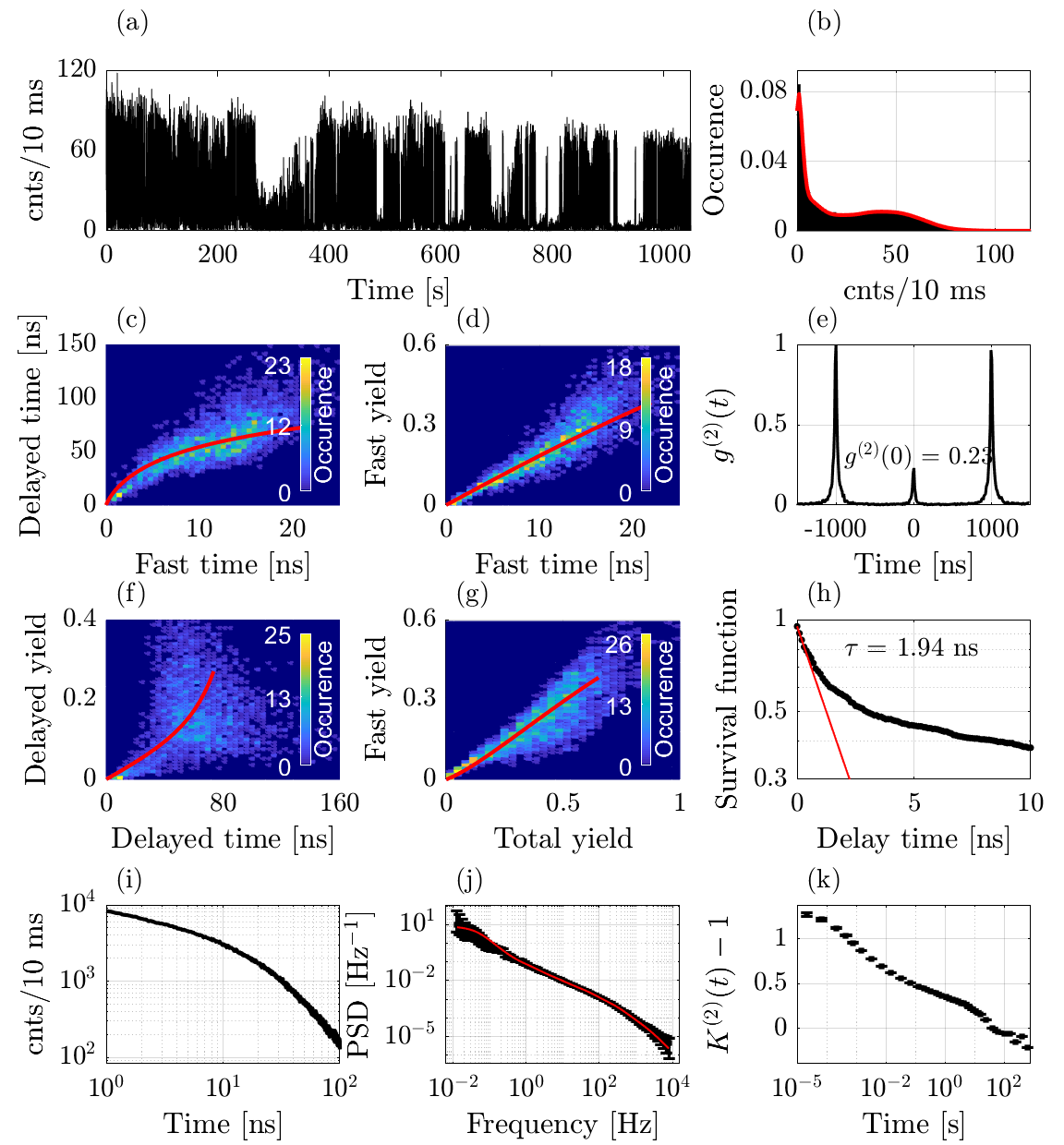}
    \caption{PNC \#6}
    \label{figS6}
\end{figure}

Figures \ref{figS1}~-~ \ref{figS6} present the processing results for  the studied PNCs. All figures  have the same structure.\\
 Panel (a) represents the PL intensity blinking trace. Panel (b) shows a photon distribution function with a theoretical fit (red line). \\
 Panels (c), (d), (f), and (g) represent the four two-dimensional distributions described in the previous sections with  corresponding fits (red lines).\\
Panel (e) shows the photon cross-correlation function on short times, normalized to the maximum. The value of $g^{(2)}(0)$ is indicated in each figure.\\
Panel (h) shows the first photon delay time characteristics, obtained according to the procedure described in the Supplementary Note 3.
Note that the panel shows the survival probability, not the distribution function Eq.(\ref{pXX}) presented in Fig.~5a in the main text.
 The survival probability is defined as
\begin{equation}
    S^{XX}_i = 1 - \frac{ \sum\limits_{k=0}^{i} N^{XX}_i} {\sum\limits_k N^{XX}_i}
\end{equation}
 Panel (h) shows the guide-to-eye exponential decay function (red line). The characteristic time is indicated in each figure.\\
  Panel (i) shows the integrated PL decay. \\
  Panel (j) presents the estimated power spectral density with its corresponding theoretical fit (red line).\\
   Panel (k) shows the long-term correlation function of the PL intensity.

\newpage

\newpage

\newpage

\newpage

\newpage

\newpage

\newpage
\section*{Supplementary Note 7. Model parameters obtained from fitting procedures.}

\begin{center}
\begin{tabular}{ |p{1.56cm}|p{1.49cm}|p{1.49cm}|p{1.49cm}|p{1.49cm}|p{1.49cm}| p{1.49cm}|}
 \hline
   & \multicolumn{6}{|c|}{PNC number} \\
 \hline
  & \#1 & \#2 &  \#3 &  \#4 &  \#5 &  \#6\\
 \hline
 $k_r $ [ns$^{-1}$] & 0.0325 & 0.343 & 0.0431 & 0.0286 & 0.0265  & 0.0238 \\
 \hline
  $k_r$' [ns$^{-1}$] & 0.0095 & 0.0133 & 0.0025 & 0.0074 & 0.0011 & 0.0103 \\
 \hline
  $k_t$ [ns$^{-1}$] & 0.0065 & 0.0069 & 0.0210 & 0.0057 & 0.0145 & 0.0048 \\
 \hline
  $k_0$ [ns$^{-1}$] & 5.64 $\times 10^{5}$ & 3.37 $\times 10^{5}$ & 7.06 $\times 10^{5}$ & 3.52 $\times 10^{5}$ & 7.54 $\times 10^{5}$ & 5.42 $\times 10^{5}$ \\
 \hline
  $k_0'$ [ns$^{-1}$] & 4.3 $\times 10^{3}$ & 2.05 $\times 10^{3}$ & 5.84 $\times 10^{3}$ & 5.29 $\times 10^{3}$ & 8.13 $\times 10^{3}$ & 3.96 $\times 10^{3}$ \\
 \hline
  $\Delta E$ [meV] & 25.77 & 25.07 & 22.99 & 25.45 & 21.97 & 27.20 \\
 \hline
  $p_0$ & 0.788 & 0.837 & 0.501 & 0.6659 & 0.5 & 0.7977 \\
 \hline
  $s_0$ & 0 & 0.0326 & 0.0844 & 0 & 0.0349 & 0.0196 \\
 \hline
  $s_1$ & 0.0121 & --- & 0.0211 & 0.0262 & 0.0097 & --- \\
 \hline
  $s_2$ & 0.0577 & --- & --- & --- & 0.0149 & 0.1159 \\
 \hline
  $s_3$ & 0.0254 & 0.0206 & 0.0676 & 0.0289 & --- & 0.0203 \\
 \hline
  $s_4$ & 0.0184 & 0.1280 & --- & 0.1157 & 0.0567 & 0.1039 \\
 \hline
  $s_5$ & 0.2667 & 0.0504 & 0.0587 & 0.0650 & 0.0198 & 0.0885 \\
 \hline
  $s_6$ & 0.0766 & 0.1243 & 0.0391 & --- & 0.1028 & 0.0451 \\
 \hline
  $s_7$ & 0.0450 & 0.0367 & 0.0281 & 0.1791 & 0.0551 & 0.0458 \\
 \hline
  $s_8$ & 0.0199 & 0.1354 & 0.0749 & 0.0380 & 0.0203 & 0.0735 \\
 \hline
  $s_9$ & 0.0587 & 0.0537 & 0.0742 & 0.0503 & 0.0956 & 0.0848 \\
 \hline
  $s_{10}$ & 0.0470 & --- & 0.0743 & 0.0642 & 0.0925 & 0.0861 \\
 \hline
  $p_1$ & 0.4218 & --- & 0.6027 & 0.4477 & 0.453 & --- \\
 \hline
  $p_2$ & 0.0398 & --- & --- & --- & 0.353 & 0.7418 \\
 \hline
  $p_3$ & 0.194 & 0.6054 & 0.3559 & 0.7497 & --- & 0.3583 \\
 \hline
  $p_4$ & 0.6883 & 0.2852 & --- & 0.9866 & 0.206  & 0.4254 \\
 \hline
  $p_5$ & 0.0098 & 0.3374 & 0.8660 & 0.2308 & 0.689 & 0.9737 \\
 \hline
  $p_6$ & 0.9832 & 0.0876 & 0.4701 & --- & 0.9777 & 0.5316 \\
 \hline
  $p_7$ & 0.9678 & 0.6663 & 0.6008 & 0.0846 & 0.124 & 0.5512 \\
 \hline
  $p_8$ & 0.4667 & 0.9178 & 0.1599 & 0.5823 & 0.470 & 0.2190 \\
 \hline
  $p_9$ & 0.0891 & 0.2467 & 0.2000 & 0.8456 & 0.016 & 0.1696 \\
 \hline
  $p_{10}$ & 0.921 & --- & 0.091 & 0.024 & 0.020 & 0.0573 \\
 \hline
  $\Gamma_1$ [s$^{-1}$] & 1.6$\times 10^{-3}$ & --- & 1.6$\times 10^{-3}$ & 1.6$\times 10^{-3}$ & 1.6$\times 10^{-3}$ & --- \\
 \hline
  $\Gamma_2$ [s$^{-1}$]&  9.3$\times 10^{-3}$ & --- & --- & --- & 9$\times 10^{-3}$ & 9$\times 10^{-3}$ \\
 \hline
  $\Gamma_3$ [s$^{-1}$]& 5.4$\times 10^{-2}$ & 5.4$\times 10^{-2}$ & 5.4$\times 10^{-2}$ & 5.4$\times 10^{-2}$ & --- & 5.4$\times 10^{-2}$ \\
 \hline
  $\Gamma_4$ [s$^{-1}$]& 3.2$\times 10^{-1}$ & 3.2$\times 10^{-1}$ & --- & 3.2$\times 10^{-1}$ & 2.9$\times 10^{-1}$ & 2.9$\times 10^{-1}$ \\
 \hline
  $\Gamma_5$ [s$^{-1}$]& 1.8 $\times 10^{0}$ & 1.8 $\times 10^{0}$ & 1.8 $\times 10^{0}$ & 1.8 $\times 10^{0}$ & 1.6 $\times 10^{0}$ & 1.6 $\times 10^{0}$ \\
 \hline
  $\Gamma_6$ [s$^{-1}$]& 10.8 $\times 10^{0}$ & 10.8 $\times 10^{0}$ & 10.8 $\times 10^{0}$ & --- & 9.5$\times 10^{0}$ &  9.5$\times 10^{0}$ \\
 \hline
  $\Gamma_7$ [s$^{-1}$]& 6.3 $\times 10^{1}$ &  6.3 $\times 10^{1}$ & 6.3 $\times 10^{1}$ & 6.3 $\times 10^{1}$ & 5.4$\times 10^{1}$ & 5.4$\times 10^{1}$ \\
 \hline
  $\Gamma_8$ [s$^{-1}$]& 3.7$\times 10^{2}$ & 3.7$\times 10^{2}$ & 3.7$\times 10^{2}$ & 3.7$\times 10^{2}$ & 3.1$\times 10^{2}$ & 3.1$\times 10^{2}$ \\
 \hline
  $\Gamma_9$ [s$^{-1}$]& 2.15$\times 10^{3}$ & 2.15$\times 10^{3}$ & 2.15$\times 10^{3}$ &  2.15$\times 10^{3}$ & 1.8$\times 10^{3}$ & 2.27$\times 10^{3}$ \\
 \hline
  $\Gamma_{10}$ [s$^{-1}$]& 1.26$\times 10^{4}$ & --- & 1.26$\times 10^{4}$ & 1.26$\times 10^{4}$ & 1$\times 10^{4}$ & 1$\times 10^{4}$ \\
 \hline

\end{tabular}
\end{center}

\newpage
\section*{Supplementary Note 8. Single PNC's emission spectra.}

\begin{figure}[h!]
\begin{minipage}[h]{0.5\linewidth}
\center{\includegraphics[width=1\linewidth]{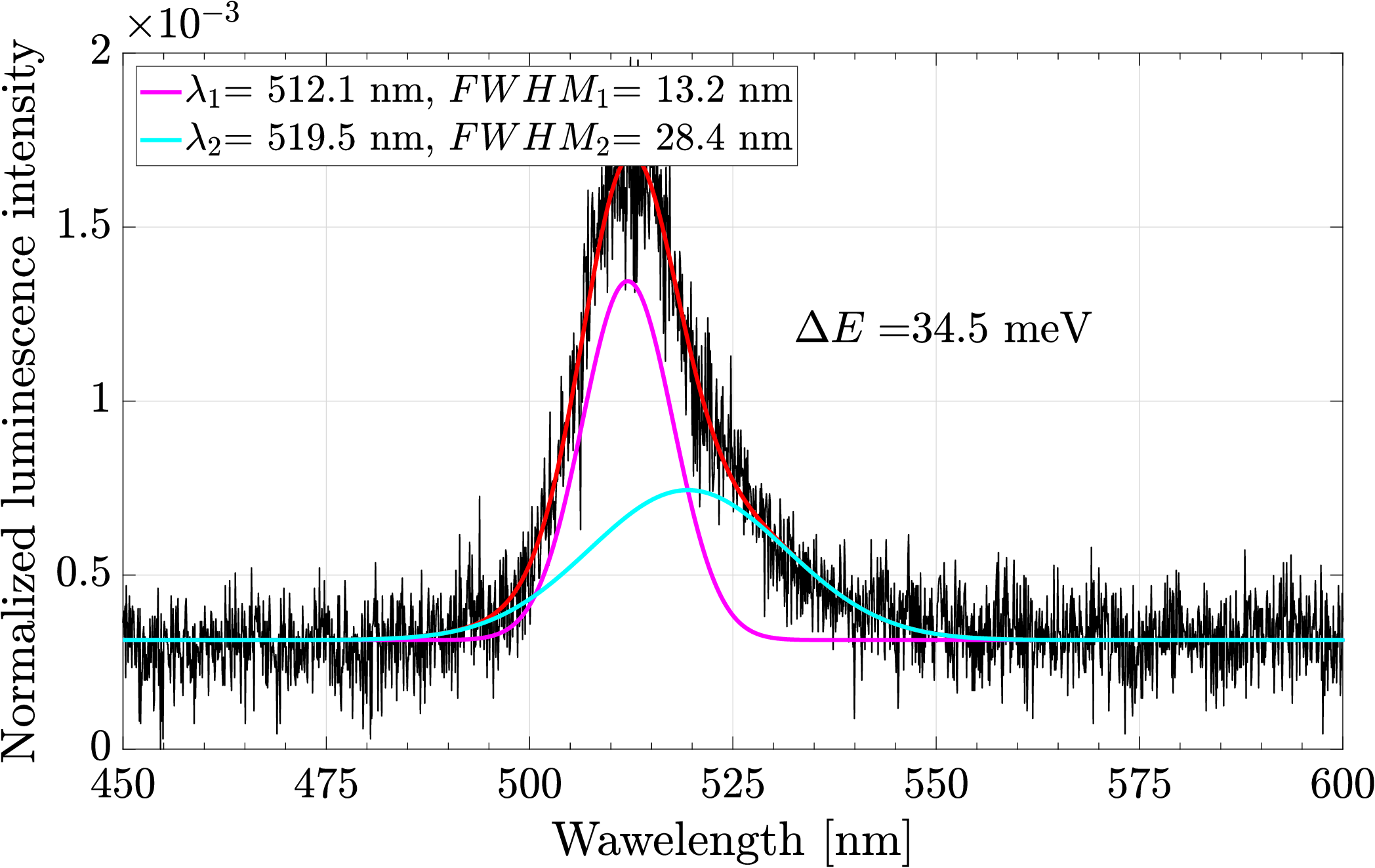}}
\end{minipage}
\hfill
\begin{minipage}[h]{0.5\linewidth}
\center{\includegraphics[width=1\linewidth]{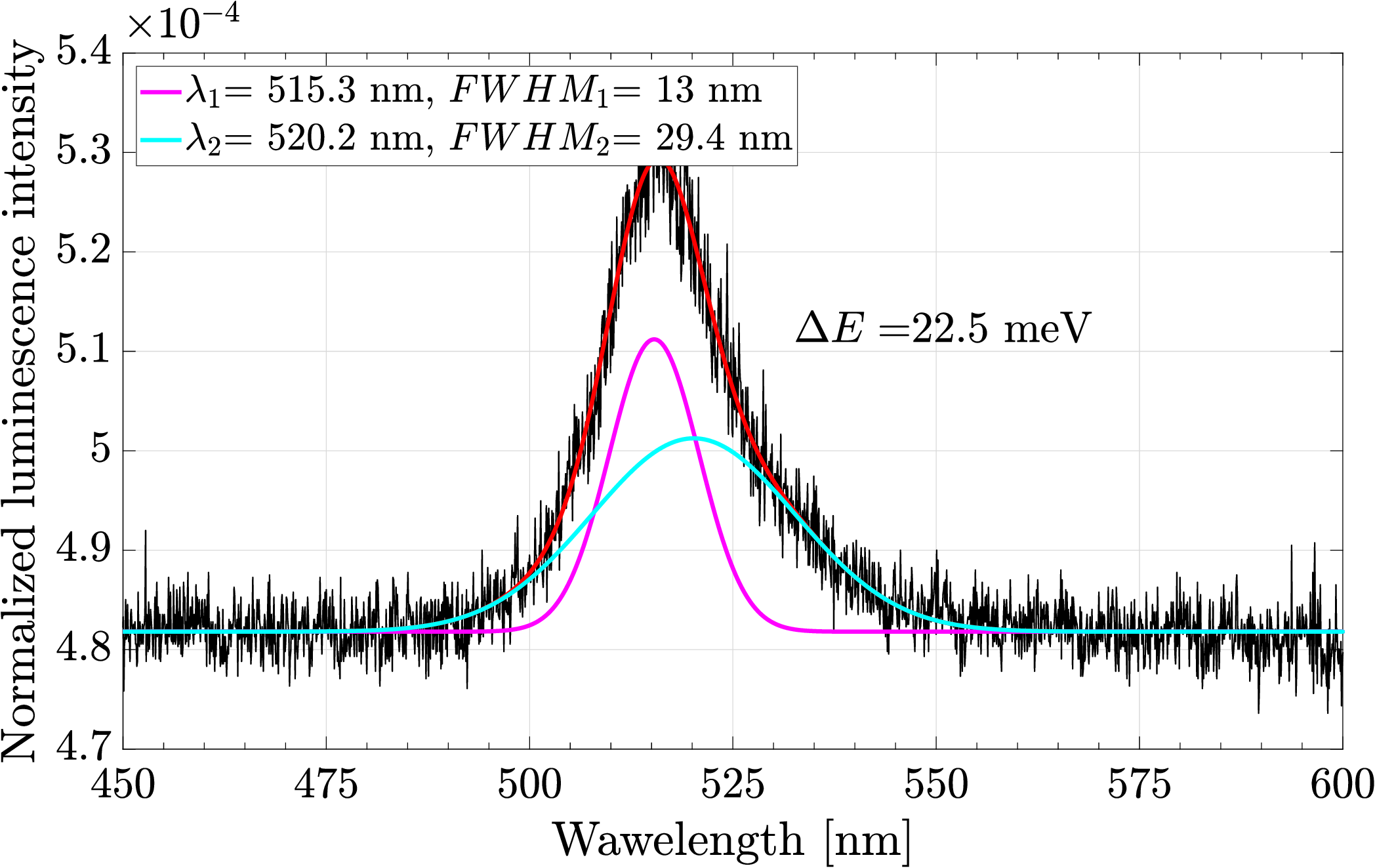}}
\end{minipage}
\vfill
\begin{minipage}[h]{0.5\linewidth}
\center{\includegraphics[width=1\linewidth]{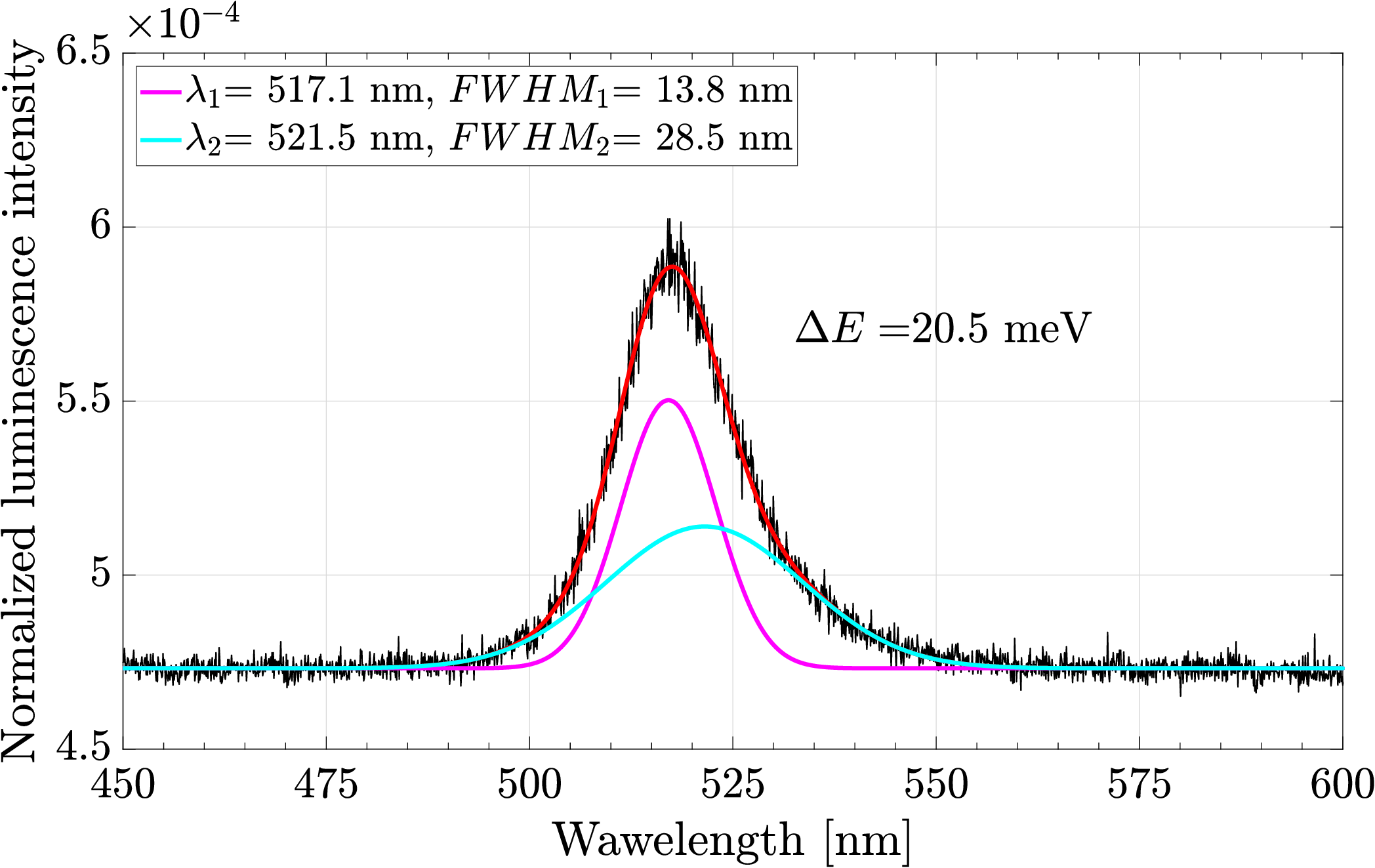}}
\end{minipage}
\hfill
\begin{minipage}[h]{0.5\linewidth}
\center{\includegraphics[width=1\linewidth]{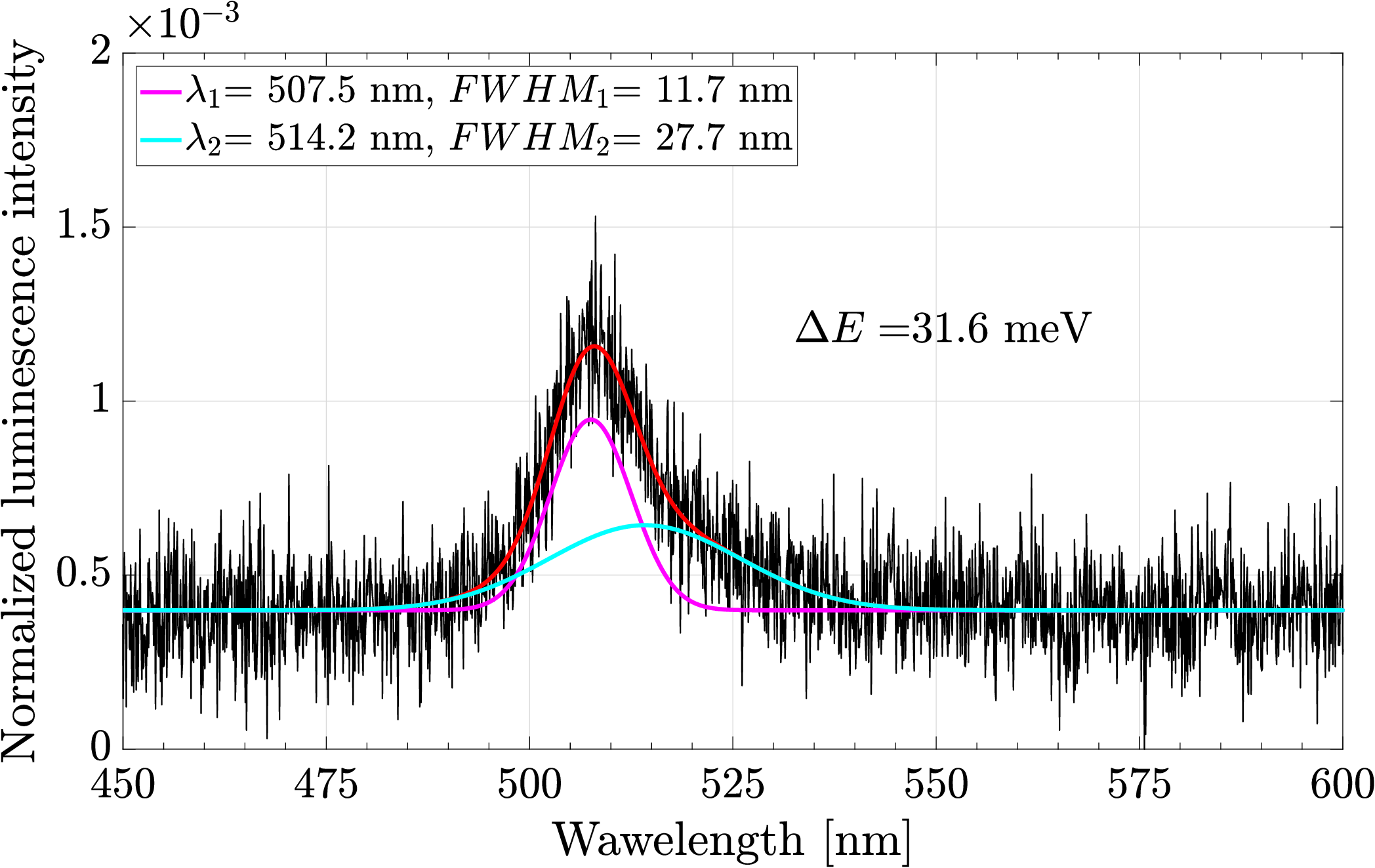}}
\end{minipage}
\vfill
\begin{minipage}[h]{0.5\linewidth}
\center{\includegraphics[width=1\linewidth]{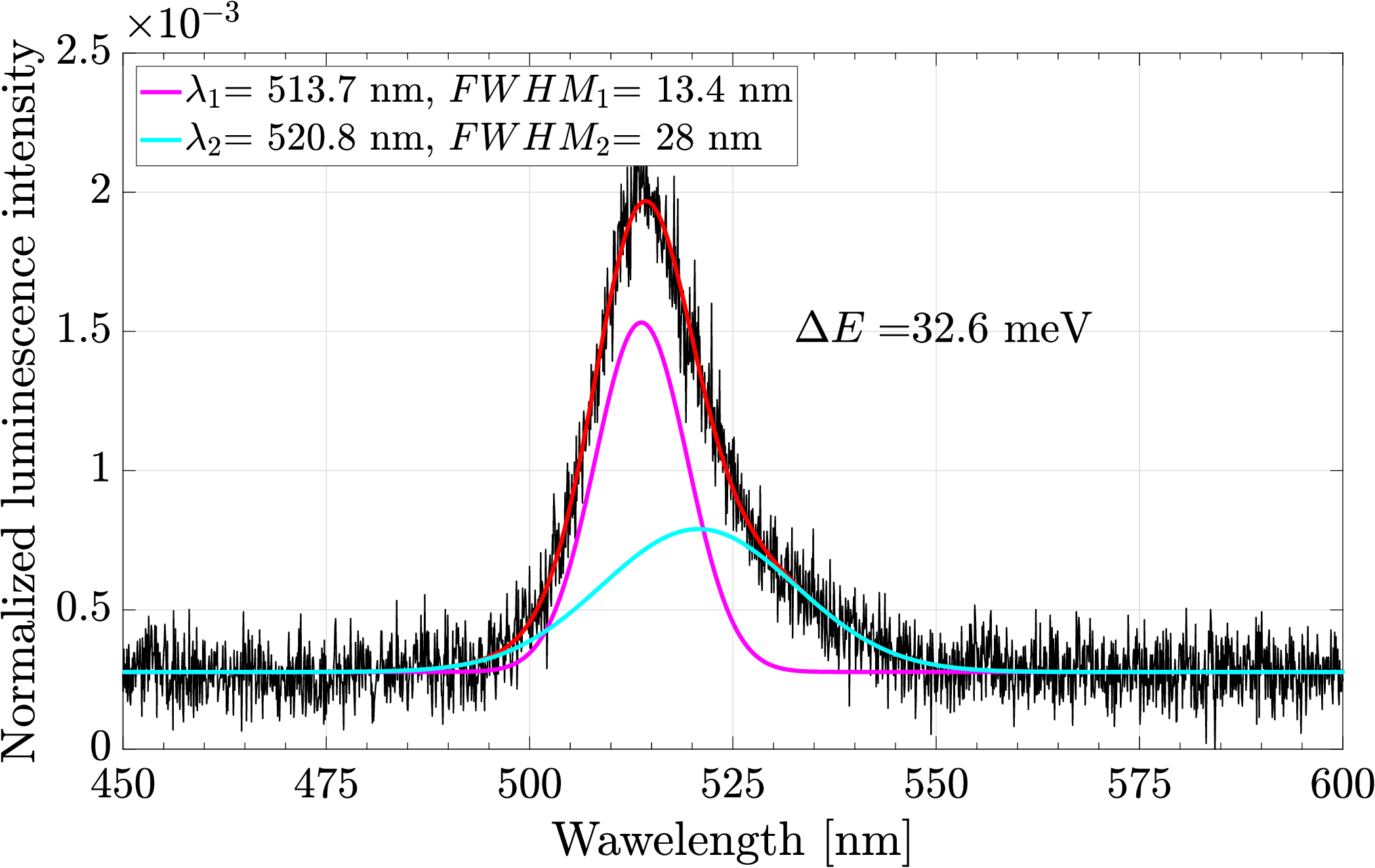}}
\end{minipage}
\hfill
\begin{minipage}[h]{0.5\linewidth}
\center{\includegraphics[width=1\linewidth]{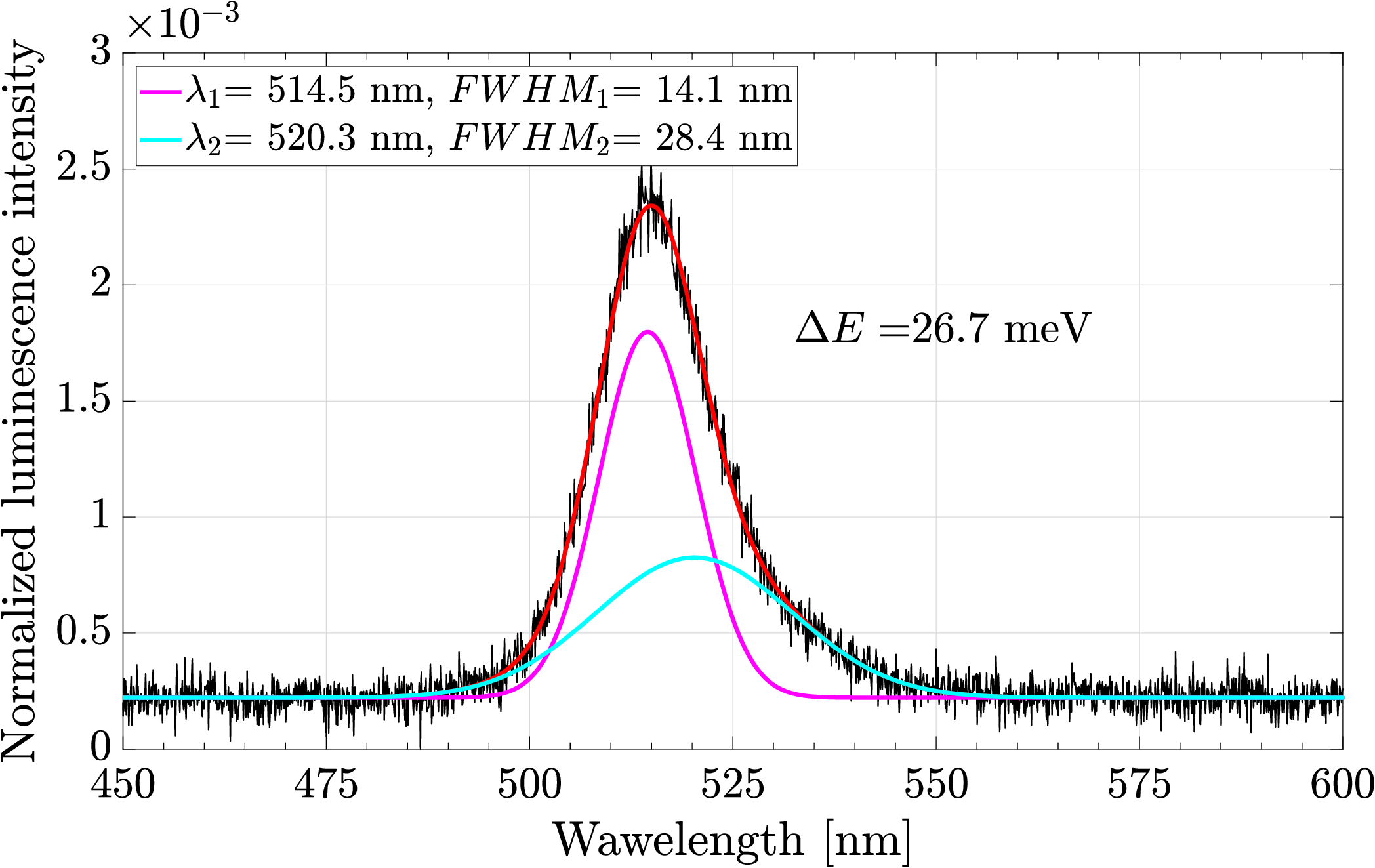}}
\end{minipage}
\vfill
\begin{minipage}[h]{0.5\linewidth}
\center{\includegraphics[width=1\linewidth]{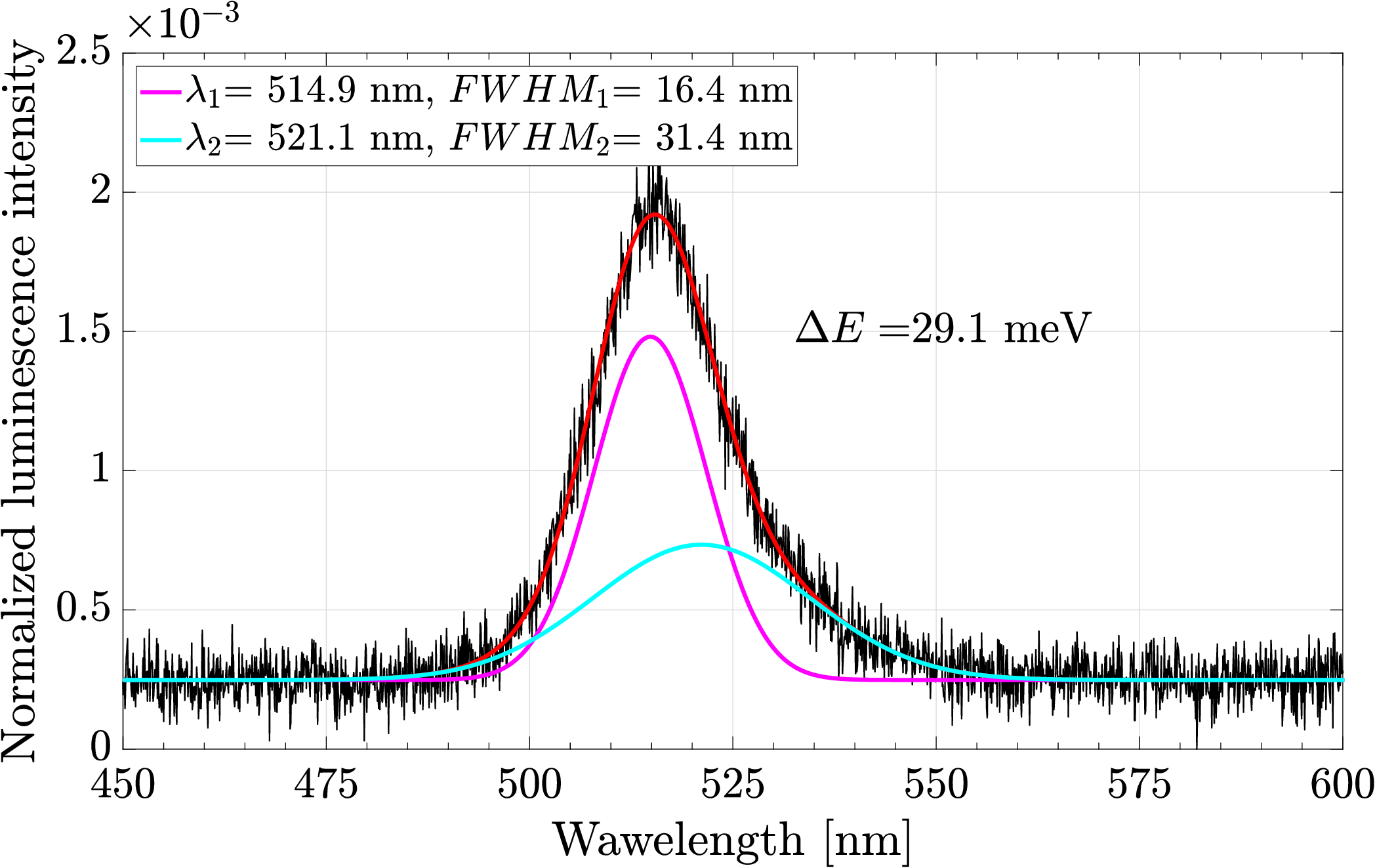}}
\end{minipage}
\hfill
\begin{minipage}[h]{0.5\linewidth}
\center{\includegraphics[width=1\linewidth]{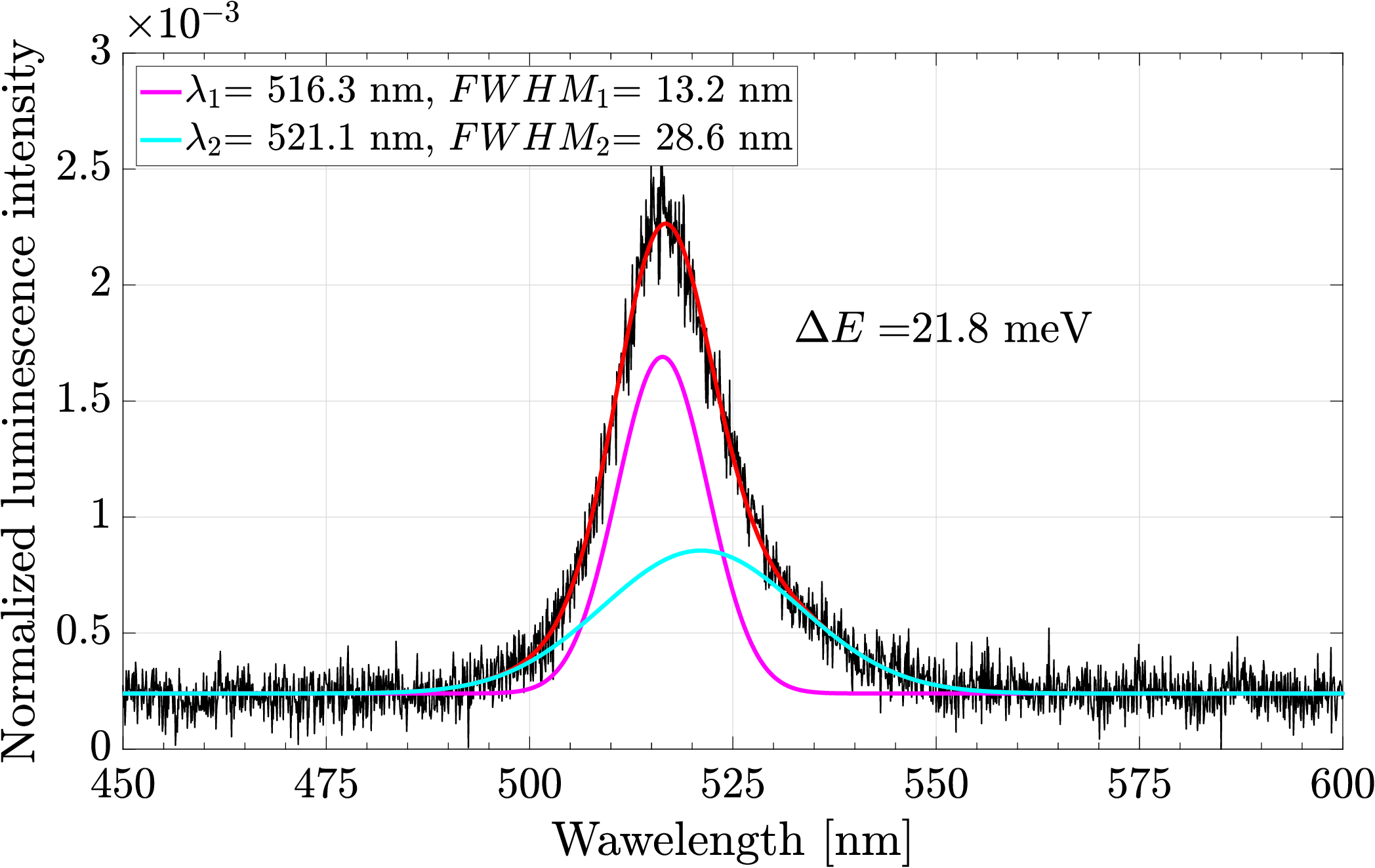}}
\end{minipage}
\caption{Experimental emission spectra of the single PNC (black line) and its fit using two gaussians (red line). Short wavelength and long wavelength components are marked with magenta and cyan lines, respectively.}
\label{Spec}
\end{figure}

\bibliography{Ref}
\bibliographystyle{rsc}